\PassOptionsToPackage{dvipsnames}{xcolor}
\documentclass[11pt]{article}
\usepackage{jheppub}

\usepackage[space]{grffile}
\usepackage{soul}
\usepackage{orcidlink}
\usepackage{pdfpages}
\usepackage{adjustbox}
\usepackage{slashed,physics}
\usepackage{graphicx}
\usepackage{amsmath,amssymb,graphicx}
\usepackage{textcomp} 
\usepackage{gensymb} 
\usepackage{epsf,color}
\usepackage{cancel}
\usepackage{framed}
\usepackage{hyperref}
\usepackage{float}
\usepackage{multirow}
\usepackage{subfigure}
\usepackage[compat=1.1.0]{tikz-feynman}
\usepackage{contour}
\usepackage{listings}

\lstnewenvironment{mg}[1][]{
    \lstset{
        basicstyle=\ttfamily,
        keywordstyle=\color{blue},
        commentstyle=\color{gray},
        stringstyle=\color{green},
        showstringspaces=false,
        breaklines=true,
        postbreak=\mbox{\textcolor{red}{$\hookrightarrow$}\space},
        #1
    }
}{}

\usepackage[normalem]{ulem}
\usetikzlibrary{shapes.geometric}
\tikzfeynmanset{warn luatex=false}
\usepackage{dsfont}

\usepackage{diagbox}
\usepackage{makecell}
\usepackage{colortbl}
\definecolor{colora1}{rgb}{0.996,0.890,0.569}
\definecolor{colora2}{rgb}{0.996,0.769,0.310}
\definecolor{colora3}{rgb}{0.984,0.604,0.161}
\definecolor{colora4}{rgb}{0.925,0.439,0.0784}
\definecolor{colora5}{rgb}{0.800,0.298,0.00784}

\definecolor{nicered}{rgb}{0.5,0.,0.}
\definecolor{nicegreen}{rgb}{0.,0.5,0.}
\definecolor{niceblue}{rgb}{0.,0.,0.5}
\hypersetup{colorlinks,citecolor=nicegreen,linkcolor=nicered,urlcolor=niceblue}
\numberwithin{equation}{section}
\newcommand{\beq}{\begin{equation}}
\newcommand{\eeq}{\end{equation}}
\newcommand{\bea}{\begin{eqnarray}}
\newcommand{\eea}{\end{eqnarray}}
\newcommand{\bear}{\begin{eqnarray}}
\newcommand{\eear}{\end{eqnarray}}

\newcommand{\ba}{\begin{array}}
\newcommand{\ea}{\end{array}}

\newcommand\aNLO{{\sc\small MadGraph5\_aMC@NLO}}
\newcommand\WZ{\sc\small Whizard}

\def\beq{\begin{equation}}
\def\eeq{\end{equation}}
\def\beqn{\begin{eqnarray}}
\def\eeqn{\end{eqnarray}}

\newcommand{\bqa}{\begin{eqnarray}}
\newcommand{\eqa}{\end{eqnarray}}
\chardef\MyArticleWithColor=\pdfcolorstackinit page direct{0 g}

\def\cCode#1{\begin{lstlisting}[mathescape,basicstyle=\small
\ttfamily,frame=leftline,aboveskip=4mm,belowskip=4mm,xleftmargin=20pt,framexleftmargin=10pt,
numbers=none,framerule=2pt,abovecaptionskip=0.0mm,belowcaptionskip=3.5mm #1]}

\newcommand{\gev}{\,\textrm{GeV}}

\newcommand\aNLOs{{\sc\small MG5\_aMC}}

\newcommand\sherpa{{\sc\small Sherpa}}

\newcommand\OL{{\sc\small OpenLoops}}

\newcommand{\LO}{{\rm LO}}

\newcommand{\ord}{{\cal O}}

\def\NLOEW{\rm NLO_{EW}}

\def\beq{\begin{equation}}
\def\eeq{\end{equation}}
\def\beqar{\begin{eqnarray}}
\def\eeqar{\end{eqnarray}}
\def\barr#1{\begin{array}{#1}}
\def\earr{\end{array}}
\def\bfi{\begin{figure}}
\def\efi{\end{figure}}
\def\btab{\begin{table}}
\def\etab{\end{table}}
\def\bce{\begin{center}}
\def\ece{\end{center}}

\newcommand{\M}{{\cal{M}}}

\def\mathswitchr#1{#1}
\newcommand{\PW}{\mathswitchr W}

\newcommand{\PZ}{\mathswitchr Z}

\newcommand{\PH}{\mathswitchr H}

\newcommand{\Pt}{\mathswitchr t}

\def\mathswitch#1{\relax\ifmmode#1\else$#1$\fi}
\newcommand{\MW}{\mathswitch {M_\PW}}
\newcommand{\MZ}{\mathswitch {M_\PZ}}
\newcommand{\MH}{\mathswitch {M_\PH}}

\newcommand{\Mt}{\mathswitch {m_\Pt}}

\newcommand{\lrMwithabs}{l(|r_{kl}|,M^2)}

\newcommand{\lsW}{l(s,\MW^2)}

\newcommand{\LrM}{L(|r_{kl}|,M^2)}

\newcommand{\Lrs}{L(|r_{kl}|,s)}

\newcommand{\LsW}{L(s,\MW^2)}

\newcommand{\lrsalpha}{l(|r_{kl}|,s)}

  \newcommand{\TO}{\rightarrow}

\newcommand{\denpoz}{{\sc\small DP}}

\newcommand{\deltaEW}{\delta^{\rm EW}_{\rm LA}}
\newcommand{\deltaQCD}{\delta^{\rm QCD}_{\rm LA}}

\newcommand{\SDKz}{{\rm SDK_0}}
\newcommand{\SDKw}{{\rm SDK_{weak}}}

\newcommand{\SigmaLO}{\sigma_{\rm LO}}
\newcommand{\SigmaNLO}{\sigma_{\rm NLO_{EW}}}
\newcommand{\SigmaNLOtwo}{\sigma_{\rm NLO_{2}}}

\newcommand{\Exp}{\rm EXP_{EW}}

\newcommand{\SigmaHBR}{\sigma_{\rm HBR}}
\newcommand{\SigmaSDKw}{\sigma_{\SDKw}}
\newcommand{\SigmaSDKz}{\sigma_{\SDKz}}
\newcommand{\SigmaExp}{\sigma_{\Exp}}

\newcommand{\SigmaNLOHBR}{\sigma_{\rm NLO_{EW}+HBR}}
\newcommand{\SigmaNNLOHBR}{\sigma_{\rm nNLO_{EW}+HBR_{\rm NLO}}}

\newcommand{\deltaNLOEW}{\delta_{\rm NLO_{EW}}}
\newcommand{\deltaHBR}{\delta_{\rm HBR}}
\newcommand{\deltaSDKw}{\delta_{\SDKw}}
\newcommand{\deltaSDKz}{\delta_{\SDKz}}
\newcommand{\deltaDL}{\delta_{\rm DL}}
\newcommand{\deltaSL}{\delta_{\rm SL}}
\newcommand{\deltastorkl}{\delta_{s\TO r_{kl}}}
\newcommand{\deltaExp}{\delta_{\Exp}}

\newcommand{\deltaNLOHBR}{\delta_{\rm NLO_{EW}+HBR}}
\newcommand{\deltaNNLOHBR}{\delta_{\rm nNLO_{EW}+HBR_{\rm NLO}}}

\newcommand{\EWjet}{j_{\rm EW}}
\newcommand{\EWjeto}{j_{\rm EW,1}}
\newcommand{\EWjett}{j_{\rm EW,2}}

\title {EW corrections and Heavy Boson Radiation at a high-energy muon collider}

\date{\today}

\author[a,b,\orcidlink{0000-0002-9419-6598}]{Yang Ma,}
\affiliation[a]{INFN,
Sezione di Bologna, via Irnerio 46, 40126 Bologna, Italy}
\affiliation[b]{Center for Cosmology, Particle Physics and Phenomenology, Universit\'e catholique de Louvain, B-1348 Louvain-la-Neuve, Belgium}

\author[a,\orcidlink{0000-0002-0553-1105}]{Davide Pagani,}

\author[c,\orcidlink{0000-0002-3279-7355}]{Marco Zaro}
\affiliation[c]{TIFLab, Universit\`a degli Studi di Milano \& INFN, Sezione di Milano, Via Celoria 16, 20133 Milano, Italy}

\emailAdd{yang.ma@bo.infn.it}
\emailAdd{davide.pagani@bo.infn.it}
\emailAdd{marco.zaro@mi.infn.it}

\preprint{\\TIF-UNIMI-2024-14, COMETA-2024-23, IRMP-CP3-25-08}
\arxivnumber{2409.09129}

\abstract{In this work we investigate several phenomenological and technical aspects related to electroweak (EW) corrections at a high-energy muon collider, focusing on direct production processes (no VBF configurations). We study in detail the accuracy of the Sudakov approximation, in particular the Denner-Pozzorini algorithm, comparing it with exact calculations at NLO EW accuracy. We also assess the relevance of resumming EW Sudakov logarithms (EWSL) at 3 and 10 TeV collisions. Furthermore, we scrutinise the impact of additional Heavy Boson Radiation (HBR), namely the weak emission of $W, Z$, and Higgs bosons in inclusive and semi-inclusive configurations. All results are obtained via the fully automated and publicly available code {\sc\small MadGraph5\_aMC@NLO}. }

\makeatletter
\gdef\@fpheader{}
\makeatother

\begin{document}
\maketitle

\section{Introduction}
\label{sec:intro}

In recent years a novel interest for a muon collider has arisen, motivated by its great potential for the investigation of the fundamental interactions of Nature \cite{MuonCollider:2022xlm,Aime:2022flm,Black:2022cth,Maltoni:2022bqs,Belloni:2022due,Accettura:2023ked}. A key aspect of a muon collider is the possibility of accelerating elementary particles at energies of several TeV's \cite{Delahaye:2019omf,Bartosik:2020xwr,Schulte:2021hgo,Long:2020wfp,MuonCollider:2022nsa,MuonCollider:2022ded,MuonCollider:2022glg}, leading to the possibility to probe fundamental interactions at unprecedented energies. 
It offers the potential to significantly advance our understanding of fundamental particle physics, enabling in-depth studies of the Standard Model (SM) Higgs boson \cite{Han:2020pif,Buttazzo:2020uzc,Han:2021lnp,Reuter:2022zuv,Celada:2023oji,Forslund:2022xjq,Chen:2022yiu,Ruhdorfer:2023uea,Han:2023njx,Dermisek:2023rvv,Liu:2023yrb,Li:2024joa,Cassidy:2023lwd}, 
searches for beyond-the-SM (BSM) heavy Higgs bosons \cite{Han:2022edd,Bandyopadhyay:2020otm,Han:2021udl,Bandyopadhyay:2024plc,Ouazghour:2023plc,Jueid:2023qcf}, 
investigations of dark matter \cite{Han:2020uak,Capdevilla:2021fmj,Han:2022ubw,Cesarotti:2024rbh,Asadi:2023csb,Belfkir:2023vpo,Jueid:2023zxx},
constraints on lepton-universality violation \cite{Huang:2021nkl,Huang:2021biu,He:2024dwh},
exploration of the muon $g-2$ anomaly \cite{Buttazzo:2020ibd,Capdevilla:2020qel,Yin:2020afe,Capdevilla:2021rwo}, 
and tests of a wide range of new physics scenarios \cite{Gu:2020ldn,Costantini:2020stv,Bandyopadhyay:2021pld,Qian:2021ihf,Bao:2022onq,Ghosh:2023xbj,Dermisek:2023tgq,Altmannshofer:2023uci,Liu:2023jta,Sun:2023cuf,Kwok:2023dck,Cesarotti:2023sje,Ema:2023buz,Li:2023tbx,Bhattacharya:2023beo,Das:2024ekt}.

A muon collider is therefore both a discovery and precision machine. In particular, precision physics relies on both a clean experimental environment (no QCD in the initial state) and precise and reliable theoretical predictions. Thus, as has been the case for LEP, the calculation of electroweak (EW) corrections will be paramount for theoretical predictions in muon collisions. In fact, unlike other possible future electron-positron colliders, {\it e.g.}~the Future Circular Collider (FCC-ee)~\cite{FCC:2018evy,Bernardi:2022hny} 
and the Circular Electron-Positron Collider (CEPC) \cite{CEPCStudyGroup:2018rmc,CEPCStudyGroup:2018ghi,An:2018dwb,CEPCAcceleratorStudyGroup:2019myu,CEPCPhysicsStudyGroup:2022uwl,CEPCStudyGroup:2023quu},
 at a high-energy muon collider the scope and relevance of EW corrections will be  much broader, especially for a 10 TeV (or higher) collision machine.\footnote{Similar effects, although of smaller sizes due to the associated lower energies, are expected also for high-energy $e^+e^-$ colliders (see {\it e.g.}~Ref.~\cite{Kuhn:2007ca}) as CLIC, ILC and $\rm C^3$ \cite{ILC:2013jhg,Behnke:2013lya,ILCInternationalDevelopmentTeam:2022izu,Linssen:2012hp, Lebrun:2012hj, CLIC:2016zwp, Brunner:2022usy,Bai:2021rdg, Vernieri:2022fae}. For such colliders, a more prominent effect originates from QED initial-state-radiation (ISR), which however in our set up would be automatically resummed within the PDF formalism.} EW corrections at high energies can be very large, even of $\ord(1)$ w.r.t.~the leading-order (LO) prediction, and therefore are expected to be unavoidable in any phenomenological study, not only those regarding precision. 
The origin of such enhancements is the so-called EW Sudakov logarithms (EWSL), which involve logarithms of ratios of the form $Q/\MW$, where $Q$ is any of the scale of the process, like the energy of the collider $\sqrt{S}$, and $\MW$ is the $W$ boson mass.

 On the one hand, such logarithms emerge from the real emission of heavy bosons $V=W, Z$ (and $H$). It has been shown that such mechanism can be exploited in order to leverage the sensitivity on BSM effects in the hard process \cite{Buttazzo:2020uzc, Chen:2022msz} and in general it has become a widespread notion that these effects will be ubiquitous in the muon collider physics \cite{AlAli:2021let}. Above all, the idea of the muon collider as a vector-boson collider has emerged, where Vector-Boson-Fusion (VBF) processes can be modelled directly via $VV$ initiated processes and the convolution of universal and process-independent parametrisations of the $V$ emission from initial-state muons \cite{Costantini:2020stv, Ruiz:2021tdt}. Different groups have already calculated and provided EW PDFs of $V$ bosons (and the other particles of the SM spectrum) in the muon \cite{Han:2020uid,Han:2021kes,Garosi:2023bvq}, resumming such effects.
 
On the other hand, EWSL originate from ``genuine'' EW corrections, {\it i.e.}, loop diagrams \cite{Sudakov:1954sw}. The calculation of such logarithms has also received a novel interest in the past years, independently from the muon-collider physics. An algorithmic procedure for the evaluation of EWSL at one- \cite{Denner:2000jv,Denner:2001gw} and two-loop \cite{Denner:2003wi,Denner:2004iz,Denner:2006jr,Denner:2008yn} accuracy,  the so-called Denner and Pozzorini ({\denpoz}) algorithm, has been available for a long time. Such algorithm has been automated for the first time \cite{Bothmann:2020sxm} in the {\sherpa} framework~\cite{Sherpa:2019gpd} and extended to the case of  multijet merging at the next-to-leading order (NLO). Afterward, the {\denpoz} algorithm has been revisited and improved in particular features~\cite{Pagani:2021vyk} and automated within the {\aNLO} framework  \cite{Alwall:2014hca, Frederix:2018nkq} and matched to NLO+PS simulations in QCD \cite{Pagani:2023wgc}. Very recently \cite{Lindert:2023fcu}, it has been automated also in the {\OL} framework \cite{Cascioli:2011va, Buccioni:2019sur}, and adapted for a dynamical treatment of resonances. 

The resummation of EWSL has also been studied and in Refs.~\cite{Chiu:2007yn,Chiu:2008vv, Manohar:2018kfx} a general method to resum
such logarithms for an arbitrary process was developed, based on the framework of 
soft-collinear effective
theory (SCET)~\cite{Bauer:2000ew,Bauer:2000yr,Bauer:2001ct,Bauer:2001yt}.  Very recently, in Ref.~\cite{Denner:2024yut}, NLL resummation has been implemented in realistic Monte Carlo simulations and studied both in the context of future leptonic (CLIC at 3 TeV~\cite{Aicheler:2012bya,Linssen:2012hp,Lebrun:2012hj,CLIC:2016zwp,CLICdp:2018cto,Brunner:2022usy}) and hadronic (FCC-hh at 100 TeV \cite{FCC:2018byv,FCC:2018vvp,Benedikt:2022kan})   colliders.

One of the reasons for the novel interest in the computation of EWSL is the fact that they can in principle approximate very well the exact NLO EW corrections, but their evaluation does not involve the explicit computation of loop diagrams; only tree-level amplitudes and logarithms are involved. Thus, the evaluation of EWSL is much faster and easier. Besides the SM scenario, their evaluation has been performed also for BSM scenarios, such as Dark-Matter studies \cite{Ciafaloni:2010ti}, and it is clearly relevant in the context of a high-energy muon collider, as shown {\it e.g.} in the already mentioned Refs.~\cite{Buttazzo:2020uzc, Chen:2022msz}. 

Both for the SM and BSM case, one should keep in mind that the EWSL are an approximation. Knowing how efficient is this approximation is of primary relevance for physics at a high-energy muon collider. However, nowadays also the automation of the exact NLO EW corrections is available for SM processes, both in hadronic and leptonic collisions~\cite{Alwall:2014hca, Kallweit:2014xda, Frixione:2015zaa, Chiesa:2015mya, Biedermann:2017yoi, Chiesa:2017gqx,  Frederix:2018nkq, Pagani:2021iwa,Bertone:2022ktl,Bredt:2022dmm}.\footnote{\label{foot:cNLO}More in general, the calculation of the so called Complete-NLO has been automated. This accuracy includes NLO QCD and NLO EW corrections and also formally subleading contributions in the $\alpha_s$ and $\alpha$ power expansion, see {\it e.g.}~Refs.~\cite{Frixione:2014qaa, Frixione:2015zaa, Pagani:2016caq, Frederix:2016ost, Czakon:2017wor, Frederix:2017wme, Frederix:2018nkq, Broggio:2019ewu, Frederix:2019ubd,Pagani:2020rsg, Pagani:2020mov, Pagani:2021iwa, Pagani:2021vyk, Maltoni:2024wyh}.}
Thus, it is possible to compare directly EWSL and exact NLO EW corrections in order to assess their level of accuracy for SM processes.

In this work, we precisely investigate this issue for the case of a muon collider at 3 and 10 TeV. We focus on direct production processes (also denoted in the literature as muon-muon annihilation), $\mu^+\mu^-\to F$, where the invariant mass of the final state is close to the energy of the collider and therefore VBF configurations are suppressed. In these configurations the only relevant PDFs are the ones of the (anti)muon in the (anti)muon, which accounts for effects from QED initial-state radiation (ISR). Moreover, our focus is not on precision physics but on large effects, as those expected from EWSL at high energies. We exploit the SM as a ``test case'', but our conclusions are instructive also for a general BSM scenario.\footnote{The case of SMEFT will be addressed in detail in an upcoming publication \cite{ElFaham:2024egs}.} All our calculations are performed via  the {\aNLO} framework \cite{Alwall:2014hca, Frederix:2018nkq,Pagani:2021vyk, Bertone:2022ktl}, in a completely automated approach.

First of all, we investigate in detail how accurate is the Sudakov approximation and in particular its evaluation via the  {\denpoz} algorithm, comparing it with exact NLO EW results. In particular, the  {\denpoz} algorithm has been rigorously derived for the approximation of one-loop amplitudes in the strict limit $\MW^2/s\to 0$, but its application for physical observables is less straightforward and has been revisited in  Ref.~\cite{Pagani:2021vyk}. On the one hand, it has shown that the usage of the so-called $\SDKw$ scheme, a purely weak version of the original one, can be superior to the more commonly used $\SDKz$ scheme when charged particles in the final state are recombined with photons. On the other hand, it has shown that logarithms such as $\log^2(s/|t|)$ or $\log^2(s/|u|)$, and in general logarithms involving ratios of Lorentz invariants of the process, can be numerically very large and the assumptions as $s\simeq|t|\simeq|u|$ are not efficient. We address both these aspects when comparing EWSL and exact NLO EW corrections. Moreover, we show a specific case for a process ($\mu^+\mu^-\to ZHH$ production) where in some regions of the phase space the LO predictions are numerically dominated by diagrams that are mass-suppressed in the  $\MW^2/s$ expansion, such that the {\denpoz} algorithm lies outside its range of applicability and therefore returns wrong results. Such an example is very counterintuitive, and the Sudakov approximation is very efficient in the region where one would not naively expect it, and {\it vice versa}. 

Then we approximate the resummation of EWSL via a simple exponentiation and, after additively matching them to the exact NLO EW, we compare this prediction with the exact NLO EW itself. We investigate when resummation is needed solely for precision studies and when it becomes essential to ensure sensible predictions and prevent negative cross sections.

Finally, we scrutinise the contribution of Heavy-Boson-Radiation (HBR), in other words the emission of weak bosons $W,Z$, and $H$. We consider different scenarios, where HBR is recombined or not with particles in the final state $F$ in $\mu^+\mu^-\to F$ production. We also consider as physical objects in the final state $F$ the so-called ``EW jets'', obtained by the clustering of $W$ and $Z$ bosons. We compare the impact of HBR, which is calculated exactly and taking into account phase-space cuts, with the one of NLO EW corrections, also calculated exactly,  and discuss their relative size and possible cancellations. We show how HBR leads in general to much smaller contributions than their virtual counterparts, only marginally compensating for the large effects due to the latter.
   
The paper is structured as described in the following. In Sec.~\ref{sec:NLOandEWSL} we briefly summarise the automation of both NLO EW corrections and EWSL in {\aNLO}, focusing on the aspects relevant to the study presented in this work. In Sec.~\ref{sec:DirectTheoFrame} we describe in detail our calculation setup and the definitions of the different approximations used. Also, we better formalise the aspects that we want to investigate in this work and that we have mentioned in the previous paragraphs. All the numerical results and the discussion of the information obtained are reported in Sec.~\ref{sec:numres}. We give our conclusions in Sec.~\ref{sec:conclusions}.

\section{NLO EW corrections and EWSL in {\sc MadGraph5\_aMC@NLO}}
\label{sec:NLOandEWSL}

\subsection{The automation of NLO EW corrections \label{sec:NLOauto}}

Given a physical observable, typically a cross section, the so-called NLO QCD and NLO EW corrections correspond to the exact $\ord(\alpha_s)$ and $\ord(\alpha)$ corrections, respectively,  to its LO prediction\footnote{For the processes that we will consider in this paper, a single coupling combination contributes at LO. See Footnote~\ref{foot:cNLO} for the more general case.}. Such corrections involve the calculation of one-loop amplitudes, their renormalisation, the regularisation of IR divergencies, and the combination of virtual as well as real-emission contributions in order to cancel them.

The automatic computation of NLO QCD and EW corrections, and the matching of the former to parton-shower, 
is a well-known feature of the metacode {\sc MadGraph5\_aMC@NLO}, achieved both  for hadronic~\cite{Frederix:2018nkq,Pagani:2021iwa} and leptonic~\cite{Bertone:2022ktl} collisions.  Before delving into aspects specific to muon colliders, we remind the reader of some general features about the building blocks of the code. The computation of NLO corrections requires the local subtraction of IR singularities and the numerical evaluation of one-loop amplitudes. The first task is achieved using the FKS subtraction scheme~\cite{Frixione:1995ms,Frixione:1997np} as implemented in {\sc MadFKS}~\cite{Frederix:2009yq,Frederix:2016rdc}. The second task builds upon a number of different numerical techniques (integrand reduction~\cite{Ossola:2006us}, tensor-integral reduction~\cite{Passarino:1978jh,Davydychev:1991va,Denner:2005nn}, Laurent-series expansion~\cite{Mastrolia:2012bu}, and an in-house
implementation of the {\sc OpenLoops} method~\cite{Cascioli:2011va}), implemented in publicly-available software libraries~\cite{Ossola:2007ax, ShaoIREGI,Peraro:2014cba,Hirschi:2016mdz,Denner:2014gla,Denner:2016kdg}, all steered 
by the {\sc MadLoop} module~\cite{Hirschi:2011pa}. Matching to parton shower (not relevant for the work in this paper) is performed using the
MC@NLO method~\cite{Frixione:2002ik}.

The capabilities of {\sc MadGraph5\_aMC@NLO} in the computation of EW corrections at
hadron colliders have been documented in a number of papers~~\cite{Frixione:2014qaa, Frixione:2015zaa, Pagani:2016caq, Frederix:2016ost, Czakon:2017wor, Czakon:2017lgo, Frederix:2017wme, Frederix:2018nkq, Broggio:2019ewu, Frederix:2019ubd, Pagani:2020rsg, Pagani:2020mov, Pagani:2021iwa, Pagani:2021vyk, Frederix:2021agh, Frederix:2021zsh}. 
For what concerns EW corrections at lepton-lepton colliders, either electron-positron
or muon-antimuon ones, far fewer results are available, 
and are limited only to the case of electron-positron colliders. We will first review these results, and then comment on
how to extend them at muon colliders. For electron-positron
colliders, in {\sc MadGraph5\_aMC@NLO} effects due to initial-state radiation (ISR) are
included in a collinear-inspired picture, {\it i.e.}, using quantities
analogous to the partonic density functions (PDFs) at hadron colliders. 
At variance with their hadronic counterpart, leptonic PDFs are 
perturbative and can thus be computed
via first principles. This requires the knowledge of their initial conditions, on which one applies the DGLAP evolution. The computation
of NLO initial conditions \cite{Frixione:2019lga} has led
to the availability of leptonic PDFs whose accuracy is 
 Next-to-Leading-Logarithmic (NLL)~\cite{Bertone:2019hks,Frixione:2012wtz, Bertone:2022ktl}.
In Ref.~\cite{Bertone:2022ktl} in particular, the dependence on physical cross sections on renormalisation and factorisation schemes has been
thoroughly scrutinised for a selection of lepton-initiated processes. This required the computation of the corresponding cross sections at NLO accuracy, which has been performed using a new version (now public) of {\sc MadGraph5\_aMC@NLO}. The most relevant difference w.r.t. the 
hadronic case, which required adaptation of the phase-space integration, 
stems from the asymptotic behaviour of lepton PDFs at large values of the
Bjorken $x$ variable. While hadronic PDFs typically vanish in the 
limit $x\to 1$, leptonic PDFs feature an integrable singularity in the same limit:
\begin{equation}
\lim_{x\to 1} \Gamma_{e^-/e^-}(x,Q^2) = C (1-x)^{\beta -1}\,,
\end{equation}
where $\beta\sim 0.05$ (more details will be given
in Appendix~\ref{sec:isr}). Owing to the (distribution) identity
\begin{equation}
(1-x)^{\beta -1} = \frac{1}{\beta}\delta(1-x) + \left.\frac{1}{1-x}\right|_+ + \mathcal O (\beta)\,,
\end{equation}
one can easily see that the bulk of the cross section
comes
from regions where $x\to 1$.\footnote{We clearly assume that there are no other enhancement effects, such as the direct production of a new resonant heavy state.}
The peculiar dependence of the leptonic PDFs required some changes in {\sc MadGraph5\_aMC@NLO},
in order to have an efficient numerical integration: the first, trivial, is to flatten out the integrable divergence 
via a suitable change of integration variables~\cite{Frixione:2021zdp}. The second one
specific to the computation of NLO EW corrections,
is to devise a phase-space mapping where the event
and its soft/collinear counterterms are evaluated
at the same values of the Bjorken $x$. Such a new mapping is
documented in the appendix of Ref.~\cite{Bertone:2022ktl}.
Finally, the existence of NLL densities in different factorisation schemes requires the inclusion in the short-distance cross section of additional terms (finite contributions to the initial-state counterterms). The same applies to the case when LL-accurate PDFs are employed, in order to attain formal NLO accuracy.

Turning specifically to muon colliders, in principle most of what has been achieved for electron-positron ones can be trivially extended. Two caveats here are in order, both related to the higher energy of which muon colliders are capable: first, the
Bjorken-$x$ range is extended towards much smaller values, a fact that leads to enhancements of partonic channels that
would be otherwise suppressed. Indeed, at small $x$ ($x\lesssim 10^{-2}$ for a 10 TeV muon collider) on top of densities related to purely QED-interacting partons (photons and singlet contribution, related {\it e.g.}~to the positron inside the electron), also those of QCD-interacting ones (quarks and gluons) can lead to non-negligible contributions~\cite{Han:2020uid,Han:2021kes,Aime:2022flm,BuarqueFranzosi:2021wrv}. Moreover, also the contributions from $W, Z$, and Higgs bosons as well as neutrinos PDFs 
have been studied \cite{Han:2020uid,Han:2021kes,Garosi:2023bvq} and found to be relevant, especially for very high energies. Second, effects due to EW corrections are sizeable (typically much larger than 10\% of the LO and reaching even more than 100\% in absolute value), and their inclusion is mandatory even for $\ord(1)$ estimates of the cross sections.

In view of these facts, and considering that this work represents a starting point in the study, within the {\sc MadGraph5\_aMC@NLO} framework, of EW effects at muon colliders, we will focus on the kinematics region
where the non-singlet muon density dominates, {\it i.e.}~large Bjorken-$x$. This will be achieved by an invariant-mass cut
on the final-state products. Besides, also considering
the current unavailability of NLL PDFs for muon colliders,  we will not discuss 
effects due to the renormalisation- or factorisation-scheme employed in the PDFs. Given their size at electron-positron
colliders, see Ref.~\cite{Bertone:2022ktl}, their effects are expected to be negligible w.r.t.~the size of the EW corrections that we calculate and discuss in this work.

\subsection{EWSL: the implementation of the {\denpoz} algorithm}

We recall in this section the main features of the {\denpoz} algorithm \cite{Denner:2000jv,Denner:2001gw}, and its revisitation presented in Ref.~\cite{Pagani:2021vyk}, as implemented in {\aNLO}. Many more details can be found in Ref.~\cite{Pagani:2021vyk} and, part of them, also in the Appendix A of Ref.~\cite{Pagani:2023wgc}.

\subsubsection{Amplitude level}
 \label{sec:EWSLamp}

When  the high-energy limit $s\gg \MW^2$ is considered,
the {\denpoz} algorithm allows for the calculation of the leading contributions of the one-loop EW corrections of a generic SM scattering amplitude. These contributions are denoted as the ``leading approximation'' (LA), which  consists of double-logarithmic (DL) and single-logarithmic (SL) corrections, both from IR and UV origin, of the form 
\beq
\LrM\equiv\frac{\alpha}{4\pi}\log^2{\frac{|r_{kl}|}{M^2}} \qquad{\rm and}\qquad 
\lrMwithabs\equiv\frac{\alpha}{4\pi}\log{\frac{|r_{kl}|}{M^2}}\,. \label{eq:generallogs}
\eeq
Such logarithms are precisely what we denote as the EWSL.
In Eq.~\eqref{eq:generallogs}, $r_{kl}$ is a generic kinematic invariant $r_{kl}\equiv(p_k+p_l)^2$ involving the momenta of a pair of external particles (all momenta defined as incoming) and $M$ is any of the masses of the SM heavy particles ($\MW, \MH,$  $\Mt$ and $\MZ$) or the IR-regularisation scale $Q$, for the case of purely QED contributions involving photons. 

Via the {\denpoz} algorithm it is possible to calculate in LA one-loop EW corrections of a generic SM scattering amplitude $\M$, which are typically denoted as $\delta \M$, to the Born approximation, which is instead typically denoted as  $\M_0$. For any individual helicity configuration of the amplitude $\M$, the {\denpoz} algorithm allows to write $\delta \M$ as a function of the logarithms in Eq.~\eqref{eq:generallogs}, the couplings of each external field to the gauge bosons (and another possible field)  or associated quantities such as electroweak Casimir operators, and {\it tree-level} amplitudes as $\M_0$ or similar ones with one or two of the external fields replaced by, {\it e.g.}, $SU(2)$ partners w.r.t.~the case of $\M_0$. This is precisely at the origin of the resurgence of the interest in EWSL and the {\denpoz} algorithm in the past few years: EWSL can be computed in a much faster and more stable way than the exact NLO EW corrections and this approach can be (supposedly) extended to the BSM case, capturing the leading corrections at high energies.

However, there are a few crucial assumptions that underly the derivation of the  {\denpoz} algorithm and we list them in the following:

\begin{itemize}

\item External legs must be on-shell.

\item All the invariants are much larger in absolute value than the typical EW scale, namely,
\beq \label{eq:Sudaklim} 
|r_{kl}|\equiv|(p_k+p_l)^2| \simeq |2p_kp_l| \gg \MW^2 \simeq \MH^2,\Mt^2,\MW^2,\MZ^2.
\eeq
Therefore the case of resonant decays is excluded.\footnote{In fact, in the case of resonances, the process before decays should be considered, and in order to cover the full on-shell and off-shell region an approach as the one presented recently in Ref.~\cite{Lindert:2023fcu} should be used.}

\item For the helicity configuration considered, in the high-energy limit, the tree-level amplitude $\M_0$ must {\it not} be mass-suppressed  by powers of the form $(\MW/\sqrt{s})^k$ with $k>0
$. In other words, by dimensional analysis, a $2\TO n$ process requires that $[\M]=(\gev)^{2-n}$ and therefore 
\beq
 \M\propto s ^{\frac{2-n}{2}}\, ,  \label{eq:masssuppr}
\eeq
with no extra $(\MW/\sqrt{s})^k$ powers. 
\end{itemize}

An additional assumption is also present in the strict LA as derived in Refs.~\cite{Denner:2000jv,Denner:2001gw}, namely, if a specific $r_{k'l'}$ in a given process is considered, then the condition 
\beq
|r_{k l}|/| r_{k' l'} |\simeq 1\,,
\eeq
is always assumed, such that logarithms of the form $\log(r_{k l}/ r_{k' l'} )$ are always discarded unless they multiply other logarithms of the kind in \eqref{eq:generallogs}.

The last point has been addressed in detail in Ref.~\cite{Pagani:2021vyk}, where it has been shown~\footnote{For what concerns amplitudes, this is one of the two main innovations presented in Ref.~\cite{Pagani:2021vyk}. The other is the identification of a missing imaginary component, which is relevant for processes of the form $2\TO n$ with $n\ge3$ and was omitted in the original derivation of the {\denpoz} algorithm.} that not only the logarithms of the form $\LsW$ and $\lsW$, but also those of the form
$L(r_{k l},r_{k' l'})$ and $l(r_{k l},r_{k' l'})$
can be relevant, especially when 
\beq
|r_{k l}|\gg |r_{k' l'}|\gg \MW^2\,.
\eeq

The former two kinds of logarithms yield the formal LA as presented in Refs.~\cite{Denner:2000jv,Denner:2001gw}, while the latter ones have been reintroduced in the {\denpoz} algorithm in the revisitation in Ref.~\cite{Pagani:2021vyk} and they have been denoted as the $\Delta^{s\TO r_{kl}}$ contribution therein. Afterward, they have been employed in the literature also in Ref.~\cite{Bothmann:2021led} and in Ref.~\cite{Lindert:2023fcu}, where they have been denoted as the sub-subleading soft collinear corrections (S-SSC)  beyond the strict LA. 
Unlike the strict LA,  the $\Delta^{s\TO r_{kl}}$ has not been derived via formal arguments and in principle some logarithms of the same form may be missed; they have to be checked case by case, but so far all the comparisons with the exact calculation of virtual corrections, presented {\it e.g.}~in Refs.~\cite{Pagani:2021vyk,Lindert:2023fcu}, have shown a (sometimes dramatic) improvement in the agreement of exact calculation and the EWSL approximation when they are included.
Therefore, for simplicity, we will refer in the following to LA regardless of the inclusion or not of the $\Delta^{s\TO r_{kl}}$ terms.

Before considering the case of squared matrix elements and cross sections it is important to note that, as we have already said, the logarithms in Eq.~\eqref{eq:generallogs} can be of the form $\log(|r_{k l}|/Q^2) $ or,  using a fictitious photon mass $\lambda$ as an infrared regulator as done in Refs.~\cite{Denner:2000jv,Denner:2001gw}, of the form $\log(\MW^2/\lambda^2)$. Needless to say, such quantity and consequently $\delta \M$ is IR-divergent and therefore non-physical, similar to the virtual corrections without any approximation. A prescription or further additional steps are therefore necessary and discussed in the next subsection.

\subsubsection{Cross-section level}
\label{sec:SudXS}
What has been discussed in the previous section is here extended and projected to the case of squared matrix elements and especially to the cross-section level. We will focus here on the case of the muon collider and will consider only processes that are of purely EW origin. 

The squared amplitude $|\M|^2$ of a given process can be directly linked to the fully differential cross section $\Sigma$. For brevity, we will consider only $\Sigma$ in the following discussion, again more details can be found in  Refs.~\cite{Pagani:2021vyk,Pagani:2023wgc}.
If we denote  the LO prediction of $\Sigma$ as $\Sigma^{}_{\LO}$ and  its  purely virtual NLO EW corrections as $\Sigma^{\rm virt}_{\NLOEW}$, in the LA  we obtain
\beq
(\Sigma^{\rm virt}_{\NLOEW})\Big|_{\rm LA}=\Sigma^{}_{\LO} \deltaEW \,, \label{eq:LONLOviadelta}
\eeq
with
\beq
\deltaEW\equiv \frac{2\Re(\M_0\delta\M^*)}{|\M_0|^2}\,, \label{eq:deltaEW}
\eeq
where  $\M_0$ is  the amplitude that once squared leads precisely to  $\Sigma^{}_{\LO}$.

Since $\deltaEW$ is an approximation of the relative virtual EW corrections to the LO, it involves photons and therefore it is IR-divergent and non-physical as the quantity $\delta\M$. Thus, it cannot be used, as it is, for a comparison with the exact NLO EW corrections, which instead involve also real emission contributions and are IR safe. Such comparison is however one of the main aspects that we want to investigate in this paper in the context of muon-collider physics. For this purpose, it is first of all useful to distinguish  three different schemes for the calculation of  the EWSL, specifying their relation  with  NLO EW corrections for physical cross sections:

\begin{itemize}
\item SDK: The SDK scheme is a very good approximation at high energies for one-loop amplitude and virtual contributions, but it cannot be used alone for phenomenological predictions. It corresponds to the usage of {\denpoz} algorithm, which was derived for amplitudes and not directly for cross sections. It may or may not include the $\Delta^{s\TO r_{kl}}$ contributions for approximating the logarithms of the form $\log(|r_{k l}|/| r_{k' l'} |)$. In practice, it is what has been discussed so far in this section.

\item $\SDKz$: It corresponds to a procedure that in the past has been used in the literature in order to remove IR singularities from the SDK scheme, allowing for predictions for physical observables. The notation $\SDKz$ has been introduced in Ref.~\cite{Pagani:2021vyk} and, as explained therein, this approach is mostly driven by simplicity. The problem of IR finiteness is bypassed by removing some QED logarithms that involve $\MW$ and the IR scale. However,  such logarithms arise due to the conventions used in Refs.~\cite{Denner:2000jv,Denner:2001gw}. In first approximation, it is equivalent to include QED radiation up to the scale $\MW$, which is not a physical argument unless the simulation of such radiation above this scale is also included, as done for instance in Ref.~\cite{Lindert:2023fcu}. It may or may not include the $\Delta^{s\TO r_{kl}}$ contributions for approximating the logarithms of the form $\log(|r_{k l}|/| r_{k' l'} |)$.
\item $\SDKw$: This scheme has been presented in Ref.~\cite{Pagani:2021vyk} precisely with the aim of solving the problematics of the $\SDKz$ scheme. The main underlying idea is that at very high energies, such as in a high-energy muon collider or a 100 TeV proton-proton collider, collinear photons will be clustered together with the charged particles that emit them, even if these charged particles are massive ($W$ bosons and top quarks). In this way, for sufficiently-inclusive observables, the contribution from real photon emissions cancels the virtual EWSL of QED  origin and therefore the IR divergences. In practice,   the $\SDKw$ scheme consists of a purely weak version of the SDK approach where almost all contributions of QED IR origin are removed.\footnote{In Ref.~\cite{Pagani:2021vyk} details on the modifications to the {\denpoz} algorithm for switching among the three schemes have been provided.} Also in this scheme,  $\Delta^{s\TO r_{kl}}$ contributions for approximating the logarithms of the form $\log(|r_{k l}|/| r_{k' l'} |)$ may or may not be included.

\end{itemize}

Assuming a realistic scenario where high-energy electrically charged particles are clustered with (quasi-)collinear photons, in Ref.~\cite{Pagani:2021vyk} it has been clearly shown that the $\rm SDK_{\rm weak}$ is superior to the $\rm SDK_{0}$ one; comparisons with exact NLO EW corrections indicate that EWSL are in general correctly captured only in the $\rm SDK_{\rm weak}$. This will be shown also in the context of muon-collider physics in Sec.~\ref{sec:SDKweak}.

\section{Direct production at muon colliders: theoretical framework}
\label{sec:DirectTheoFrame}
\subsection{Calculation set-up}
\label{sec:CalcSetUp}

At a high-energy muon collider,  the inclusive production of a final state $F$ with zero total electric charge  ({\it e.g.}~$F=t\bar t, W^+ W^-$, {\it etc.}) mainly originates from two distinct production mechanisms: the direct production, $\mu^+\mu^- \TO F$, and the Vector-Boson-Fusion (VBF) mechanism, $\mu^+\mu^- \TO F + (\mu^+\mu^-/\nu_\mu \bar\nu_\mu)$, where the hard scattering process is in fact $VV\TO F$ with the  $V=\gamma, W, Z$ radiated from the initial-state muons. 

The two classes of processes entail completely different kinematics, especially in the bulk of the associated cross sections. Direct production is dominated by the phase-space region $s\simeq S$, where $\sqrt s$ is the total energy of the partonic process in its rest frame while $ \sqrt S$ is the collider energy. Configurations with $s\ne S$ are induced by the emissions of photons, in particular the initial-state-radiation (ISR), which is taken into account in the collinear limit directly via PDF evolution of the (anti)muon in the (anti)muon $\Gamma_{\mu^\pm/\mu^\pm}$ or otherwise via the NLO EW corrections. VBF production is instead dominated by the phase-space region $\sqrt s\simeq m(F) \simeq \sum_{i \in F} m_i$, where $i$ is any particle that is part of the final state $F$ and $m(F)$ is the invariant mass of the final state $F$. In other words,  The $\mu\mu^-/\nu_\mu \bar\nu_\mu$ additional pair in the final state carries away most of the energy of the colliding muons and it typically does it along the beam pipe axis. The hard process is in fact  $VV\TO F$, and the leading contributions, especially at very high energies, can be simulated in the so-called Effective-Vector-boson-Approximation (EVA) \cite{Dawson:1984gx,Kane:1984bb}, which has already been implemented in {\aNLO} \cite{Ruiz:2021tdt}. 
Instead of simulating $2\TO n+2$ processes, where $n$ is the multiplicity of $F$, $2\TO n$ matrix elements are sufficient ($VV\TO F$) and the ``$V$ in the muon'' can be modeled similarly to what is done in the case of the photon in the lepton in the Weizs\"{a}cker--Williams approximation \cite{vonWeizsacker:1934nji,Williams:1934ad}.
Not only, as already mentioned Sec.~\ref{sec:NLOauto}, these effects can also be resummed and taken into account in a PDF formalism, as shown in Refs.~\cite{Han:2020uid,Han:2021kes,Garosi:2023bvq}.\footnote{It is interesting to note that since, at least for low multiplicities,  $m(F)\simeq \sum_{i \in F} m_i\simeq \MW$, not only logarithms of the form $\log(s/\MW^2)$ entering the PDFs are important but also power corrections of the form $(m(F)/\MW)^n$  to the matrix elements cannot be neglected. However, this aspect is beyond the scope of this paper and it will not be investigated here. }

In this work we want to study the phenomenology of EW corrections at muon colliders for hard scattering processes at high energies, {\it i.e.}, for the direct production mechanism. \footnote{The calculation of EW corrections for the direct production of multi-boson final states has also been performed in the {\WZ} framework \cite{Kilian:2007gr} in Ref.~\cite{Bredt:2022dmm}.} In particular, we consider final states with massive particles only. It is important to notice that unless a very high final-state multiplicity is chosen, in the SM for both  3 and 10 TeV collisions the two scales $m(F)$ and $S$ are very well separated, with $m(F)\ll S$. Thus, also when NLO EW corrections are taken into account, the two classes of processes (direct production and VBF) can be studied independently. In order to do so, unless differently specified, in all the results of the paper we apply the following cut on the invariant mass of the final state $F$:
\beq
m(F)\ge 0.8\, \sqrt{S}\,,
\label{eq:minvcut}
\eeq
which means that at least 80\% of the collider energy is carried away by the final state $F$. This cut has important consequences on the set-up of our calculation. In particular, the VBF contribution is completely negligible and we can safely focus on the direct-production mode. As a consequence, the only relevant PDFs for our calculations are those of the muon in the muon, $\Gamma_{\mu^-/\mu^-}$, and of the antimuon in the anti muon, $\Gamma_{\mu^+/\mu^+}$. As can be easily seen in Refs.~\cite{Han:2020uid,Han:2021kes,Garosi:2023bvq}, when 3 or 10 TeV collisions are considered any luminosity different than $\Gamma_{\mu^+/\mu^+}\ast\Gamma_{\mu^-/\mu^-}$ is strongly suppressed, being several orders of magnitude smaller than $\Gamma_{\mu^+/\mu^+}\ast\Gamma_{\mu^-/\mu^-}$.  The choice of the parameterisation of the quantities $\Gamma_{\mu^\pm/\mu^\pm}$ deserves a detailed discussion and we postpone it to Appendix~\ref{sec:isr}, while we continue on the description of the calculation set-up.

\medskip 

Since we will consider direct-production processes at high energies, the particles in the final-state $F$ will be typically very boosted. Therefore the heavy particles of the SM (the bosons $W,Z$, and $H$ and the top quark) will be experimentally identified as (fat) jets. In this work, we will study NLO EW corrections, and therefore it is important to think about how to treat the real emission of a photon, which is a contribution of the NLO EW corrections themselves. Similarly what is done at lower energies (present and past colliders) for bare leptons and photons, which are recombined into ``dressed'' leptons, here we recombine photons also with heavy charged particles: top quarks and $W$ bosons.  Not only we believe this procedure is going to mimic, for what concerns the treatment of photon emissions at high energies, a realistic analysis where decays and jet clustering are considered, but it has an impact also on the size of the EW corrections themselves. Indeed, similarly to the case of leptons at lower energies ($s\simeq m_W^2 \gg m_\ell^2$), at high energies ($s\gg m_W^2 $) the recombination of photons with heavy particles leads to the cancellations of part of the EWSL from virtual corrections. Unless differently specified, when considering the exact NLO EW corrections we will always cluster photons with any electrically charged particle $X$ if 
\beq
\Delta R(X,\gamma) <0.2\,,
\label{eq:Rphoton}
\eeq
where $\Delta R \equiv \sqrt{(\Delta \phi)^2+(\Delta \eta)^2}$ and  $\Delta \phi$ is the azimuthal angle between $X$ and $\gamma$ and $\Delta \eta$ is the difference between their pseudorapidities.

In order to mimic a realistic experimental set-up, since we consider only undecayed particles that are typically boosted, we require that they are both separated in angle among them and also with the beam pipe axis. In particular, for any particle $X$ and $Y$ that are part of the final state $F$ we require that 
\beq
    \eta(X) < 2.44\,, \qquad p_T(X) > 100\, {\gev}\,, \qquad \Delta R(X,Y)> 0.4\,, \qquad\,
    \label{eq:cuts}
\eeq
 where cuts are applied after the recombination of photons described before. These cuts should be considered as illustrative: for example, the $p_T(X)$ cut may be larger depending on the collider energy. Our overall conclusions are not affected by the specific values employed.

\medskip

When energies of several TeV's are reached, one may wonder if not only photons but also $H$, $Z$ and $W$ bosons have to be clustered together with heavy particles. Indeed, at such energies EW radiation is expected to lead to large effects. Their origin is the real-emission counterpart of the virtual EWSL: soft and/or collinear enhancements in the radiation. On the one hand, from an experimental point of view, if massive particles are very close in angle, their decay products are expected to be clustered in a single fat jet. On the other hand, from a theoretical point of view, real $H, Z$, and $W$ radiation induces $\ord(\alpha)$ corrections to the inclusive direct production of a given final state $F$, which therefore are of the same perturbative order of NLO EW corrections or EWSL. In fact, they can be formally considered as part of the NLO EW corrections to the hard process. This aspect has already been discussed in the literature in the context of present and future hadron-colliders, {\it e.g.}, in Refs.~\cite{Frixione:2014qaa,Frixione:2015zaa,Czakon:2017wor}, and this new contribution has been denoted in these references as Heavy-Boson-Radiation (HBR).

For many processes, the effect of HBR at hadron colliders has been found to be much smaller than the NLO EW corrections. Also, it has been understood that only the details of the experimental analysis employed on the specific signature targeted to detect a specific final state $F$ can determine how much of the HBR contribution will be actually part of the signal. In view of these considerations and since the direct production at high energy muon colliders involves a completely different kinematics w.r.t.~the case of hadron colliders, it is interesting to explore the impact of HBR also in our study.  Thus, when we will study the HBR we will consider three different approaches:
\begin{itemize}

\item {\bf No recombination}: We consider the final state $F$ and we take into account also the direct production $F+B$ with $B=H,Z$. We do not cluster the HBR of $B$ with any of the particles in the final state $F$ and we do not set any cut on $B$. While for the results presented in Sec.~\ref{sec:HBRwwtt} one never has the case where $B\in F$, should this condition be realised, meaning that the particle $B$ appears $k$ times in $F$, the cuts \eqref{eq:minvcut}--\eqref{eq:cuts} are intended to be imposed inclusively on the HBR process, {\it i.e.}, by asking that at least $k$ $B$-type particles pass them. 
\item {\bf Recombination}:  Same as the previous point, but we cluster any particle $X$ in $F$ with $B=H,Z$ if   
\beq
\Delta R(X,B) <0.2\,,
\label{eq:Rboson}
\eeq
and we denote the clustered object as the original $X$ (for instance, clustering a ``bare'' top-quark with a $Z$ we call it a ``dressed'' top). \\
For the processes considered this procedure is rather intuitive since in these cases there always exists a vertex $X\bar XB$ in the SM. For more general cases, one may decide whether to be agnostic or not about the underlying theory (in the latter case, one would not {\it e.g.} cluster a pair of $Z$ bosons together).

\item {\bf EW Jets}: This approach can be used in our study only for final states $F$ that contain only $V=W,Z$.
Starting from the direct production of the final states $F=nV$ (notice that in this case $V\ne H$ but it can be instead a $W$ boson) the HBR contribution from the  $(n+1)V$ final states is considered. The physical objects are EW jets, which are obtained (via {\sc \small FastJet}~\cite{Cacciari:2011ma}) clustering any $V$ and using the Cambridge/Aachen algorithm \cite{Dokshitzer:1997in} with a jet radius of 0.2 and requiring the presence of at least $n$ jets passing the cuts in \eqref{eq:minvcut}--\eqref{eq:cuts}.\footnote{The choice of the Cambridge/Aachen algorithm is due to the fact that we opt for a purely-geometric clustering (in the $\eta$--$\phi$ plane), in order to have a more clear picture of the results. Other choices are of course possible.}

\end{itemize}
For all three approaches, similarly to the case of the photon recombination, the cuts are applied only after the recombination is performed.

\medskip 

Finally, we specify the input parameters that have been used for obtaining the numerical results that are reported in Sec.~\ref{sec:numres}.

Input parameters
are defined in the $G_\mu$ scheme, which is what we employ for the computation of EW corrections and in particular for the renormalisation. The numerical values are:
\begin{equation}
M_Z = 91.188 \text{ GeV}, \quad M_W = 80.419 \text{ GeV}, \quad G_{\mu} = 1.16639 \times 10^{-5} \text{ GeV}^{-2},
\end{equation}
and the top quark and Higgs boson masses are set to
\begin{equation}
M_H = 125 \text{ GeV}, \quad m_t = 173.3 \text{ GeV} .   \\
\end{equation}
The renormalisation scale is not relevant and the factorisation scale has been set equal to $\sqrt {s}$.

\subsection{Definition of different approximations}
\label{sec:approxs}

In Sec.~\ref{sec:numres} we will present several numerical results and we will employ different approximations for the evaluation of EW corrections. In this section we properly define them introducing the notation that will be used within the rest of the paper.

\subsubsection{Exact EW corrections and HBR}
\label{sec:approxs-ewhbr}

In this work, we consider the inclusive $\mu^+ \mu^- \to F (+X)$ production,  for SM processes that feature tree-level amplitudes only involving  EW interactions. For such processes, the LO is of $\ord(\alpha^n)$, where the value of $n$ is process dependent, and we stress that there are no other perturbative orders that can be present at LO, {\it i.e.} starting from tree-level diagrams. Thus, the only perturbative orders involving one-loop corrections, the NLO corrections, are the NLO QCD corrections, which are of $\ord(\alpha_S\alpha^{n})$ and are also denoted in the literature as $\rm NLO_1$, and the NLO EW corrections, which are of $\ord(\alpha^{n+1})$ are also denoted in the literature as $\rm NLO_2$.\footnote{This means that the so-called Complete-NLO predictions do not involve other perturbative orders and coincide with LO+NLO QCD+NLO EW. No $\rm NLO_i$ with $i>2$ is present.} In this work we focus on the latter and we notice that unless QCD interacting particles are present in $F$, such as the top quark, the NLO QCD corrections are not even present for the processes considered, as {\it e.g.}~for multiboson final states. On top of this, the presence of a single coupling combination at LO (LO$_1$) allows for a more direct comparison between the exact NLO EW corrections and their Sudakov approximation.~\footnote{In particular, no contribution due to QCD corrections on top of subleading LO contributions is present (in the Sudakov approximation, this corresponds to the term denoted $\deltaQCD$ in Ref.~\cite{Pagani:2021vyk}).}

The main purpose of this paper is the study of such NLO EW corrections for the inclusive direct production $\mu^+\mu^-\TO F (+X)$ at a high-energy muon collider. On the one hand, we want precisely to analyse the validity of the Sudakov approximations and compare them with the exact NLO EW. On the other hand, we want to study the contribution of the corresponding real-emission counterpart, the HBR, and compare virtual and real corrections. In the following, we properly define the quantities and the notation that we will use in our study.

First of all, considering that for the processes we consider no QCD coupling enters at LO, we introduce the following  quantities 

 \bea
\SigmaLO&\propto& \alpha^n \,,\\
\SigmaNLOtwo&\propto& \alpha^{n+1}\,. \\
\SigmaNLO&\equiv &\SigmaLO +\SigmaNLOtwo \,. 
\eea

In other words, $\SigmaLO$ is the LO prediction, $\SigmaNLO$ is the prediction at NLO EW accuracy and the NLO EW corrections, $\SigmaNLOtwo$, correspond to $\SigmaNLO-\SigmaLO$. Thus, the relative impact of NLO EW corrections corresponds to the quantity
\beq
\deltaNLOEW\equiv\frac{\SigmaNLO-\SigmaLO}{\SigmaLO}\,\propto \alpha, \label{eq:deltaNLOEW}
\eeq
where we have made explicit that this quantity is proportional to $\alpha$. NLO EW corrections, and therefore $\deltaNLOEW$, account for the exact contributions at $\ord(\alpha)$ from one-loop corrections and the tree-level emission of photons. The (LO) contribution from HBR to the inclusive cross sections is also of $\ord(\alpha)$, but traditionally is treated separately because of two reasons. First, it is {\it per se} IR finite, hence it can be computed independently. Second, at the typical LHC energies, the signature from HBR is distinguishable from the process without emissions of $B=H, W, Z$. Also in this work, the two contributions will be separately treated, unless differently specified.
The cross section associated with the HBR is denoted as $\SigmaHBR$ and the relative impact on the LO prediction as
\beq
\deltaHBR\equiv\frac{\SigmaHBR}{\SigmaLO}\,\propto \alpha\,. \label{eq:deltaHBR}
\eeq

The quantity $\deltaNLOEW$ as well as $\deltaHBR$ are expected to be dominated by large EWSL. In order to investigate the degree of cancellation of $\ord(\alpha)$ EW corrections between the standard NLO EW corrections and the HBR  we introduce also the quantity
\beq
\SigmaNLOHBR\equiv\SigmaNLO+\SigmaHBR\, , 
\eeq
and the corresponding
\beq
\deltaNLOHBR\equiv\frac{\SigmaNLOHBR - \SigmaLO}{\SigmaLO}\,\propto \alpha\,. \label{eq:deltaNLOHBR}
\eeq

\subsubsection{EWSL and their approximate resummation}
\label{sec:approxs-ewsl}

As already mentioned, in this work we want to investigate how the {\denpoz}~algorithm and its revisitation presented in Ref.~\cite{Pagani:2021vyk} accurately catches the virtual EWSL within $\deltaNLOEW$ and consequently how efficiently works as an approximation of it. Our default option is denoted as $\SigmaSDKw$ and corresponds to  $\SigmaNLO$ where the NLO EW corrections $\SigmaNLO-\SigmaLO$ are approximated via the EWSL in the $\SDKw$ approach and taking into account also the $\Delta^{s\TO r_{kl}}$ contributions, see Sec.~\ref{sec:SudXS} and Ref.~\cite{Pagani:2021vyk} for more details.

Analogously to the NLO EW case we define 
\beq
\deltaSDKw \equiv\frac{\SigmaSDKw-\SigmaLO}{\SigmaLO}=\deltaDL+\deltaSL+\deltastorkl\, \label{eq:deltaSDKw}
\eeq
where
\bea
\deltaDL&\propto& \LsW\,\\
\deltaSL&\propto& \lsW\,\\
\deltastorkl &=& f\left[ \Lrs, \lrsalpha \right]\,.
\eea

A few comments on the previous formula can be useful. The terms $\deltaDL$ and $\deltaSL$ correspond to the double and single logarithms (see also Eq.~\eqref{eq:generallogs}) of the EWSL in the strict LA expansion for $s\gg \MW^2$; they are exactly evaluated via the {\denpoz} algorithm.\footnote{Here we are understanding the additional imaginary terms introduced in Ref.~\cite{Pagani:2021vyk} and especially that the range of applicability of the algorithm is satisfied, especially: no resonances and no mass-suppressed Born amplitudes. }  The $\deltastorkl$ term accounts for large logarithms of ratios of kinematic invariants of the process, as explained in Sec.~\ref{sec:SudXS}, and correspond to the  $\Delta^{s\TO r_{kl}}$ contributions, which are a linear combination all the possible $\Lrs$ and $ \lrsalpha$ that can appear starting from all the possible invariants $r_{kl}$. Unlike, $\deltaDL$ and $\deltaSL$ they have not been formally derived and there is no guarantee that all the possible logarithms of this form are captured; case by case has to be checked. However, as already mentioned in Sec.~\ref{sec:EWSLamp}, for several processes it has already been observed that they are very effective. We remind again the reader that all the components of $\deltaSDKw$ are evaluated in the $\SDKw$ approach.

With such a definition of $\SigmaSDKw$, at high energies and expanding in powers of $\MW^2/s$ one gets that
\beq
\deltaNLOEW-\deltaSDKw \propto \alpha  \left(\MW^2/s\right)^n~~{\rm with}~~n\ge0\, \label{eq:LAok}, 
\eeq
and in general, if $\Delta^{s\TO r_{kl}}$ is an efficient approximation, as observed in many cases, if a given invariant $r_{kl}$ is such that $|r_{kl}|\ll s$, expanding in powers of  $|r_{kl}|/s$ one gets
\beq
\deltaNLOEW-\deltaSDKw \propto \alpha  \left(|r_{kl}|/s\right)^n~~{\rm with}~~n\ge0\, \label{eq:storklOK}. 
\eeq
In other words, Eqs.~\eqref{eq:LAok} and \eqref{eq:storklOK} say that if EWSL are correctly calculated, at high energies they should correctly capture the bulk of the NLO EW corrections and only percents effects could be missed. When we study this aspect in Sec.~\ref{sec:numres} we will also introduce the quantities $\SigmaSDKz$ and $\deltaSDKz$, that are analogous to $\SigmaSDKw$ and $\deltaSDKw$, respectively, but based on the $\SDKz$ approach. Also, we will study the impact of $\deltastorkl$, by setting it to zero, as in the original formulation of the {\denpoz} algorithm.  

\medskip 
At 10 TeV, but also at lower energies, the EWSL due to $\deltaDL$ as well as to $\deltaSL$
can be very large and up to the point, as we will see in Sec.~\ref{sec:numres}, that in some kinematic regimes $\deltaSDKw < -100\%$,  which implies $\SigmaNLO<0$. In these cases, resummation is therefore not a procedure for improving the precision and accuracy of the predictions but for obtaining sensible results, {\it i.e.}, positive cross sections. Resummation of EWSL has already been studied in the literature \cite{Kuhn:1999nn,Fadin:1999bq,Ciafaloni:1999ub,Beccaria:2000jz,Hori:2000tm,Ciafaloni:2000df,Denner:2000jv,Denner:2001gw,Melles:2001ye,Beenakker:2001kf,Denner:2003wi,Pozzorini:2004rm,Feucht:2004rp,Jantzen:2005xi,Jantzen:2005az,Jantzen:2006jv,Chiu:2007yn,Chiu:2008vv,Manohar:2012rs, Bauer:2017bnh,  Manohar:2018kfx}  and recently a detailed study on its limitations and subtleties, considering terms up to Next-to-Leading-Logarithmic (NLL) accuracy have been discussed in detail in Ref.~\cite{Denner:2024yut}. Here we do not aim to reach such a precision or investigate the resummation procedure; we want to simply asses when resummation is either desirable or mandatory 
in order to obtain meaningful predictions in the case NLO EW corrections lead to a vanishing or negative cross section. To this purpose, we define the following quantity:

\beq
\SigmaExp \equiv \left(\SigmaLO \,e^{\deltaSDKw} \right )+ \left( \SigmaNLO-\SigmaSDKw \right ) = \SigmaNLO + \ord(\alpha^2) \times  \SigmaLO . \label{eq:resumdef}
\eeq
The r.h.s.~of Eq.~\eqref{eq:resumdef} says that if $\SigmaExp$ is expanded in powers of $\alpha$ the NLO EW prediction is captured exactly, while beyond  $\ord(\alpha)$ the resummed tower of EWSL of order $\alpha^n \log^k(s/\MW^2)$ with $n>1$ and $k=2n, 2n-1$ is approximated via simple exponentiation. We stress again that we do {\it not} claim we are doing NLL resummation of EWSL. We instead want to study when and if this procedure is necessary, by comparing $\deltaNLOEW$ with the relative corrections induced by $\SigmaExp$, namely
 \beq
\deltaExp \equiv\frac{\SigmaExp-\SigmaLO}{\SigmaLO}= \deltaNLOEW+\ord(\alpha^2)\,. \label{eq:deltaExp}
\eeq

In the exponentiation procedure, we do not include the contributions from HBR. As it will be manifest in Sec.~\ref{sec:numres}, the effects due to the HBR (real) are in general much smaller than the one induced by the virtual loops. Thus, the resummation of such contributions is clearly not necessary as their virtual counterpart. However, we do see a case where both NLO EW corrections and HBR are relevant,  the multi EW jet ($\EWjet$) production, for which we calculate additional quantities. 

\subsubsection{Quantities relevant to EW jets}
\label{sec:approxs-ewjet}

The definition of EW jets has been provided in Sec.~\ref{sec:CalcSetUp}, and in Sec.~\ref{sec:HBREWj} we will use it for studying inclusive EW-dijet production, $\mu^+\mu^-\TO 2\EWjet(+X)$. For such a process we introduce additional quantities.
First of all,
\beq
\sigma_{X}(2\EWjet)\equiv\sigma_{X}(2V) \qquad{\rm for} ~X={\rm LO,~NLO~EW,~}\SDKw \,,
\eeq
which means that the LO prediction, $\SigmaLO(2\EWjet)$, is given by the prediction for the production of $2V=W^+W^-, ZZ$ at LO and applying the clustering for obtaining the EW jets. Similar considerations apply for $X={\rm NLO~EW},~\SDKw$.  It is also clear that 
\bea
\SigmaHBR(2\EWjet)&\equiv&\SigmaLO(3V) \,, \\
\SigmaNLOHBR(2\EWjet)&\equiv& \SigmaNLO(2V) + \SigmaLO(3V)\,, \label{eq:NLOHBREWj}
\eea
and in addition we also define:
\bea
\SigmaNNLOHBR(2\EWjet)&\equiv&  \SigmaLO(2V) \left(1+\deltaNLOEW +\frac{\deltaSDKw^2}{2} \right)  \nonumber \\
 &+& \SigmaNLO(3V)+ \SigmaLO(4V)\,. \label{eq:NNLO}
\eea

The prediction $\SigmaNLOHBR$ takes into account all the corrections of $\ord(\alpha)$: the NLO EW corrections to $2V$ and HBR, meaning $3V$ production at LO. 
The prediction $\SigmaNNLOHBR$ instead takes into account  all the corrections of $\ord(\alpha)$, as $\SigmaNLOHBR$,  and those of $\ord(\alpha^2)$, where the two-loop corrections to $2V$  are approximated via their Sudakov component in the $\SDKw$ scheme;\footnote{The first line of Eq.~\eqref{eq:NNLO} corresponds to $\SigmaExp$ truncated at $\ord(\alpha^2)$ w.r.t.~LO prediction.} it corresponds to  $\SigmaNLOHBR$ plus NLO EW corrections to HBR, double HBR, and the approximation of the two-loop corrections that we have just mentioned.

For all these quantities we understand, consistently with the notation already used before:
\beq
\delta_X\equiv \frac{\sigma_X-\sigma_{\rm LO}}{\sigma_{\rm LO}}\,. \label{eq:deltaX}
\eeq
One should notice the exception of the case of HBR, Eq.~\eqref{eq:deltaHBR}.

\subsection{List of aspects investigated in this work}
\label{sec:list}
 In this section, we list the different aspects that we want to investigate, which are all related to EW corrections to direct-production process at high-energy muon colliders.

\begin{enumerate}
\item First of all we want to give an overview of how large EW corrections can be, especially when differential distributions are considered. Our work considers only SM processes and therefore total rates can be very small for some of them. However, the features of EW corrections that we will discuss in Sec.~\ref{sec:numres} are not specific to the SM itself but can be extended, in principle, to any BSM  theory involving EW-charged particles. Thus, we will focus on relative corrections rather than the rates. The SM can be considered as a test case for a more general  EW-interacting theory.

\item We want to show how at the differential level EW corrections can be very different in a high-energy lepton collider w.r.t.~an hadronic one. Direct production processes at a hadron collider are dominated, similarly as in VBF at lepton colliders, by kinematic regions with partonic $\sqrt{s}\simeq\sum_{i \in F} m_i$. On the contrary, as already said, in a high-energy muon collider they are dominated by kinematic regions with $ \sqrt s \simeq \sqrt S$, where $\sqrt S$ is the collider energy, due to the very different (opposite in fact) Bjorken-$x$ dependence of the PDFs. The DL of the EWSL in a high-energy muon collider is therefore typically of the form $\simeq L(S,\MW^2)$, regardless of the differential distribution considered, unlike in hadron collisions.
  
\item When NLO EW corrections reach, or even worse surpass, the relative size of $-100\%$ of the LO, resummation of EWSL does not concern precision but the physical sense of predictions.
We want to investigate when we should expect such situations at 3 and 10 TeV collisions.

\item The EW Sudakov approximation is expected to be a very good approximation for the processes we are studying in this work. We want to scrutinise under which conditions  EWSL are or are not in fact a good approximation of the NLO EW corrections as expected; using the notation of Sec.~\ref{sec:approxs}, it means verifying that the quantity $|\deltaNLOEW-\deltaSDKw|$ is at most at the percent level and a constant at the differential level. This aspect is of particular relevance in the BSM context, since at the moment, not only for EFT theories but also for UV-complete models, NLO EW exact calculations are not yet available. On the other hand, EWSL can be in principle calculated in an easier way, therefore it is important to know if and when one can use this approximation. In particular, using the SM as a test case, we want to check the following aspects for direct production at a high-energy muon collider:
\begin{enumerate} 
\item How relevant is the choice of the scheme $\SDKw$ or $\SDKz$ for a correct approximation of NLO EW corrections?
\item How relevant is the inclusion of the $\deltastorkl$ term for a correct approximation of NLO EW corrections?
\item What happens if the {\denpoz} algorithm is used also in cases where the LO prediction is mass-suppressed (see Eq.~\eqref{eq:masssuppr} and text around it)?
\end{enumerate}

\item The common lore is that EW radiation, HBR in our notation, will be a ubiquitous and large effect at a high-energy muon collider, as it is for QCD radiation at the LHC or at a future high-energy hadron collider. We want to check if this is always true and compare the size of HBR with the one from EW loops and real emission of photons, {\it i.e.}, the NLO EW corrections. By comparing $\deltaNLOEW$ and $\deltaHBR$, we will check which one of the two effects is dominant and, knowing also the EWSL component of the former ($\deltaSDKw$), we will check if both the contributions need to be ``cured'' by resummation. In the case of HBR, we will also investigate the impact that the mass $M_B$, $B=H,W,Z$, has and we will consider in specific cases (EW jets) the impact of double HBR.

\end{enumerate}

\section{Numerical results}
\label{sec:numres}

In this section, we present numerical results that have been obtained with the purpose of investigating the points listed in Sec.~\ref{sec:list}. The setup for the computations, including the settings of input parameters, can be found in Sec.~\ref{sec:CalcSetUp}. \\
In particular, in Sec.~\ref{sec:NLOvsEWSL} we study the accuracy of the EWSL approximation via comparisons with exact $\NLOEW$ predictions. The section takes care of presenting different aspects, and it is divided into three parts, Sec.~\ref{sec:SDKweak}, Sec.~\ref{sec:angular} and Sec.~\ref{sec:failMS}. They present, respectively, the comparison of the $\SDKw$ and $\SDKz$ approaches, the importance of logarithms involving the ratio of invariants (the $\deltastorkl$ term), and a case when mass-suppressed terms arise in certain kinematics regions and spoil the applicability of the {\denpoz} approach method. The discussion of results continues in Sec.~\ref{sec:resum}, where we discuss under what circumstances the need for resumming EW corrections arises. Finally, the numerical importance of HBR is discussed in Sec.~\ref{sec:HBR}, where we first consider, in Sec.~\ref{sec:HBRwwtt}, the case HBR contributions to specific processes --the production of a pair of $W$ bosons or of top quarks--, while in Sec.~\ref{sec:HBREWj} we discuss the case of EW jets.

\subsection{Sudakov approximation {\it vs.} exact $\NLOEW$ predictions}

\label{sec:NLOvsEWSL}

As already said, in this section we study the accuracy of the Sudakov approximation via comparisons with the exact $\NLOEW$ predictions. In particular, In Sec.~\ref{sec:SDKweak} we show the relevance of using the $\SDKw$ scheme w.r.t.~the commonly used $\SDKz$ one. In Sec.~\ref{sec:angular} we show that not only the logarithms in Eq.~\eqref{eq:generallogs} are numerically relevant but also those involving ratios of kinematically invariants have to be taken into account. In other words, the quantity denoted as $\deltastorkl$ in Sec.~\ref{sec:approxs-ewsl} cannot be ignored. Finally, in Sec.~\ref{sec:failMS} we show an explicit example of how the presence of terms that are mass suppressed, but numerically relevant, completely invalidates the accuracy of the Sudakov approximation derived via the {\denpoz} algorithm, as expected by its range of applicability.   

In this section we will also start to describe features of the EW corrections that are distinctive for direct production processes at high-energy lepton colliders and quite different from the case of a high-energy hadronic machine such as LHC or FCC-hh (see, {\it e.g.}, Refs.~\cite{Mangano:2016jyj, Azzi:2019yne} for overviews of EW corrections at high-energy hadronic machines). We remind the reader that all calculations have been performed for SM processes, which are the only ones that can be calculated at $\NLOEW$ accuracy with automated tools. However, many of the conclusions from this study, especially those stating the limitations of some approaches,
 can be clearly generalised also to BSM scenarios. The SM has to be regarded only as a test case in this respect.

\subsubsection{Importance of $\SDKw$ versus $\SDKz$}
\label{sec:SDKweak}

In this section, we start to show comparisons of predictions at exact NLO EW accuracy, $\NLOEW$, with predictions that take into account NLO EW corrections in the Sudakov approximation. The reference scheme is the one denoted as $\SDKw$, introduced in Ref.~\cite{Pagani:2021vyk} and briefly described in Sec.~\ref{sec:SudXS}. We also show the more commonly used $\SDKz$ and compare the two schemes. 
Here, we focus only on $2\to 2$ processes. With such a choice, we minimise the interplay of the differences between the $\SDKw$ and $\SDKz$ scheme with the effects that are studied in the next section: the relevance of logarithms of ratios of invariants, which we always retain in our predictions unless differently specified.

\begin{figure}[!t]
  \includegraphics[width=0.48\textwidth]{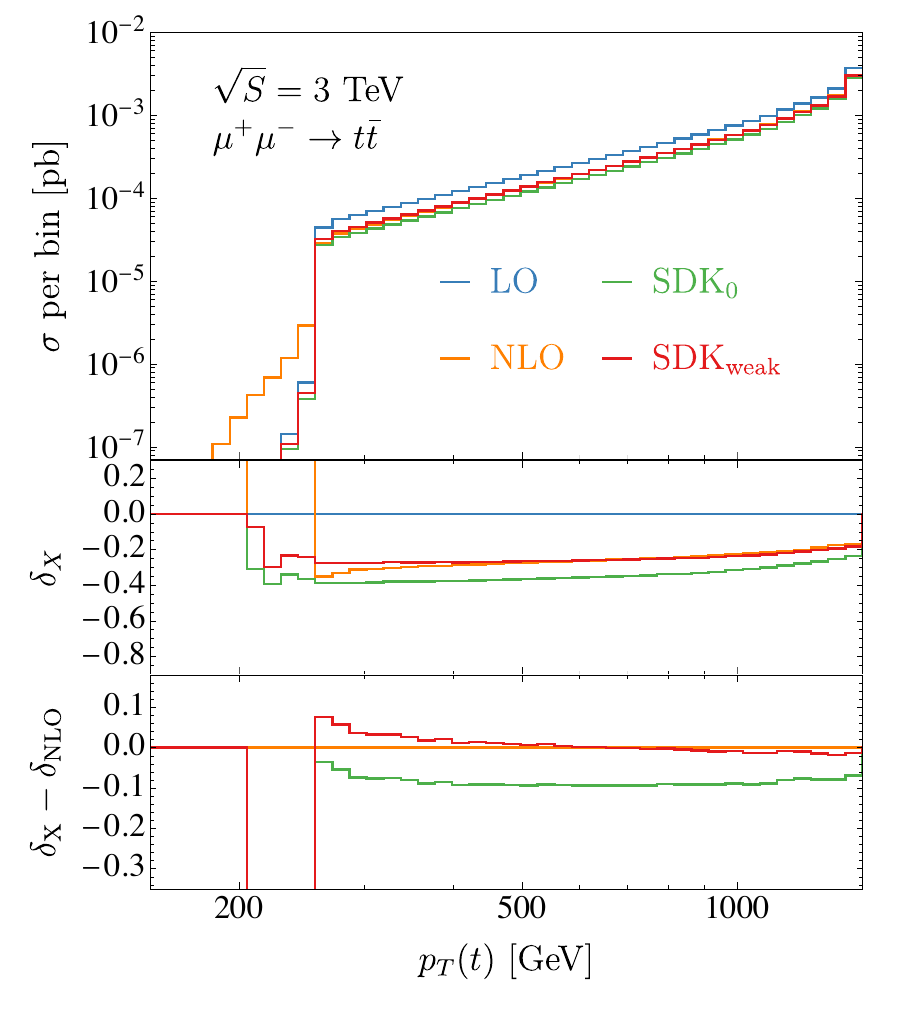}
  \includegraphics[width=0.48\textwidth]{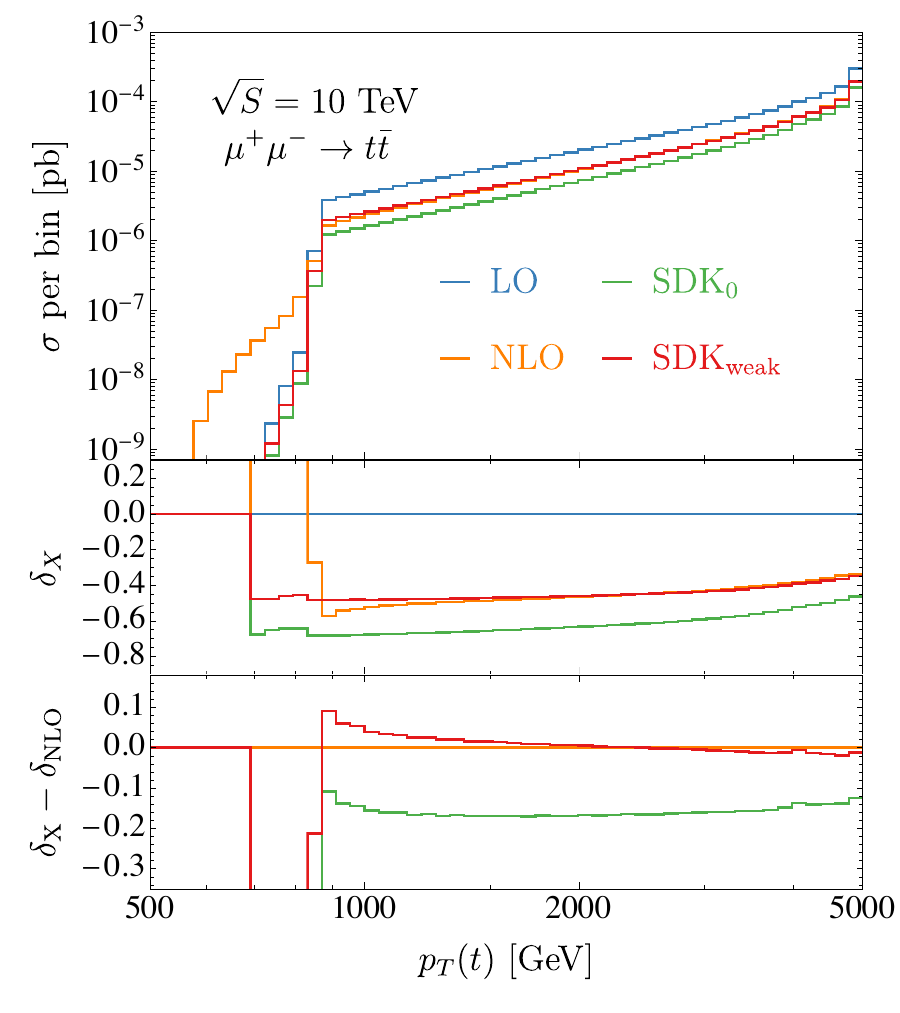}
  \caption{The top quark $p_T$ distribution in $\mu^+\mu^- \to t {\bar t}$. The left (right) plot shows results at $\sqrt S=3$ TeV ($\sqrt S=10$ TeV). The histograms show $\SigmaLO$ (blue) and  $\SigmaNLO$ (orange) as well as predictions in the Sudakov approximation in different approaches, {\it i.e.},  $\SigmaSDKw$ (red) and $\SigmaSDKz$ (green).}\label{fig:mmtt_ptt}
\end{figure}

We start by looking at the top-quark transverse momentum distribution, $p_T(t)$, in $\mu^+\mu^-\to t \bar t$ production. In Fig.~\ref{fig:mmtt_ptt} we show results for 3 TeV collisions in the left plot and for 10 TeV collisions in the right one. Cuts and the calculation set-up are described in Sec.~\ref{sec:CalcSetUp}. Both plots have the same layout, which will be used also for the other figures of this section and, with more or less small variations, throughout the whole paper. In the following we describe it and we also discuss how to interpret the plots.    

Using the same colour code of Ref.~\cite{Pagani:2021vyk}, in the main panel we show LO  (blue), $\NLOEW$ (orange), $\SDKw$ (red), and $\SDKz$ (green) predictions.\footnote{The rigorous definition of these quantities can be found in Sec.~\ref{sec:approxs}.} Whenever a distribution has a negative cross section, the corresponding histogram is plotted as a dotted line. In the first inset, we plot the relative corrections induced by such approximations w.r.t.~the LO predictions, {\it i.e.}, the quantities $\delta_X$ defined in Eqs.~\eqref{eq:deltaNLOEW}, \eqref{eq:deltaSDKw} and more in general \eqref{eq:deltaX}.  In such inset it is possible to appreciate the size and the sign of EW corrections, either calculated at exact $\NLOEW$ accuracy or via Sudakov approximation(s). Then, in the second inset, we plot the difference $\delta_X-\delta_{\NLOEW}$ for the two cases $X=\SDKw$ and $X=\SDKz$ in order to test their accuracy. Rather than the minimisation of this quantity, the validity of the Sudakov approximation consists in having a {\it small} constant difference ($|\delta_X-\delta_{\NLOEW}|\simeq \alpha$) over the full spectrum, {\it i.e.}~a horizontal line, see also Eqs.~\eqref{eq:LAok} and \eqref{eq:storklOK} and text around it.  Indeed, $\ord(\alpha)$ contributions are expected to be present, while non-horizontal lines indicate an (at least) logarithmic-enhanced contribution that is not captured. Such contribution may accidentally compensate the $\ord(\alpha)$ constant term and lead to $\delta_X-\delta_{\NLOEW}\simeq0$ for particular phase-space regions, but this is not to be regarded as an indication of the validity of the approximation.  That said, a {\it large} constant difference ($|\delta_X-\delta_{\NLOEW}|\gg \alpha$), however, also points to logarithms that are not correctly captured. In particular, at a high-energy lepton collider, the direct-production processes studied in this work and characterised by $\sqrt{s}\simeq \sqrt{S}$ may show such effects induced by missing large double logarithms (see Eq.~\eqref{eq:generallogs}) of the form  $L(s,\MW^2)$,\footnote{It is easy to see in Sec.~4.1 of Ref.~\cite{Pagani:2021vyk} that, at variance with the $\SDKz$ scheme, the $\SDKw$ scheme implies the substitution $C_{\rm EW}\to C_{\rm EW}-Q^2$, where $Q$ is the charge of the particle considered, in the prefactor of the double-logarithm of the form  $L(s,\MW^2)$ entering the formula of $\deltaDL$ in Eq.~\eqref{eq:deltaSDKw}. Similar effects are also present in the single logarithms and lead to large constants when a fixed $s$ is considered.} which are therefore large constants for the full phase-space.  

In Fig.~\ref{fig:mmtt_ptt} we notice that LO, $\SDKw$ and $\SDKz$ predictions quickly drop  for $p_T(t)\lesssim250~\gev$ ($p_T(t)\lesssim800~\gev$) at 3 TeV (10 TeV) collisions. This is due to the cut on pseudorapidities in \eqref{eq:cuts}. In that region, only contributions with $\sqrt{s}<\sqrt{S}$ are allowed, and therefore the large suppression from the muon PDF at Bjorken-$x<1$ is the reason for such decrease in the rates. On the contrary, $\NLOEW$ predictions, featuring also $2\to3$ configurations via real photon radiation, can allow for smaller values of $p_T(t)$ also with Bjorken-$x\simeq1$, avoiding the PDF suppression. In that region, which is very much disfavoured w.r.t.~the bulk, NLO EW corrections are much larger than the LO predictions and one should in principle also take into account effects from the photon PDF into the muon. Moreover, being dominated by photon real emissions, the comparison of $\NLOEW$ predictions with the Sudakov approximation, either $\SDKw$ or $\SDKz$ is meaningless. 

For $p_T(t)\gtrsim250~\gev$ ($p_T(t)\gtrsim800~\gev$) at 3 TeV (10 TeV) collisions, we can instead discuss for the left (right) plot the features of NLO EW corrections related to the bulk of the distributions and that can be approximated via EWSL. We describe them in the following. First of all, in the first inset we notice that the shape of the EW corrections is very different w.r.t.~the typical one observed for $p_T$ distributions at hadron colliders, for $pp\to t \bar t$ can be found {\it e.g.}~in Refs.~\cite{Pagani:2016caq,Czakon:2017wor}.  While at hadron colliders, excluding the threshold, $\deltaNLOEW$ is negative for  $t \bar t$ production and from small values at small    $p_T(t)$ constantly grows in absolute value at large $p_T(t)$, here  $\deltaNLOEW$ is in general large and negative over the full considered spectrum, $\deltaNLOEW\simeq-(20$--30\%) at 3 TeV and $\deltaNLOEW\simeq-(40$--50\%) at 10 TeV. Somehow counterintuitively, it slightly decreases in absolute values at large $p_T(t)$, the opposite of what is observed at hadron colliders. 

The origin of such behaviour has again to be ascribed to the different Bjorken-$x$ dependence of the PDFs of the proton and of the muon. In muon collisions, unlike in the case of the hadron collisions, regardless of the value of $p_T(t)$, $s\simeq S$. Thus the $\deltaDL$ component proportional to $L(s,\MW^2)$  in Eq.~\eqref{eq:deltaSDKw} is present over the full considered spectrum. In other words, double logarithms are large, especially at 10 TeV, and constant. The single logarithms, as well as the logarithms entering $\deltastorkl$ in Eq.~\eqref{eq:deltaSDKw}, conversely, do depend on the kinematic and in particular on the other two Mandelstam variables $t$ and $u$. Overall, they lead to smaller values of $\deltaSDKw$, which, as can be clearly seen in the second insets of the two plots of Fig.~\ref{fig:mmtt_ptt}, is a very good approximation of the $\NLOEW$ predictions.  

Unlike the case of  $\SDKz$, for $p_T(t)\gtrsim \sqrt{S}/10$ the $\SDKw$ approach can very well approximate the $\NLOEW$ result, with a constant discrepancy $\deltaNLOEW-\deltaSDKw$ of very few percents w.r.t.~the LO. This is manifest in the second inset of the two plots of Fig.~\ref{fig:mmtt_ptt}, where it can also be seen that in the case of $\SDKz$ this discrepancy is instead of the order $\deltaNLOEW-\deltaSDKz\simeq-(5$--10\%) at 3 TeV and $\deltaNLOEW-\deltaSDKz\simeq-(15$--20\%) at 10 TeV. Thus, $\deltaNLOEW-\deltaSDKz$ is much larger in absolute value than $\deltaNLOEW-\deltaSDKw$, and it depends much more on the value of $p_T(t)$ and especially on the energy of the collider. They are all clear signs that both double and single EWSL logarithms are not correctly captured by the $\SDKz$ scheme, unlike the case of the $\SDKw$ one. 

In the region just above $p_T(t)\simeq250~\gev$ ($p_T(t)\simeq800~\gev$) at 3 TeV (10 TeV) collisions another effect is entering, slightly altering the agreement of $\deltaSDKw$ with $\deltaNLOEW$, and similarly for the $\deltaSDKz$ case. At NLO, the real emission collinear to the initial state can alter the kinematic and therefore has an impact on the accuracy of the Sudakov approximation. On the one hand, even with Bjorken-$x\simeq1$ for the muon PDF, we can have a smaller invariant mass $m(t \bar t)$ for the $t \bar t$ pair, allowing for smaller values of $p_T(t)$ also with the cuts in \eqref{eq:cuts}. On the other hand, the boost from the recoil against the photon emission can lead to more peripheral top quarks, which cannot pass therefore the cuts. In conclusion, it is not surprising that such effects are arising close to cuts that LO, $\SDKw$ and $\SDKz$ predictions cannot pass, unless Bjorken-$x\lesssim1$, but that $\NLOEW$ predictions can instead pass also with Bjorken-$x\simeq1$. Moreover, in the case of 3 TeV collisions, in this region $p_T(t)$ is only mildly larger than $\MW$, so non-negligible power corrections of the form $\MW^2/t$ or $\MW^2/u$ cannot be excluded.

\begin{figure}[!t]
  \includegraphics[width=0.48\textwidth]{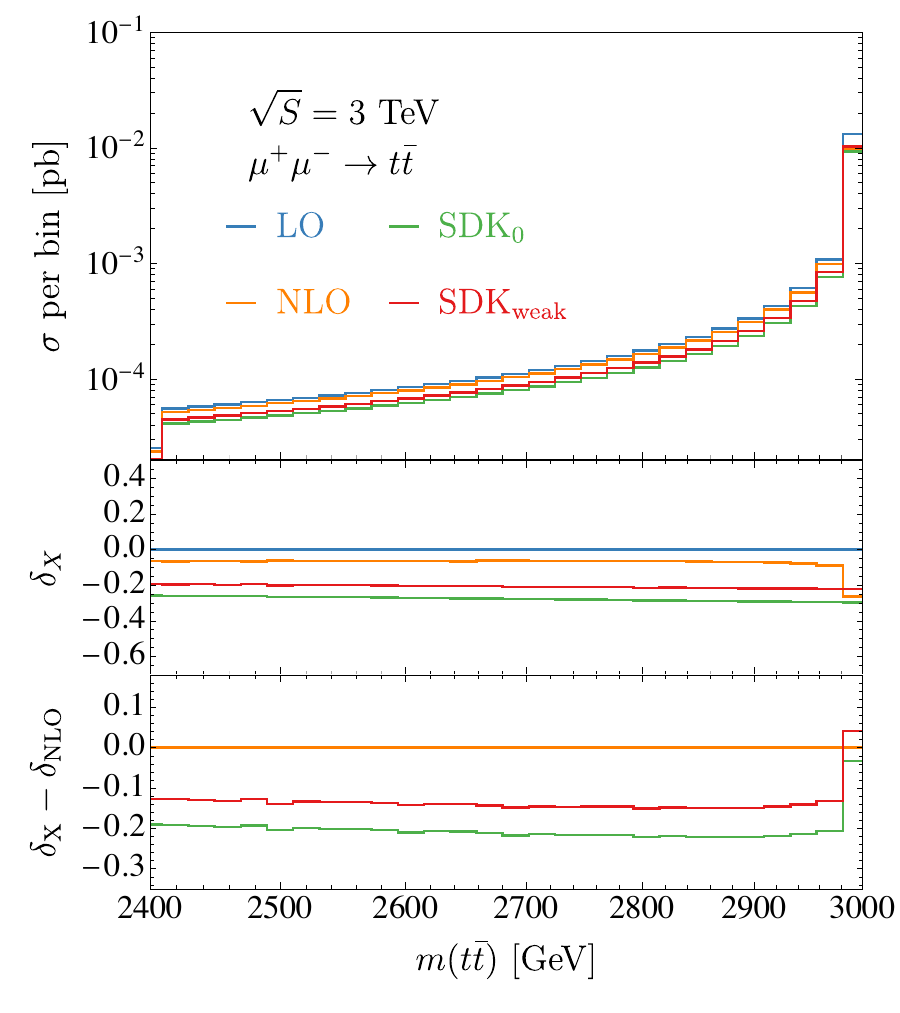}
  \includegraphics[width=0.48\textwidth]{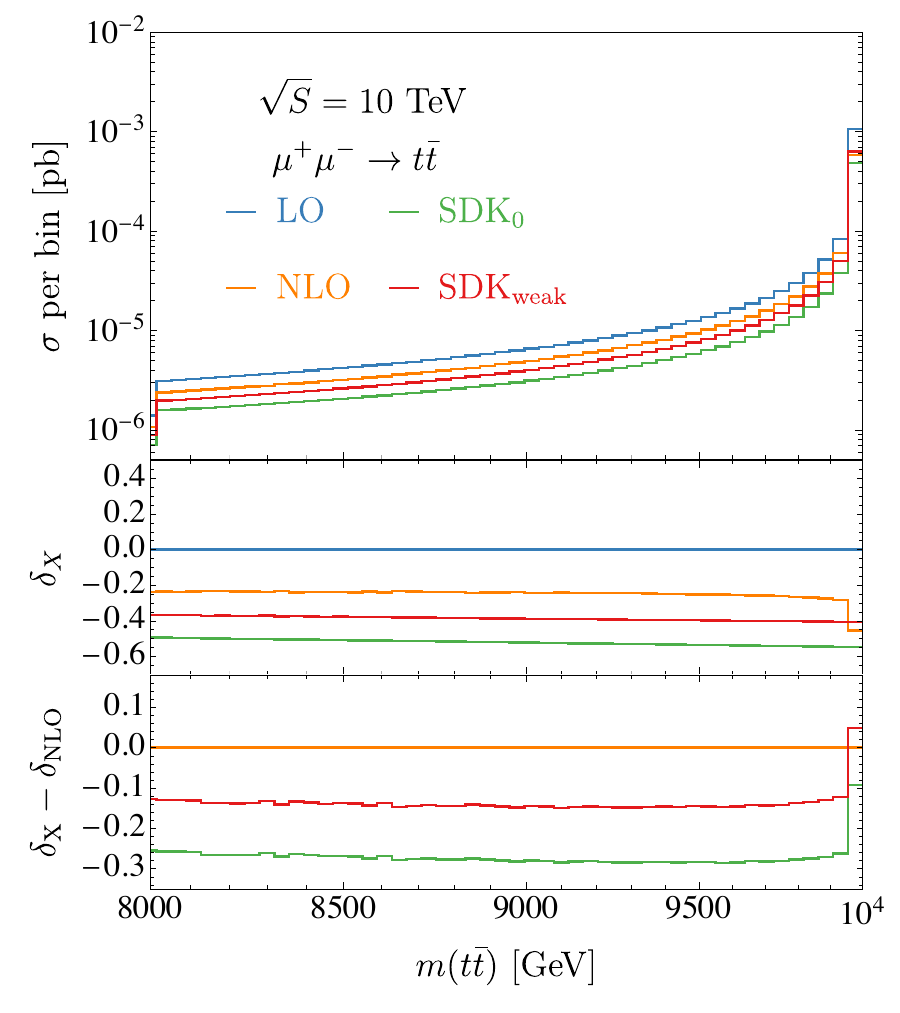} 
  \caption{Same as Fig.~\ref{fig:mmtt_ptt}, for the $m(t{\bar t})$ distribution in $\mu^+\mu^- \to t {\bar t}$.}\label{fig:mmtt_Mtt}
\end{figure}

Many of the points of the discussion of the $p_T(t)$ plots of Fig.~\ref{fig:mmtt_ptt} can be better understood by looking at the top-quark invariant mass distribution $m(t \bar t)$, which we show in Fig.~\ref{fig:mmtt_Mtt}. All the contributions from  LO, $\SDKw$ and $\SDKz$ predictions at Bjorken-$x\simeq 1$ enter the last bin at $m(t \bar t)\simeq \sqrt{S}$, while in the case of $\NLOEW$ predictions they can contribute over the full spectrum. This is the reason why, besides in the rightmost bin, the agreement between the $\NLOEW$ predictions and their Sudakov approximation is not good, regardless of the scheme choice. We remind the reader that we cluster the photon in the real emission with the top (anti)quark if they are collinear such that, for this kind of contributions, $m(t \bar t)\simeq S$ also in the presence of very hard photons.

\medskip

\begin{figure}[!t]
  \includegraphics[width=0.48\textwidth]{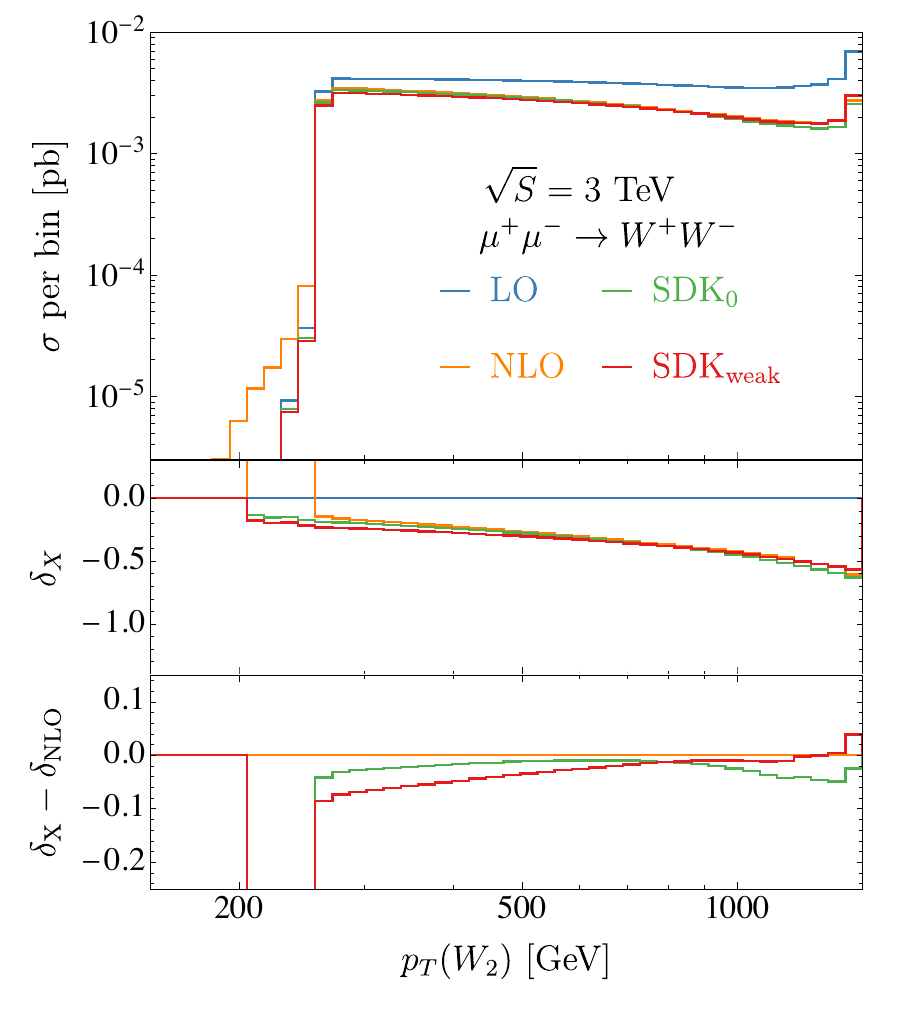}
  \includegraphics[width=0.48\textwidth]{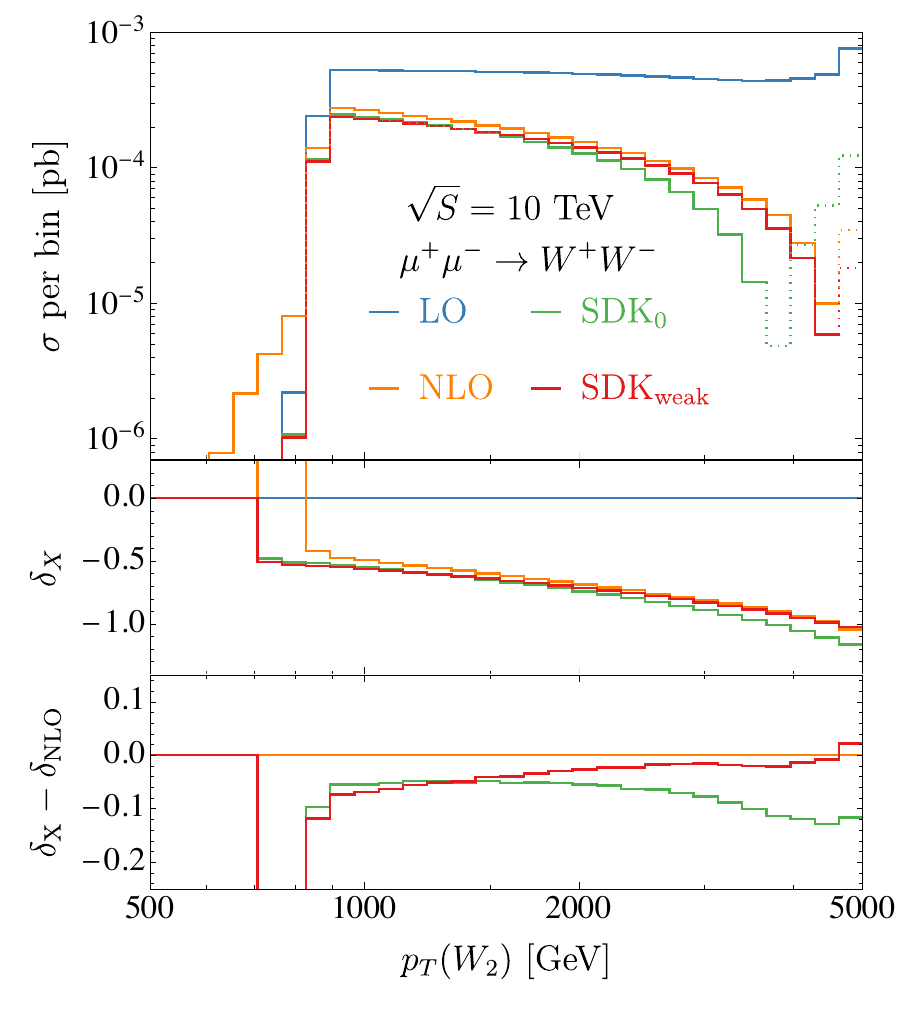} 
  \caption{Same as Fig.~\ref{fig:mmtt_ptt}, for the $p_T(W_2)$ distribution in $\mu^+\mu^- \to W^+ W^-$.}\label{fig:mmww_ptw2}
\end{figure}

We move now to the case of a different process, the $\mu^+\mu^-\to W^+  W^-$ production. In Fig.~\ref{fig:mmww_ptw2} we show the distribution of the transverse momentum of the softest $W$ boson, $p_T(W_2)$. Many features are common to the case of the $\mu^+\mu^-\to t \bar t$ production process in Fig.~\ref{fig:mmtt_ptt}. In the following, we highlight the differences rather than the similarities.

At variance with the $\mu^+\mu^-\to t \bar t$ production process, the tree-level amplitude of  $\mu^+\mu^-\to W^+  W^-$ production features $t$- and $u$-channel diagrams and consequently LO predictions are much less suppressed moving from large to small values of  $p_T(W_2)$ w.r.t.~what is observed in in Fig.~\ref{fig:mmtt_ptt}. Thus, the distributions are much flatter, excluding again the region $p_T(W_2)\lesssim250~\gev$ ($p_T(W_2)\lesssim800~\gev$) at 3 TeV (10 TeV) collisions, which is affected by the rapidity cuts in \eqref{eq:cuts}. 

The EW corrections, exact ($\NLOEW$) or in Sudakov approximation ($\SDKw$ or $\SDKz$) are much larger in the case of $W^+  W^-$, {\it cf.} Fig.~\ref{fig:mmtt_ptt} and Fig.~\ref{fig:mmww_ptw2}. At 10 TeV the EW corrections are so large that at large $p_T(W_2)$ values, where the cross section is maximal, $\deltaNLOEW<-100\%$ and therefore the $\NLOEW$ prediction becomes negative. This is a clear sign of the necessity of resumming the large EWSL and we will return to this aspect in Sec.~\ref{sec:resum}. Larger corrections are not surprising, since the couplings of the $W$ bosons with EW gauge bosons are larger w.r.t.~the top quarks.  The shape of  $\deltaNLOEW$ is also different w.r.t.~$t \bar t$ direct production. First, $\deltaNLOEW$ is much less flat, denoting a larger contribution from single logarithms, as well as the logarithms entering $\deltastorkl$ in Eq.~\eqref{eq:deltaSDKw}. Second, for larger values of $p_T(W_2)$, $\deltaNLOEW$ grows in absolute value, similarly to the typical shape observed at hadron colliders.

Moving to the comparison of the Sudakov approximation against the exact NLO EW corrections, the overall pattern is quite similar with a few differences w.r.t.~Fig.~\ref{fig:mmtt_ptt}. It is impressive how at large $p_T(W_2)$ values exact NLO EW corrections can be approximated at the level of $\ord(1\%)$ of the LO by the $\SDKw$ predictions (see second inset) when the corrections themselves are of $\ord(100\%)$ of the LO for 10 TeV collisions (see first inset). The same argument does not apply to the  $\SDKz$ predictions. Considering smaller values of $p_T(W_2)$, we see that the agreement of $\SDKw$ predictions and the exact $\NLOEW$ is less good w.r.t.~the case of $t \bar t$ production. To the best of our understanding, this is due to the logarithms involving the ratios of invariants as  $|t|/s$ or $|u|/s$. The  $\deltastorkl$ improves a lot the approximation of these contributions, see the discussion in the next section, but as already said it may miss some of such logarithms. In the case of $W^+W^-$ production, we see non-negligible effects due to these logarithms that are correctly captured only by $\NLOEW$ predictions. 

\medskip

\begin{figure}[!t]
  \includegraphics[width=0.48\textwidth]{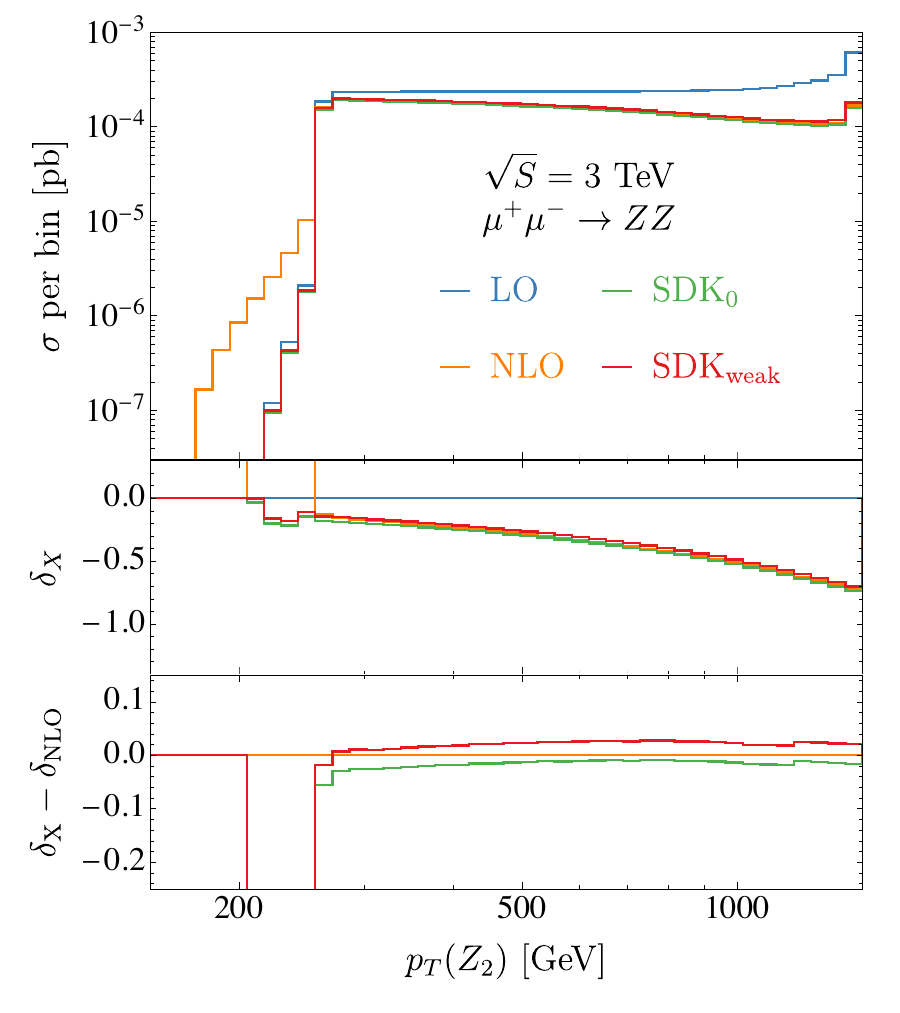}
  \includegraphics[width=0.48\textwidth]{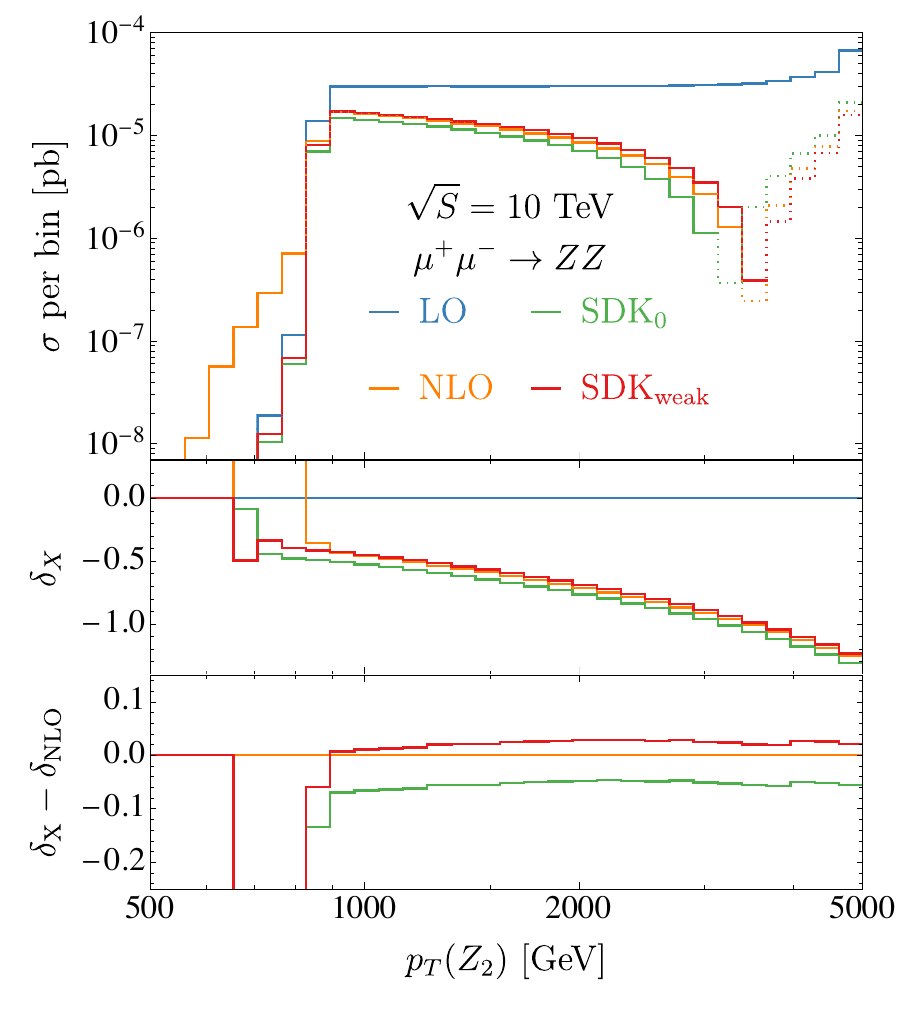} 
  \caption{Same as Fig.~\ref{fig:mmtt_ptt}, for the $p_T(Z_2)$ distribution in $\mu^+\mu^- \to Z Z$.} \label{fig:mmzz_ptz2}
\end{figure}

In Fig.~\ref{fig:mmzz_ptz2} we show the analogous distribution of Fig.~\ref{fig:mmww_ptw2} for $\mu^+\mu^-\to ZZ$ production, $p_T(Z_2)$. Here NLO EW corrections are even larger in absolute value than in the case of $WW$ production. Still, the $\SDKw$ approximation is again accurate at the level of  $\ord(1\%)$ of the LO. Since the particles in the final state are not electrically charged the choice of the $\SDKw$ scheme is not returning results that are very different w.r.t.~the $\SDKz$ one, especially for what concerns the shape of distributions, since the photon exchange between the initial and final state is not possible. Still, the muons in the initial state are electrically charged, so there are double logarithms of the form $L(s,\MW^2)\simeq L(S,\MW^2)$ that are treated differently in the two schemes and lead, especially at 10 TeV, to a constant difference between $\deltaSDKw$ and $\deltaSDKz$, degrading the agreement with $\deltaNLOEW$ for the latter. 

We have also considered the $m(W^+W^-)$ and $m(ZZ)$ distributions, which we do not show for brevity here. Similar to the case of $t \bar t$ production we see a large suppression due to PDFs for $m(W^+W^-),m(ZZ)<\sqrt{S}$. In the case of $m(W^+W^-)$ we see a similar pattern, although less dramatic when moving from the last bin, $m(W^+W^-)\simeq\sqrt S$, to the others, $m(W^+W^-)<\sqrt S$. In the case of $ZZ$ production, we do not cluster photons with $Z$ bosons, since from them no photon emissions leading to EWSL are possible. For the same reason, we do not see a discontinuity from the rightmost bin with  $m(ZZ)\simeq\sqrt S$ to the other ones with $m(ZZ)<\sqrt S$. For both processes, the contributions from hard photons collinear to the initial-state muons are subtracted by PDF counterterms in the $\NLOEW$ predictions. These are the same double logarithms mentioned in the previous paragraph and this subtraction is correctly taken into account by the $\SDKw$ scheme, which indeed exhibits for the $m(ZZ)$ distributions an $\ord(1\%)$ of the LO accuracy over the full spectrum.  

\medskip
\begin{figure}[!t]
  \includegraphics[width=0.48\textwidth]{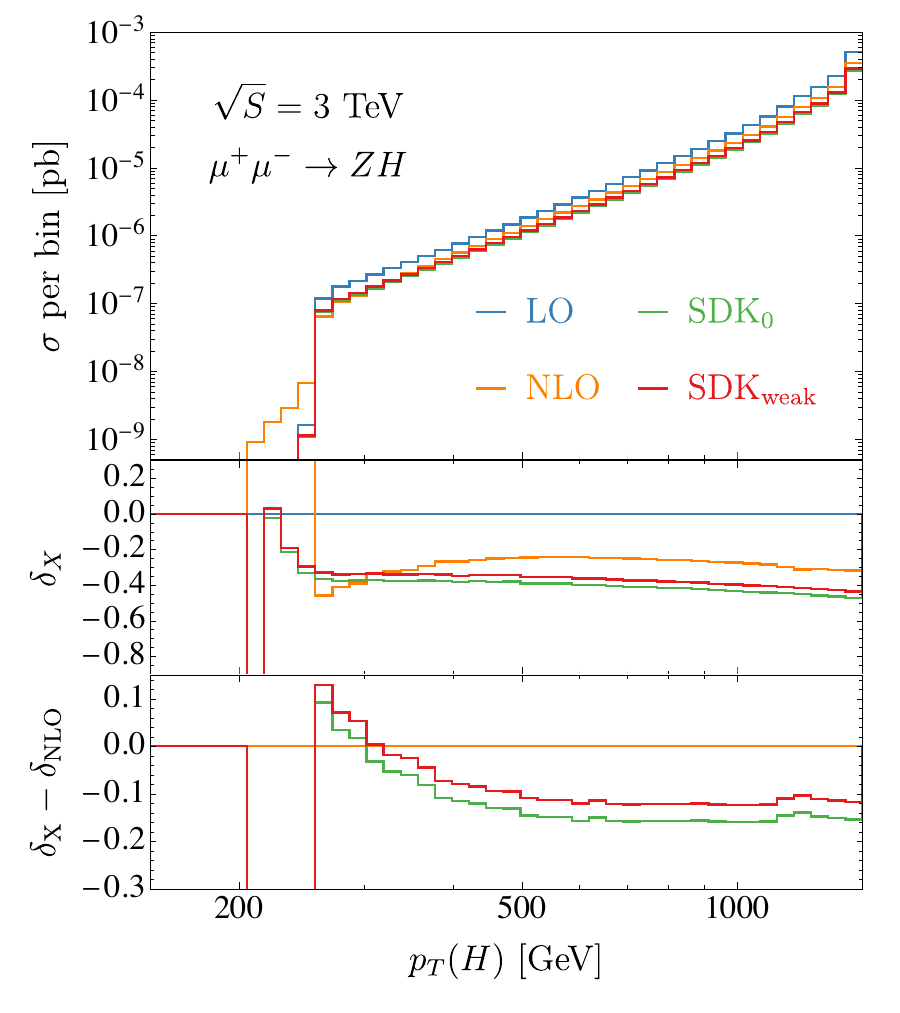}
  \includegraphics[width=0.48\textwidth]{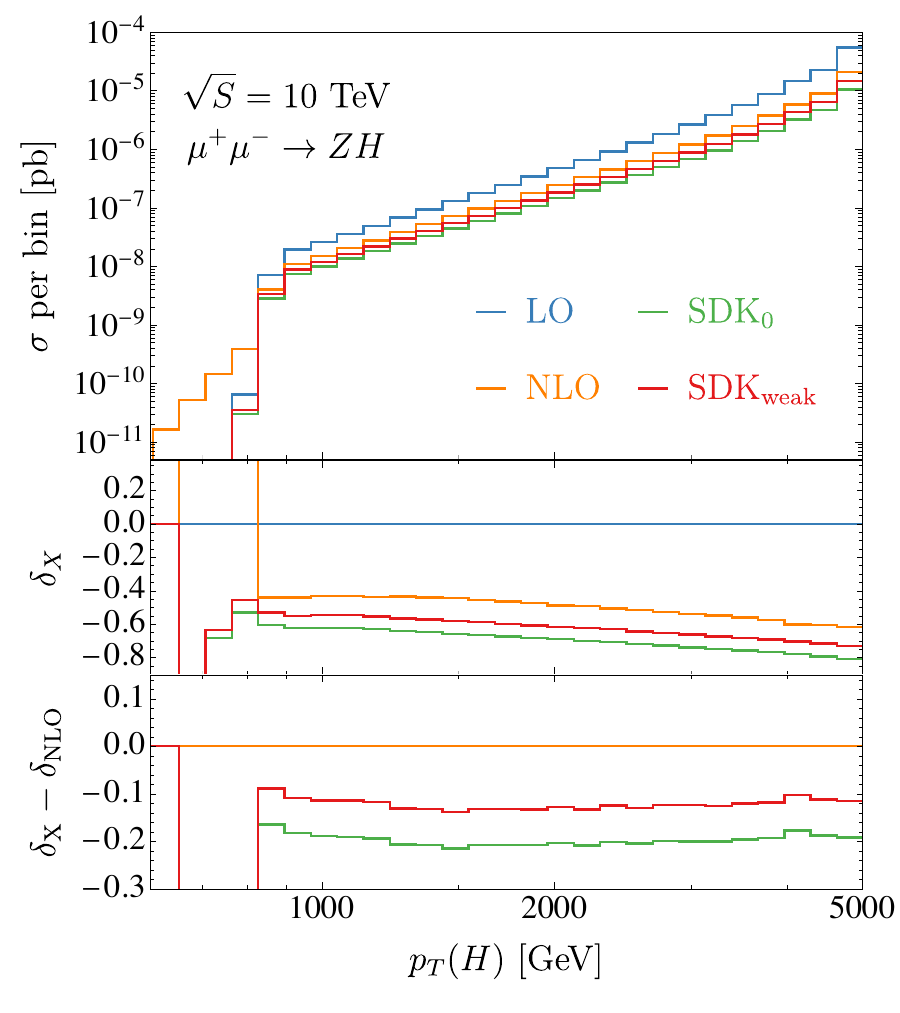}
  \caption{Same as Fig.~\ref{fig:mmtt_ptt}, for the $p_T(H)$ distribution in $\mu^+\mu^- \to Z H$.}\label{fig:mmzh_pth}
\end{figure}

Finally, in Fig.~\ref{fig:mmzh_pth} we show the transverse-momentum distribution for the Higgs boson in   $\mu^+\mu^- \to Z H$ production, $p_T(H)$. Similarly to the case of $ t \bar t$ production in Fig.~\ref{fig:mmtt_ptt}, we observe a much less flat LO prediction w.r.t.~the case of $WW$ and $ZZ$ production. As in the case of $ t \bar t$ production, and unlike $WW$ and $ZZ$, the tree-level amplitude features only an $s$-channel diagram. As in $ t \bar t$ production, and even more, $\deltaNLOEW$ also is quite flat over the full spectrum. We also observe that the difference $\deltaSDKw-\deltaNLOEW$ is larger, of the order of $10\%$ both for 3 and 10 TeV collisions. Instead, in the case of $\deltaSDKz-\deltaNLOEW$, such difference is of the order of $15\%$ at 3 TeV and  $20\%$ at 10 TeV. Thus, the accuracy of $\deltaSDKz$ is not only worse but also energy-dependent. To the best of our understanding, the $\mu^+\mu^- \to Z H$ has large non-logarithmic-enhanced contributions at $\ord(\alpha)$ and on top of that the  $\deltaSDKz$ wrongly approximates the double logarithms of the form $L(s,\MW^2)\simeq L(S,\MW^2)$ from the photon exchange among the muons in the initial state. Notice that the difference $\deltaSDKw-\deltaSDKz$ is the same observed,  both at 3 and 10 TeV, for $\mu^+\mu^-\to ZZ$ production in Fig.~\ref{fig:mmzz_ptz2}, for which the same argument was presented.

We have also calculated and analysed the $m(ZH)$ distribution. The situation is similar to the one observed for the $m(ZZ)$ distribution, but the   $\deltaSDKw-\deltaNLOEW$ remains constant at the order of $10\%$ as discussed for the $p_T(H)$ distribution.
Some of the features discussed in this section have also been observed in Ref.~\cite{Bredt:2022dmm}, where for the $ZZ$ and $ZH$ final state one can also find the analytical expression of the corresponding EWSL.

\subsubsection{Importance of logarithms involving ratios of invariants}
\label{sec:angular}

\begin{figure}[!t]
  \includegraphics[width=0.48\textwidth]{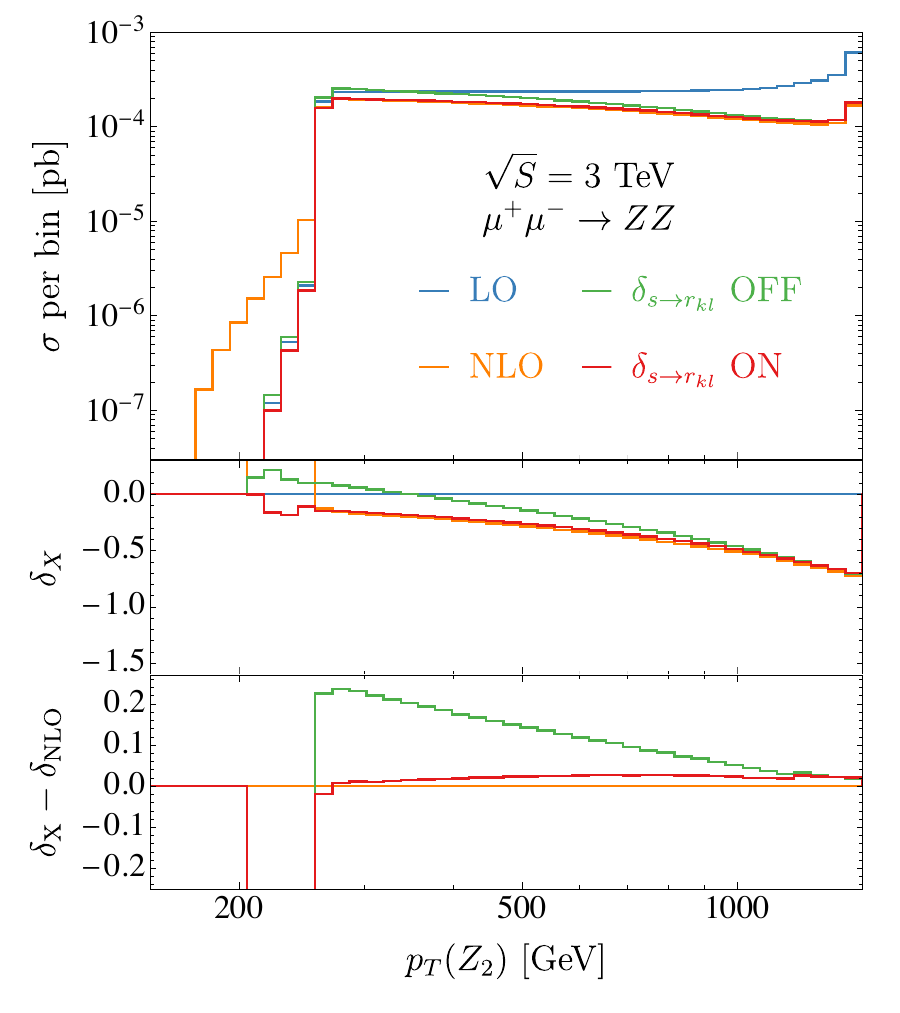}
  \includegraphics[width=0.48\textwidth]{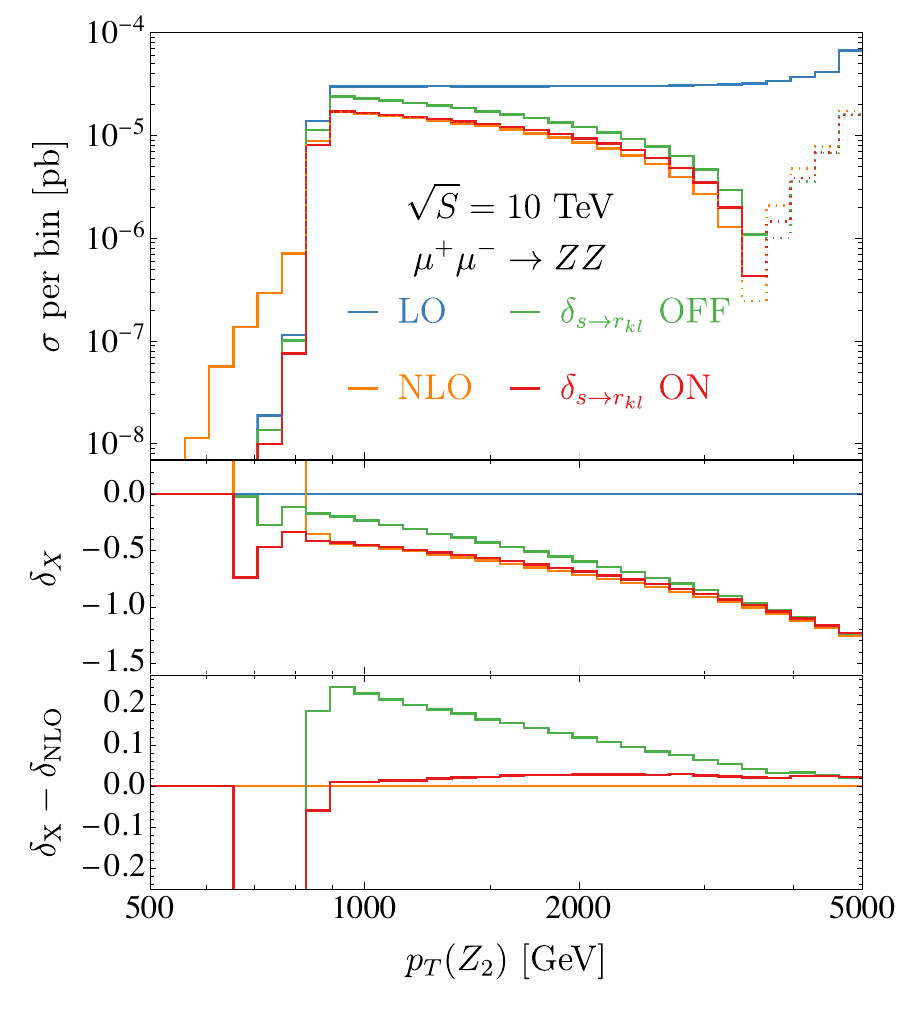}
  \caption{The $p_T(Z_2)$ distribution in $\mu^+\mu^- \to Z Z$.
   The left (right) plot shows results at $\sqrt S=3$ TeV ($\sqrt S=10$ TeV). The histograms show $\SigmaLO$ (blue) and  $\SigmaNLO$ (orange) as well as predictions in the Sudakov approximation in the $\SDKw$ approach including (red) or neglecting (green) the term $\deltastorkl$.
  }\label{fig:mmzz_Srij_ptz2}
\end{figure}

In this section, we discuss the relevance of the term $\deltastorkl$ entering Eq.~\eqref{eq:deltaSDKw} and introduced in Ref.~\cite{Pagani:2021vyk}. As already explained, this term accounts for (large part of the) logarithms of the form $\Lrs$ and $\lrsalpha$, see Eq.~\eqref{eq:generallogs}. Whenever a large hierarchy among invariants is present, these logarithms become numerically relevant. For processes as those studied in this work, where $\sqrt s \simeq \sqrt S=3$ or 10 TeV but transverse momenta can be a few hundred GeV's, $\deltastorkl$ is expected to be very relevant. One should notice that invariants can be small(large) due to small(large) angles among two particles and therefore angular distributions are very sensitive to these logarithms.

In order to minimise the overlap with the effects discussed in the previous section, $\SDKw$ {\it vs.}~$\SDKz$, we consider final states with only neutrally charged particles, in particular: the $ZZ$,  $ZH$,  $ZZZ$, and $ZZH$ production processes. The layout of the plots in this section is very similar to those shown in the previous section. The only difference w.r.t.~them is that we show here $\SDKw$ as defined in Eq.~\eqref{eq:deltaSDKw} (again displayed as a solid red line) and the same quantity where we set $\deltastorkl=0$ (solid green line) in the aforementioned equation.

We start by showing again the same observables considered for $ZZ$ and $ZH$ production in the previous section, $p_T(Z_2)$ for $ZZ$ production in Fig.~\ref{fig:mmzz_Srij_ptz2} and $p_T(H)$ for $ZH$ production in Fig.~\ref{fig:mmzh_Srij_pth}. 
In Fig.~\ref{fig:mmzz_Srij_ptz2} we clearly see that for $p_T(Z_2)\gtrsim250~\gev$ ($p_T(Z_2)\gtrsim800~\gev$) at 3 TeV (10 TeV) collisions, the very good accuracy of the $\SDKw$ prediction ($\deltaSDKw-\deltaNLOEW$ constant and of $\ord(1\%)$) is much degraded when $\deltastorkl=0$, {\it i.e.} the green line in the second inset. Indeed, while at large $p_T(Z_2)$ we see $\deltaSDKw-\deltaNLOEW\simeq 1\%$, at smaller values,   $p_T(Z_2)\simeq250~\gev$ ($p_T(Z_2)\simeq 800~\gev$) at 3 TeV (10 TeV), we notice that  $\deltaSDKw-\deltaNLOEW\simeq 30\%$. Since the $p_T(Z_2)$ distribution is quite flat, this discrepancy at small $p_T(Z_2)$ has an effect also at the level of the total cross section; setting $\deltastorkl=0$ we find that, for both 3 and 10 TeV, with the cuts considered $\deltaSDKw-\deltaNLOEW\simeq 10\%$ also for the total cross section. A similar (quite constant) discrepancy is observed in the $m(ZZ)$ distribution, too. These results are clearly dependent on the cuts in $\eqref{eq:cuts}$, in particular the one on the pseudorapidity of the $Z$ bosons. Setting $\deltastorkl=0$ logarithms of the form, {\it e.g.}, $L(|t|,s)$ are omitted and, in the proximity of the pseudorapidity cuts, such logarithms are of the same order for both the  3 and 10 TeV results.

\begin{figure}[!t]
  \includegraphics[width=0.48\textwidth]{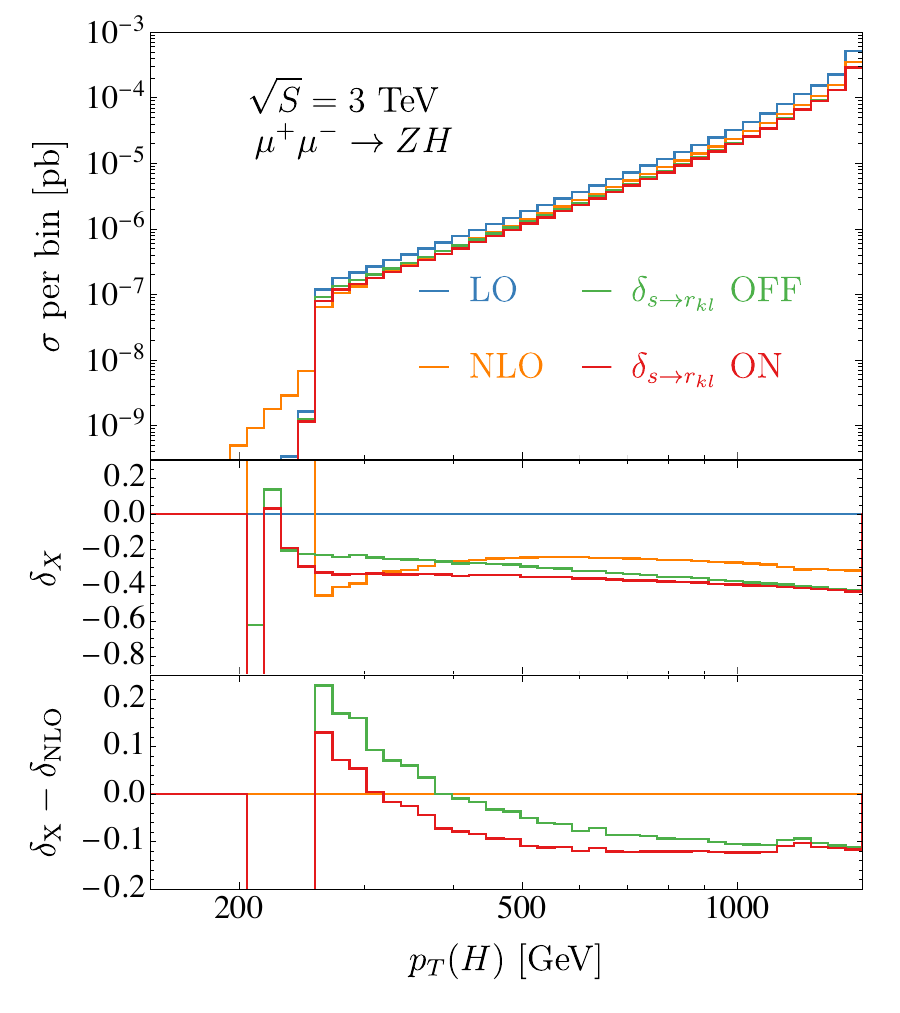}
  \includegraphics[width=0.48\textwidth]{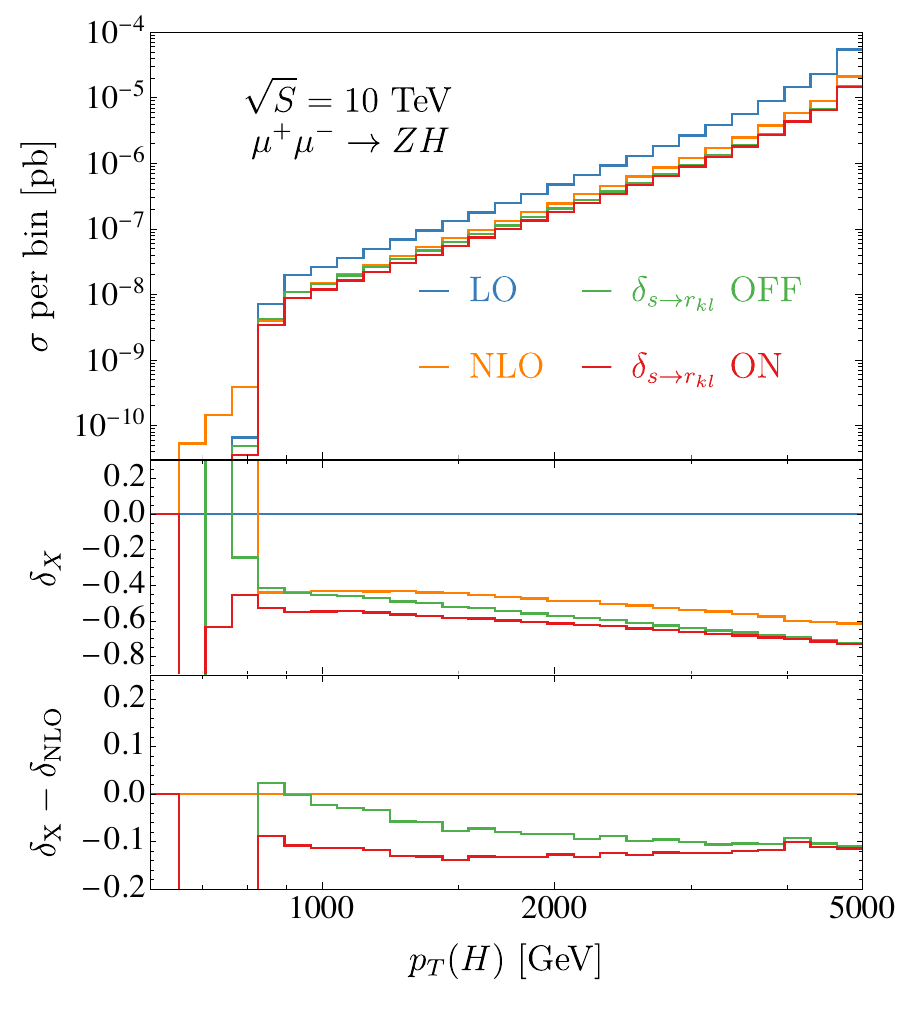}
  \caption{Same as Fig.~\ref{fig:mmzz_Srij_ptz2}, for the $p_T(H)$ distribution in $\mu^+\mu^- \to Z H$.}\label{fig:mmzh_Srij_pth}
\end{figure}

The case of $p_T(H)$ distribution in $ZH$ production, Fig.~\ref{fig:mmzh_Srij_pth}, shows a very similar pattern at the differential level, although the impact of $\deltastorkl$ is smaller. Moreover, since the $p_T(H)$ distribution is much less flat, it is manifest that at the inclusive level the accuracy for the $\SDKw$ prediction is not affected by the assumption $\deltastorkl=0$. It is interesting to note that for small values of  $p_T(H)$ at 10 TeV the case with $\deltastorkl=0$ yields smaller values of $\deltaSDKw-\deltaNLOEW$. As said at the beginning of the previous section, this is not {\it per se} a sign of better accuracy. Indeed this effect is due to missing logarithms among invariants that accidentally compensate the large non-logarithmically enhanced $\ord(\alpha)$ component already discussed in the case of Fig.~\ref{fig:mmzh_pth}.

\begin{figure}[!t]
  \includegraphics[width=0.48\textwidth]{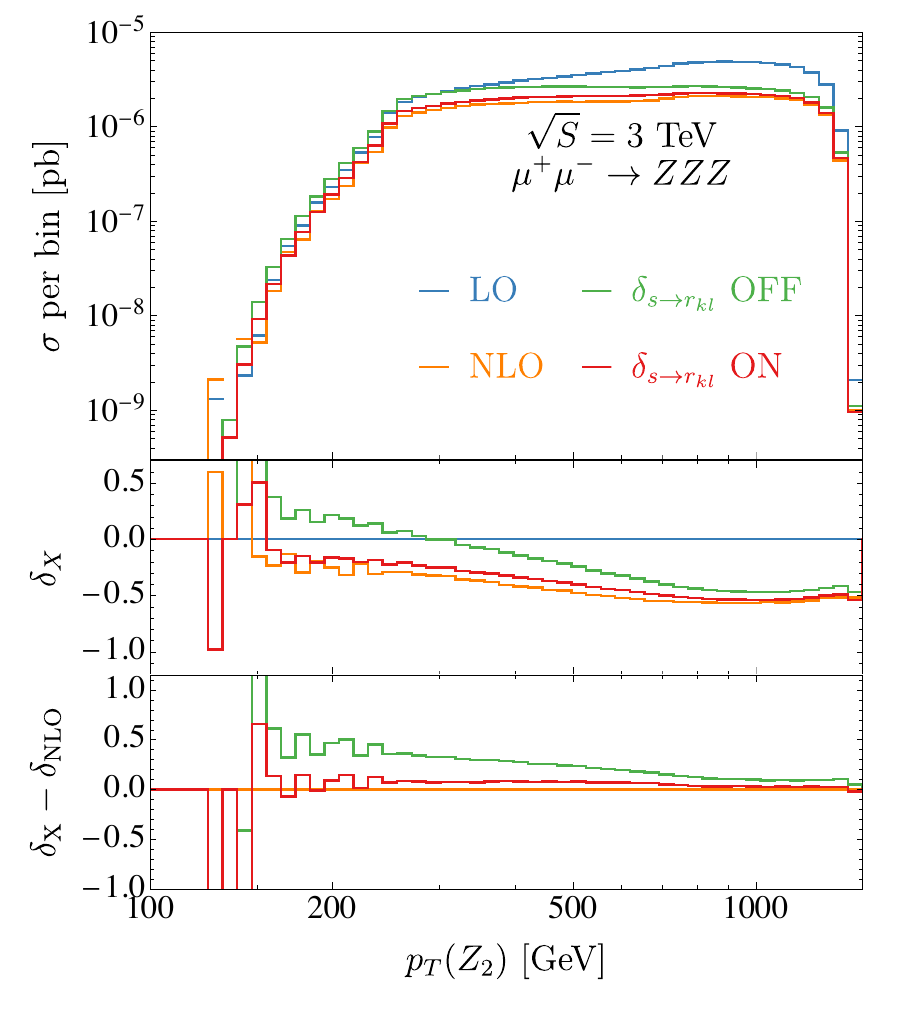}
  \includegraphics[width=0.48\textwidth]{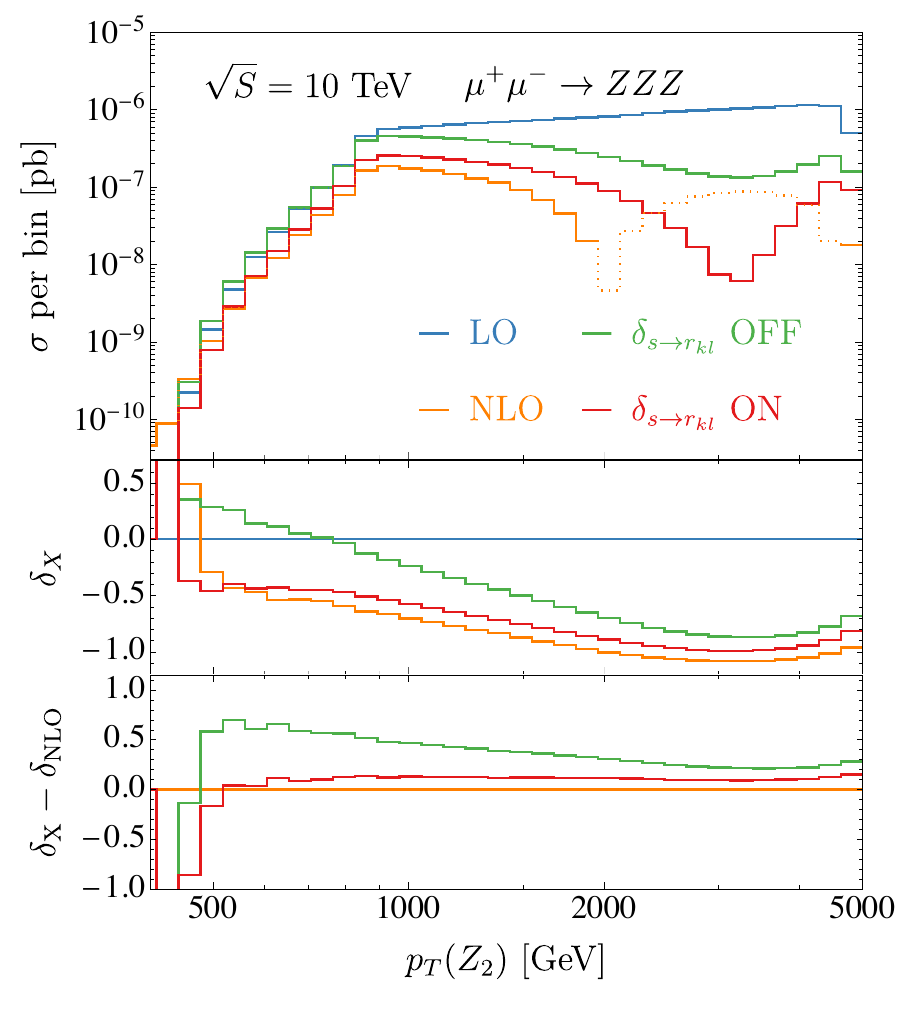} 
  \caption{Same as Fig.~\ref{fig:mmzz_Srij_ptz2}, for the $p_T(Z_2)$ distribution in $\mu^+\mu^- \to Z Z Z$.}\label{fig:mmzzz_Srij_ptz2}
\end{figure}

We move now to the case of $2\to3$ processes, $ZZZ$, and $ZZH$ production, for which more independent kinematic invariants are present. In Fig.~\ref{fig:mmzzz_Srij_ptz2} we show the transverse-momentum distribution of the second-hardest $Z$ boson, $p_T(Z_2)$. As can be noticed, EW corrections are very large ($\deltaNLOEW<-100\%$ at 10 TeV in the bulk of the distribution), and the $\SDKw$ approximation (solid line) is able to capture correctly the kinematic dependence with a constant discrepancy  $\deltaSDKw-\deltaNLOEW\simeq 10\%$. On the contrary, setting $\deltastorkl=0$ (dashed line), we observe a constant growth of $\deltaSDKw-\deltaNLOEW$ moving to small $p_T(Z_2)$ values.

\begin{figure}[!t]
  \includegraphics[width=0.48\textwidth]{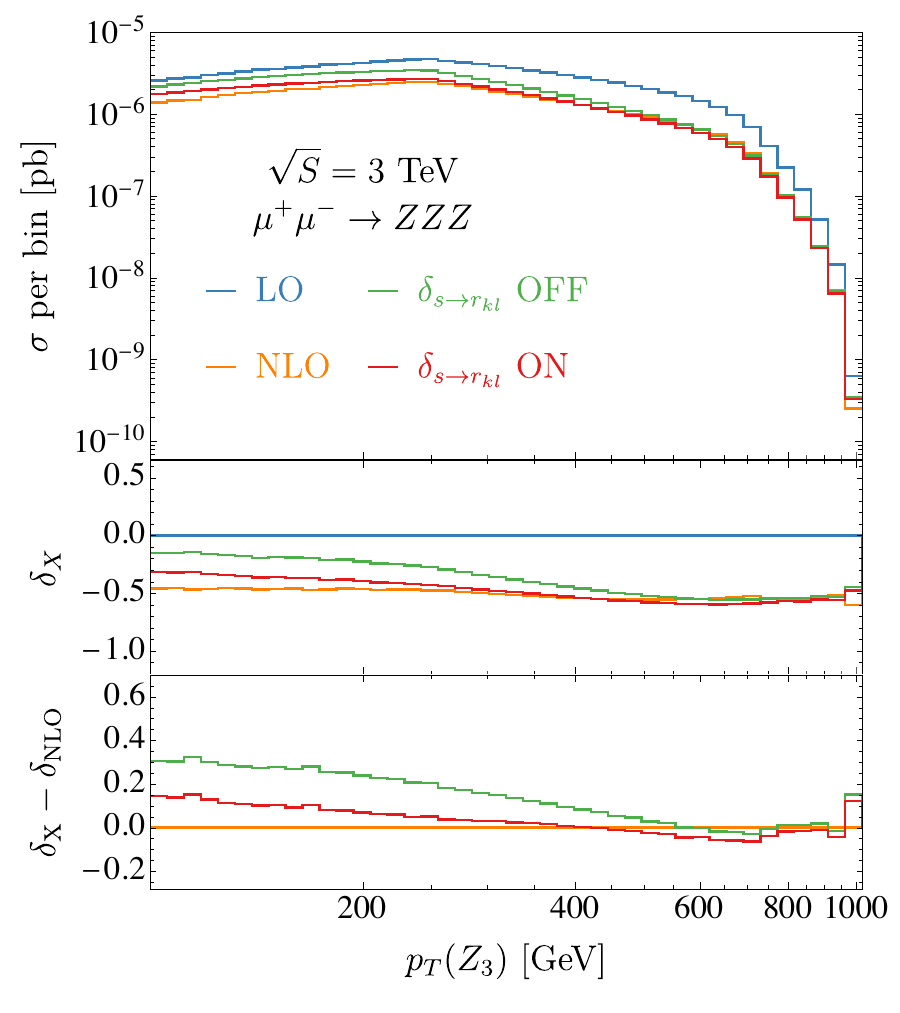}
  \includegraphics[width=0.48\textwidth]{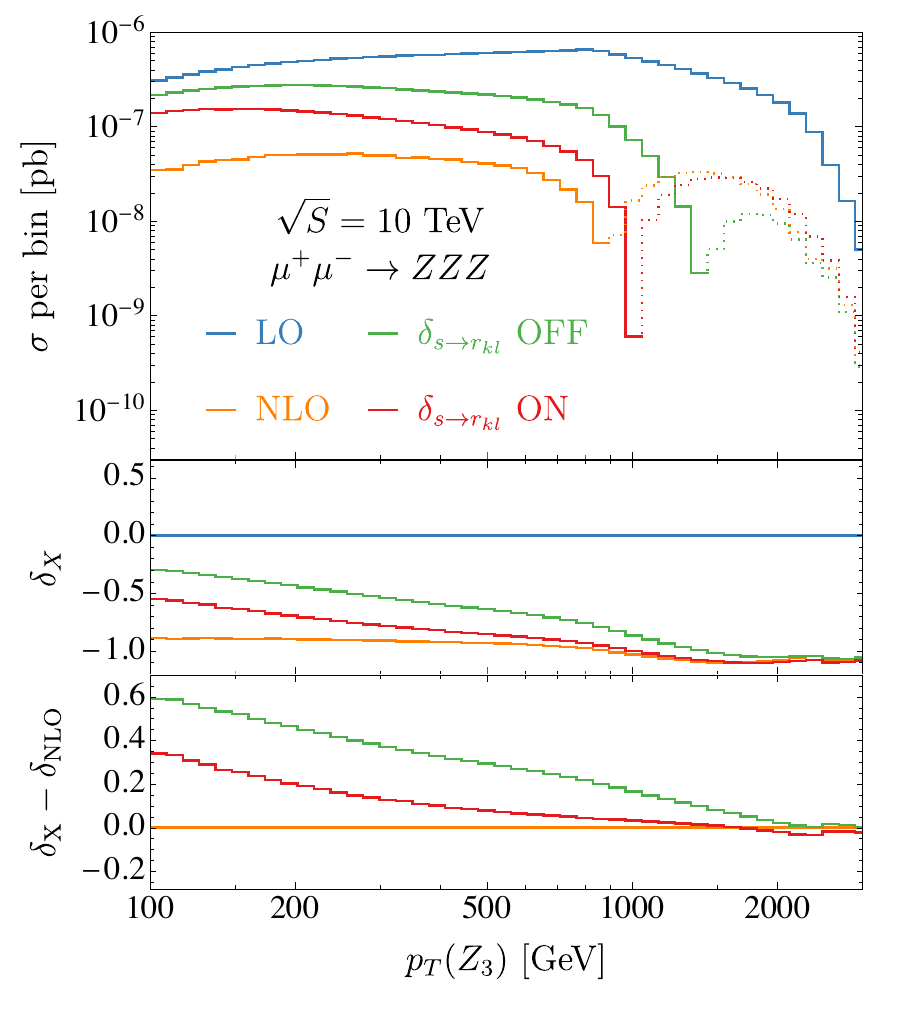}
  \caption{Same as Fig.~\ref{fig:mmzz_Srij_ptz2}, for the $p_T(Z_3)$ distribution in $\mu^+\mu^- \to Z Z Z$.}\label{fig:mmzzz_Srij_ptz3}
\end{figure}

\begin{figure}[!t]
  \includegraphics[width=0.48\textwidth]{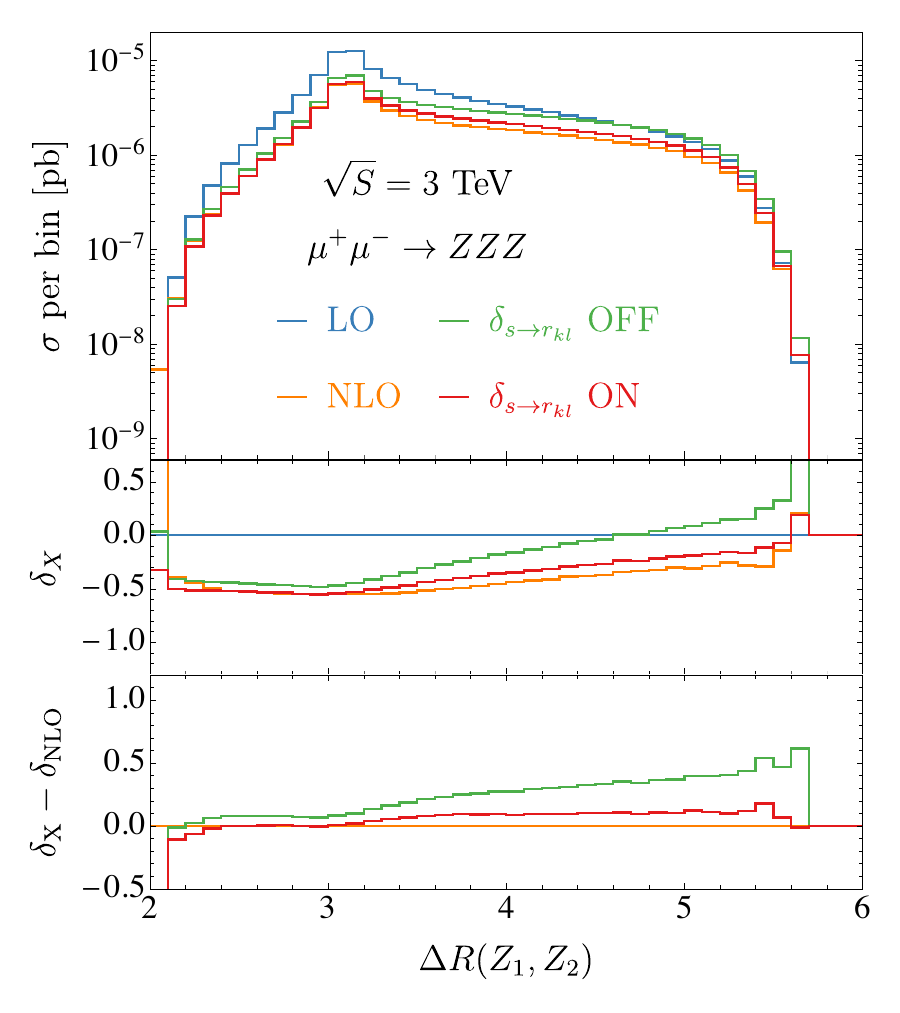}
  \includegraphics[width=0.48\textwidth]{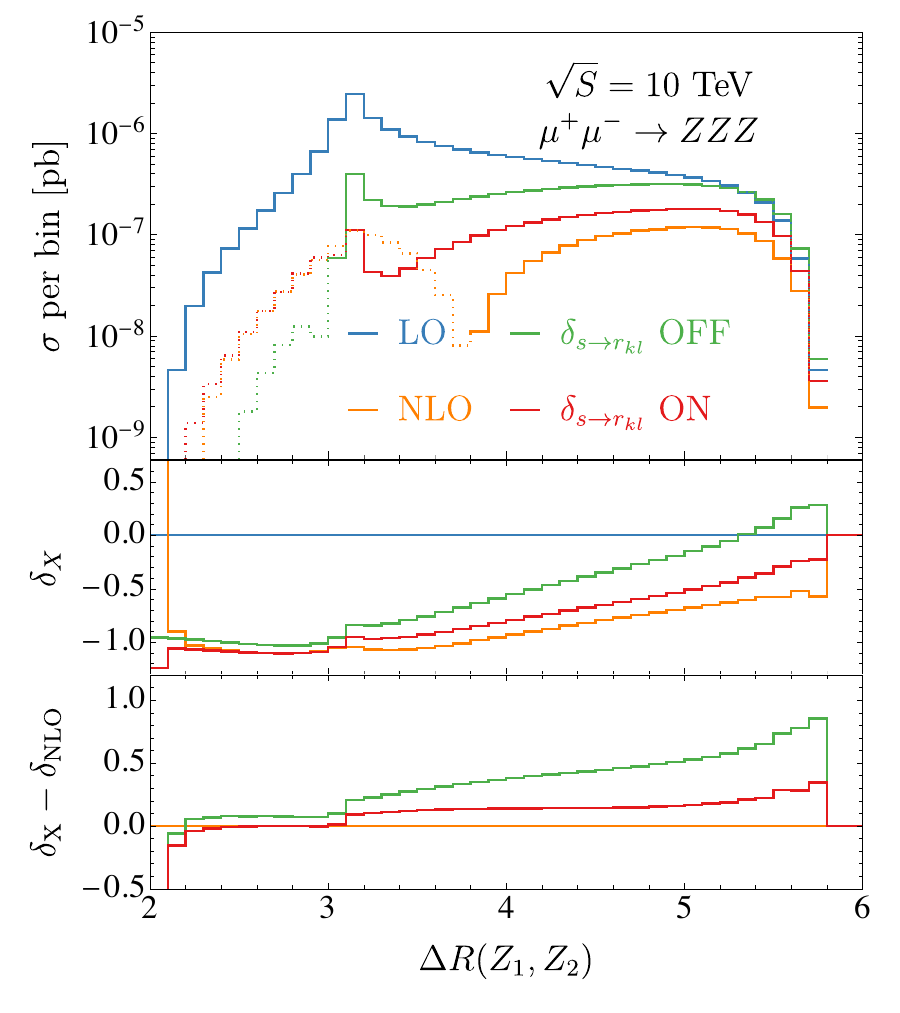}  
  \caption{Same as Fig.~\ref{fig:mmzz_Srij_ptz2}, for the $\Delta R(Z_1, Z_2)$ distribution in $\mu^+\mu^- \to Z Z Z$.}\label{fig:mmzzz_Srij_Drz1z2}
\end{figure}

To the best of our understanding, this $10\%$ discrepancy is not due to a large non-logarithmically enhanced $\ord(\alpha)$ component, as in the case of $ZH$ distributions, but to large logarithms that the $\SDKw$ scheme is not able to capture even retaining the $\deltastorkl$ term. This can be better understood by looking at the $p_T(Z_3)$ distribution, the $p_T$ of the softest $Z$ boson, in Fig.~\ref{fig:mmzzz_Srij_ptz3} and the $\Delta R (Z_1,Z_2)$ distribution in Fig.~\ref{fig:mmzzz_Srij_Drz1z2}, in particular at 10 TeV. First we observe that for very large $p_T(Z_3)$, meaning all $Z$ bosons that are hard and so all the invariants that are large, $\deltaSDKw-\deltaNLOEW\to 0$, while the same quantity constantly grows in the opposite direction.
Second, for $\Delta R (Z_1,Z_2) \lesssim \pi$, $\deltaSDKw-\deltaNLOEW\simeq 0$, while for  $\Delta R (Z_1,Z_2) \gtrsim \pi$ the same quantity jumps to $\sim 10\%$ and remains constant up to $\Delta R (Z_1,Z_2) \simeq 5$. The dominant kinematic configuration is $Z_1$ and $Z_2$ that are almost back-to-back, {\it i.e.} $\Delta R (Z_1,Z_2) \simeq \pi$. The region $\Delta R (Z_1,Z_2) \lesssim \pi$ is dominated by large values of $p_T(Z_3)$, which therefore are correlated and both show $\deltaSDKw-\deltaNLOEW\simeq 0$. Instead, in the region $\Delta R (Z_1,Z_2) \gtrsim \pi$ large contributions from the $\NLOEW$ prediction originate from the $ZZZ\gamma$ final state, which allows a further recoil and enhances $|\eta(Z_1)-\eta(Z_2)|$ and in turn $\Delta R (Z_1,Z_2)$. This dynamics is only captured by the $\NLOEW$ prediction. 
Another peculiar behaviour is observed in the first bins of the distribution, where the EWSL prediction departs from the NLO EW. This is due to the fact that, in a Born-like kinematics, the first bin can be filled only when all $Z$ bosons have equal transverse momentum. Photon radiation, captured only by the NLO EW prediction, lifts such a constraint, and can thus enhance this region
All in all, for this process, the inclusion of $\deltastorkl$ is crucial for improving the approximation. Nevertheless, non-negligible effects are not captured.

\begin{figure}[!t]
  \includegraphics[width=0.48\textwidth]{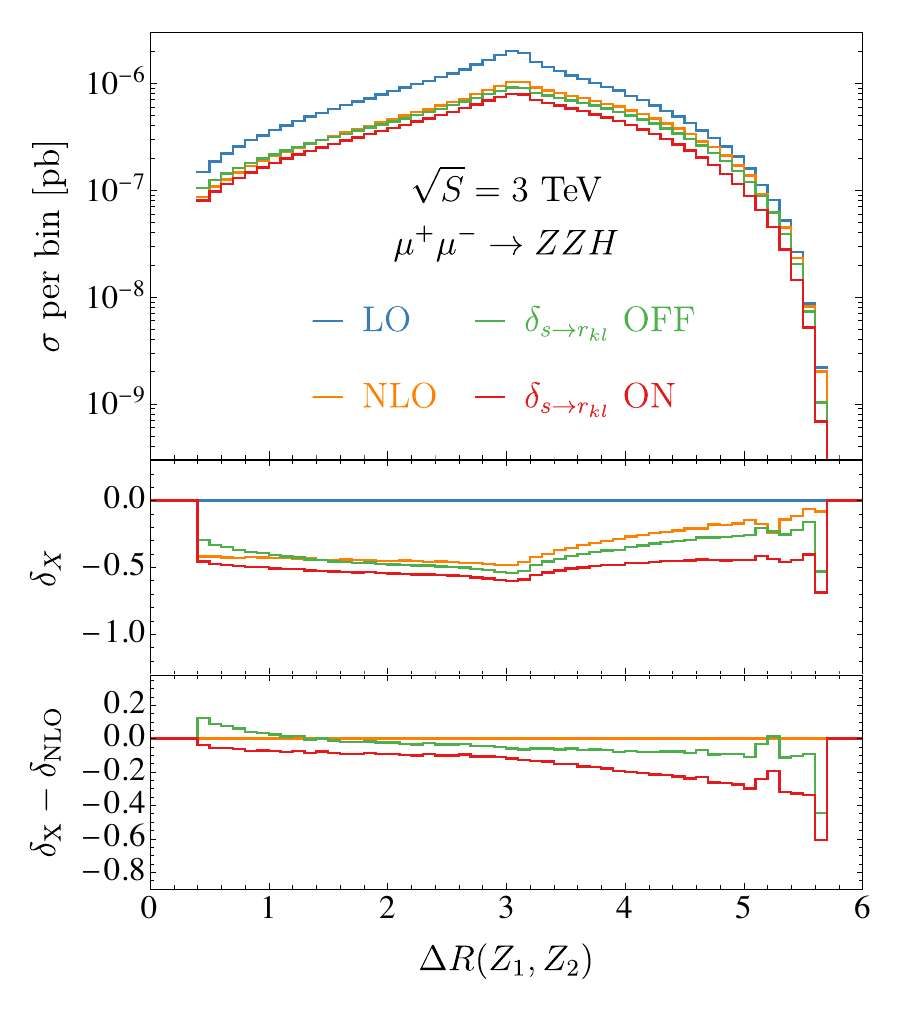}
  \includegraphics[width=0.48\textwidth]{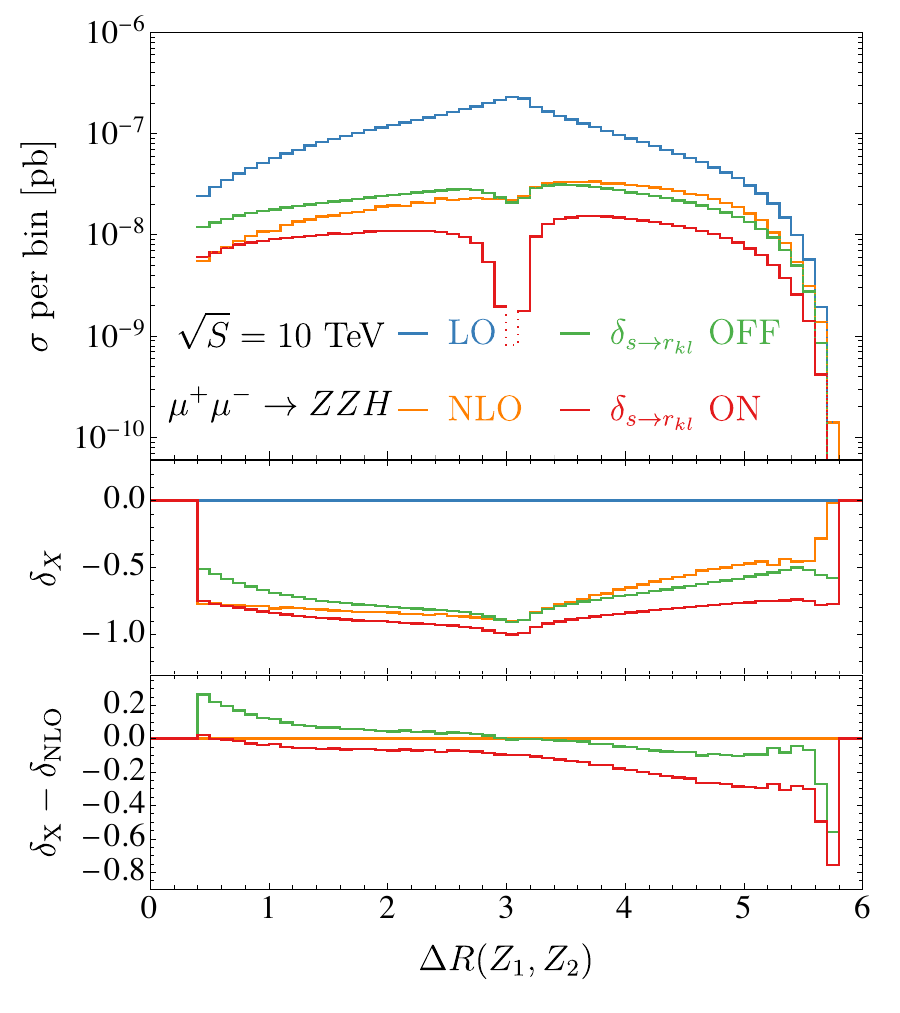}
  \caption{Same as Fig.~\ref{fig:mmzz_Srij_ptz2}, for the $\Delta R(Z_1, Z_2)$ distribution in $\mu^+\mu^- \to Z Z H$.} \label{fig:mmzzh_Srij_Drzz}
\end{figure}
In Fig.~\ref{fig:mmzzh_Srij_Drzz} we show the $\Delta R(Z_1, Z_2)$ distribution for $Z Z H$ production. In this case, the situation is a combination of effects observed already for the $ZH$ and $ZZZ$ processes. To the best of our understanding, there are both logarithms that cannot be correctly captured, as in $ZZZ$, and a large non-logarithmically enhanced $\ord(\alpha)$ component.

In conclusion, considering also other distributions for other processes that we have calculated but not shown in the paper, the quantity $\deltastorkl$ in general improves the approximation of the EWSL, but additional contributions present at $\NLOEW$ accuracy can be omitted. For precise predictions, such contributions cannot be neglected.

\subsubsection{The case of numerically large contributions from mass-suppressed terms.}
\label{sec:failMS}

As clearly stated in Refs.~\cite{Denner:2000jv,Denner:2001gw}, the {\denpoz} algorithm assumes that the helicity configuration considered is {\it not} mass suppressed, {\it i.e.}, as said in Sec.~\ref{sec:EWSLamp}, that it scales as  $\M\propto s ^{\frac{2-n}{2}}$ for a $2 \to n$ process. In general, in the SM, at least one of the helicity configurations of the processes considered is typically {\it not} mass suppressed and therefore at high energies is very enhanced  w.r.t.~the other ones.\footnote{This enhancement is very large,  by at least $\ord(s/\MW^2)$.} For this reason, even blindly applying the {\denpoz} algorithm to all helicity configurations, regardless if they are or are not mass-suppressed, the prediction for the sum over the polarisations is consistent, namely it corresponds to the high-energy limit $\MW^2/s\to 0$. In other words, even if the algorithm returns wrong results for the mass-suppressed helicity configurations, they are so suppressed that the relative impact in the sum over the helicity configurations is completely negligible. However, it is known that there can be processes where none of the helicity configurations are {\it not} mass-suppressed, such as Higgs VBF production.\footnote{In an upcoming paper, this aspect will be discussed in detail in the context of the SM Effective-Filed theory (SMEFT) \cite{ElFaham:2024egs}, where these cases are much more common.} For such cases, the {\denpoz} algorithm is known to no be working.

In this section, we give an example that is a bit more subtle. We consider the case of the $\mu^+\mu^- \to Z H H$ production process, which does feature helicity configurations that are {\it not} mass suppressed, but that in part of the phase space are not numerically the dominant ones. Thus, the {\denpoz} algorithm leads to wrong results for the evaluation of high-energy limit $\MW^2/s\to 0$  of the EW corrections.

\begin{figure}[!t]
  \includegraphics[width=0.48\textwidth]{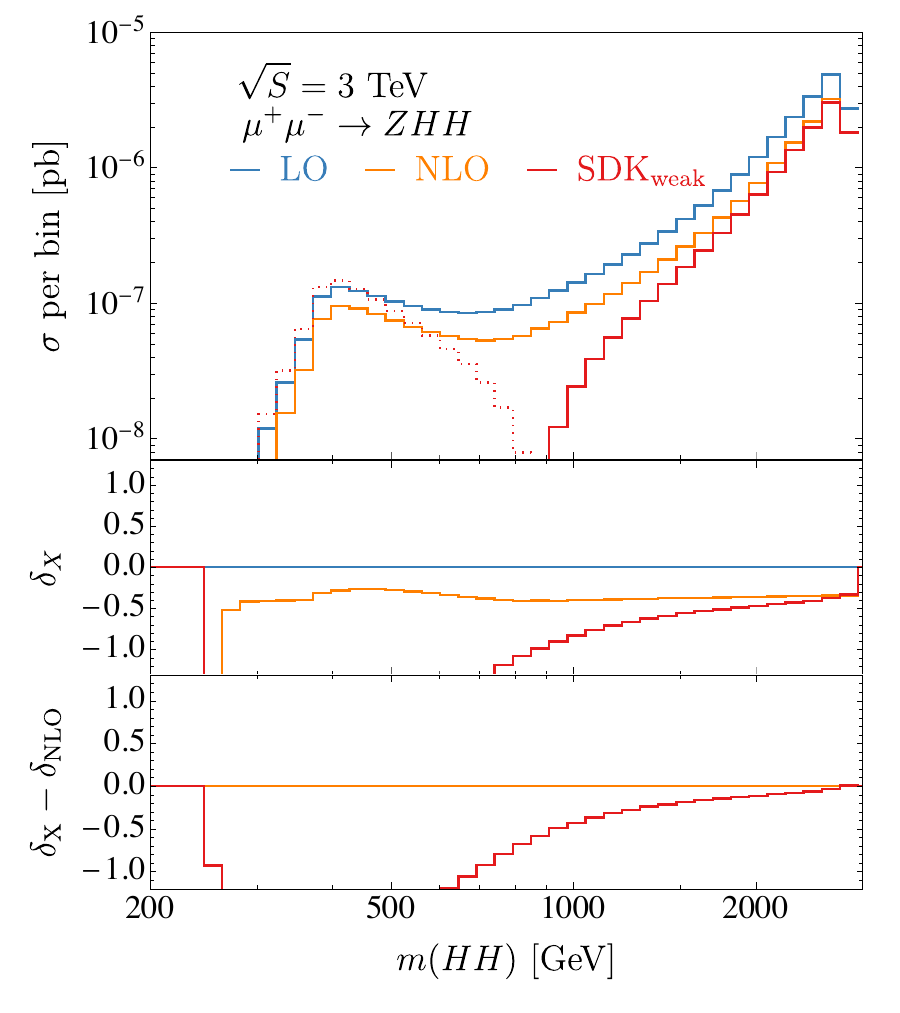}
  \includegraphics[width=0.48\textwidth]{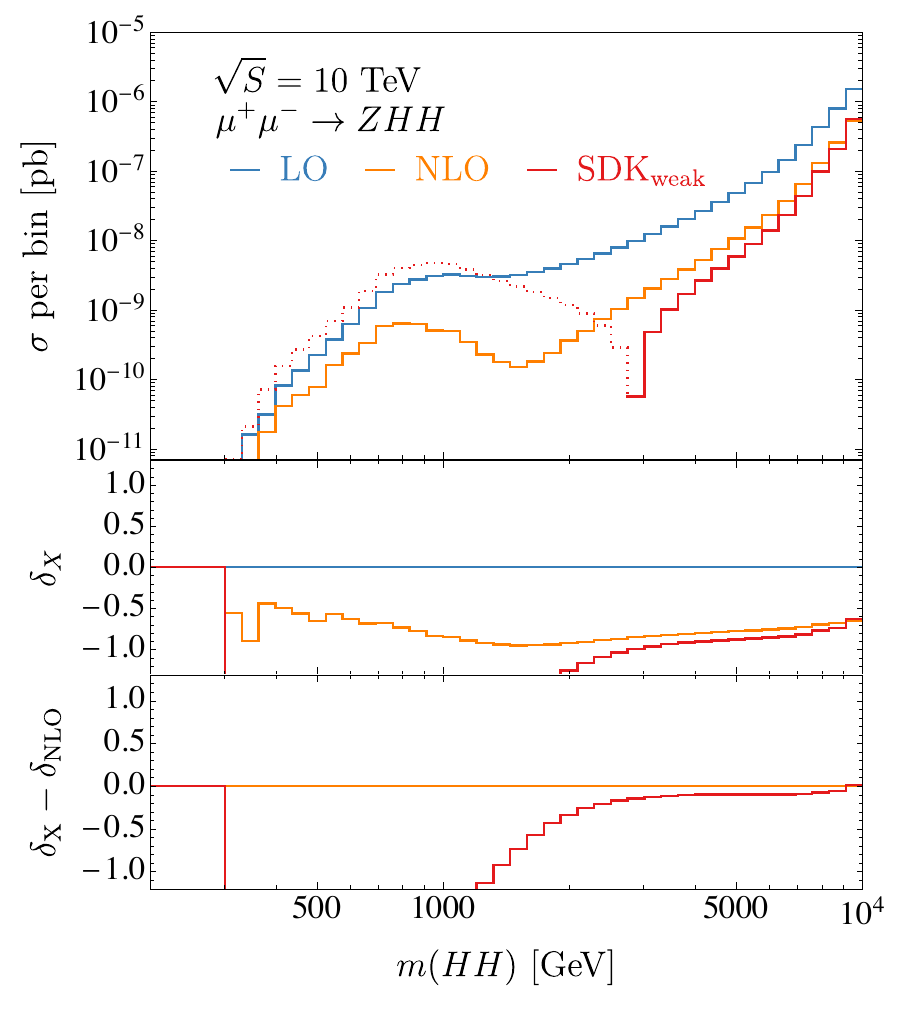} 
  \caption{The  $m(HH)$ distribution in $\mu^+\mu^- \to Z H H$.
  The left (right) panel shows results at $\sqrt S=3$ TeV ($\sqrt S=10$ TeV). The histograms show $\SigmaLO$ (blue), $\SigmaNLO$ (orange) and  $\SigmaSDKw$.
  }\label{fig:mmzhh_Mhh}
\end{figure}

\begin{figure}[!t]
  \includegraphics[width=0.48\textwidth]{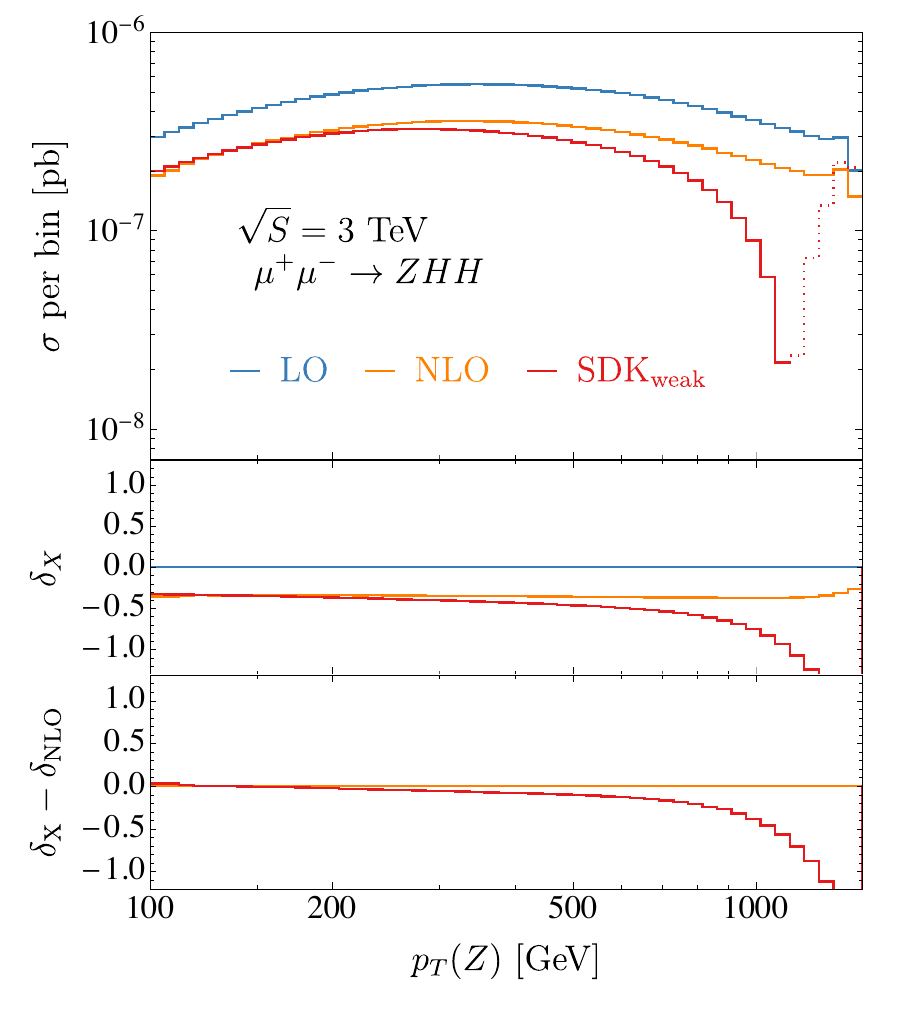}
  \includegraphics[width=0.48\textwidth]{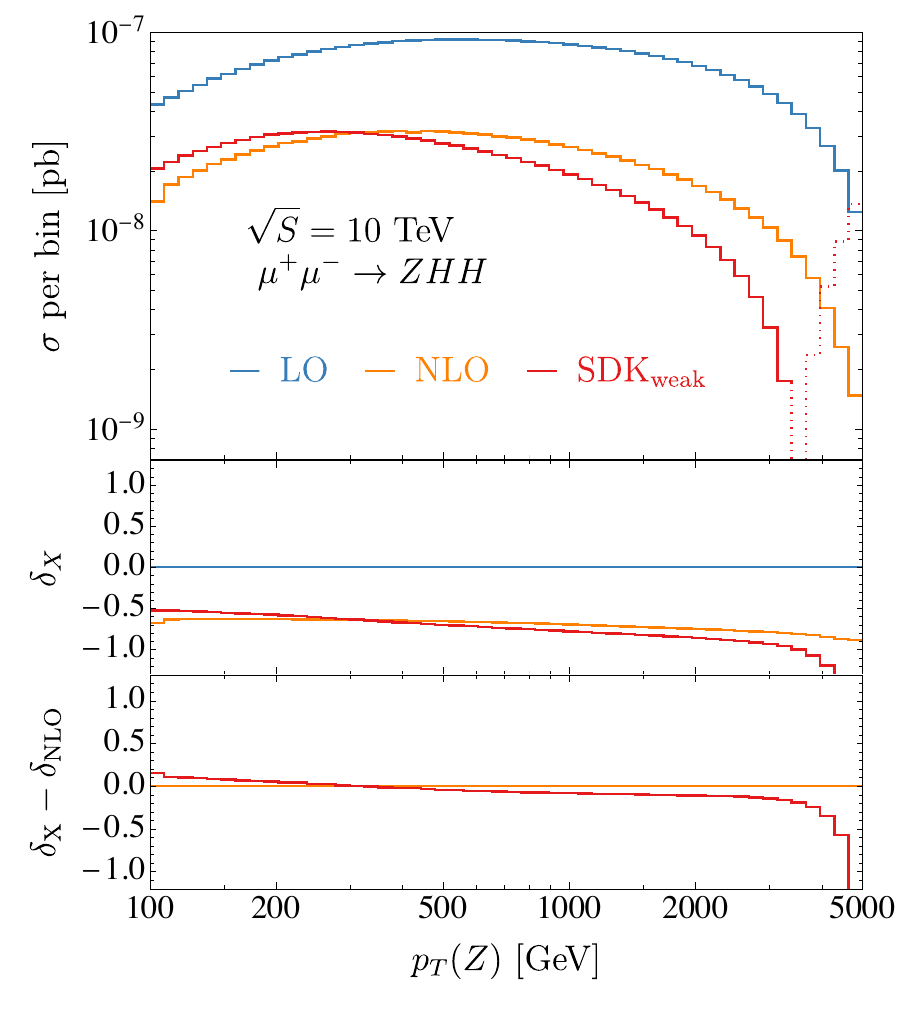} 
  \caption{Same as Fig.~\ref{fig:mmzhh_Mhh}, for the $p_T(Z)$ distribution in $\mu^+\mu^- \to Z H H$.}\label{fig:mmzhh_ptz}
\end{figure}

In Fig.~\ref{fig:mmzhh_Mhh} we show the invariant-mass distribution of the $HH$ pair, $m(HH)$, while in Fig.~\ref{fig:mmzhh_ptz} we show the $p_T(Z)$ distribution. All plots have the same layout used already in the previous sections, but we show only $\SDKw$ results for the Sudakov approximation. It is manifest that for low values of $m(HH)$, the $\SDKw$ prediction is completely off from the exact $\NLOEW$ one: $|\deltaSDKw-\deltaNLOEW|\gg 100\%$.
At first, one may think that even including the $\deltastorkl$ contribution large logarithms of the form $L(m^2(HH),s)$ are not correctly captured, but this bad agreement between $\SDKw$ and $\NLOEW$ starts to appear already at quite large $m(HH)$ values. The origin is different and we explain it in the following.

Since $\sqrt{s}\simeq \sqrt{S}$, small $m(HH)$ values are related to configurations where a hard $Z$ boson recoils against a $HH$ pair, with the two Higgs bosons hard and collinear. Indeed, the same features present at low $m(HH)$ in Fig.~\ref{fig:mmzhh_Mhh} are visible also in Fig.~\ref{fig:mmzhh_ptz} at large $p_T(Z)$.  For $m(HH)\ll \sqrt{s}$, the tree-level diagram featuring the $\mu^+\mu^- \to ZH^* (H^*\to HH)$ topology leads to numerically large contributions that formally are mass suppressed. Indeed, considering simply the $H^*\to HH$ part of the amplitude,  it leads to a contribution of order 
\beq
\M(H^*\to HH)\propto \frac{v \lambda}{\left(m(HH)\right)^2-M_H^2}\sim \frac{M_H^2}{v \left(m(HH)\right)^2}
\left[1+
\ord\left(\frac{M_H^2}{\left(m(HH)\right)^2}\right)\right]\,,
\eeq
which is not numerically small but it is mass suppressed. Indeed if it were not mass suppressed it would have scaled as $1/m(HH)$, consistent with the scaling $\M\propto s ^{\frac{2-n}{2}}$ for a $2 \to n$ process, or equivalently for a process $n \to 2$ where here $n=1$. In this scenario, the {\denpoz} algorithm is not expected to work and indeed it does not.

The most important point to keep in mind is that NLO EW corrections are not small, but they cannot be approximated via the {\denpoz} algorithm. Even more surprisingly (at least before understanding the underlying dynamics), the $\SDKw$ works well for small $p_T(Z)$ values but not for large values, which is the opposite of what one would expect.\footnote{The agreement between $\SDKw$ and $\NLOEW$ predictions at small  $p_T(Z)$ is better at 3 TeV than at 10 TeV. In the latter case, the gap between $s$ and other invariants can be so large that the $\deltastorkl$ contributions in the Sudakov approximation are not sufficient in order to approximate the $\NLOEW$ prediction at the same level observed at 10 TeV. }  Similar situations may manifest also for BSM scenarios, where rates may be much larger than the SM process considered here. This is a clear sign of the necessity of exact $\NLOEW$ corrections also in BSM studies for the physics at the muon collider.

\subsection{Beyond NLO EW: the relevance of resummation}
\label{sec:resum}

\begin{figure}[!t]
  \includegraphics[width=0.48\textwidth]{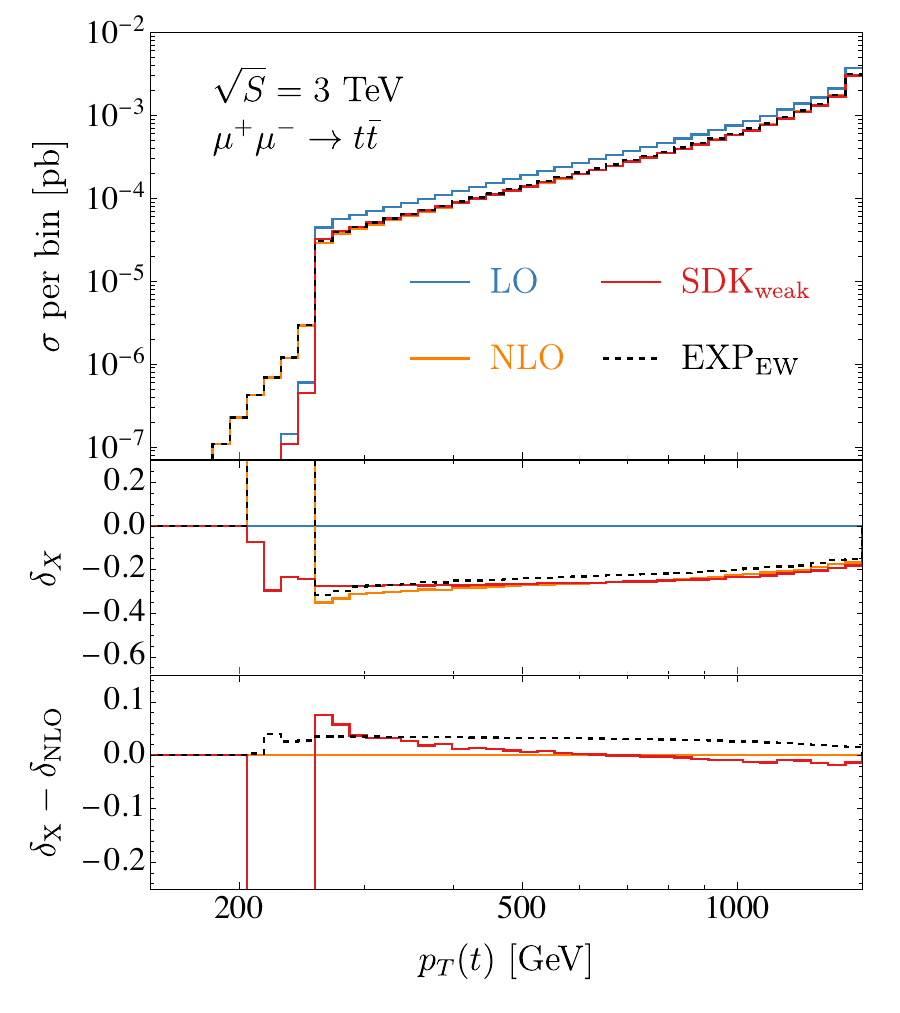}
  \includegraphics[width=0.48\textwidth]{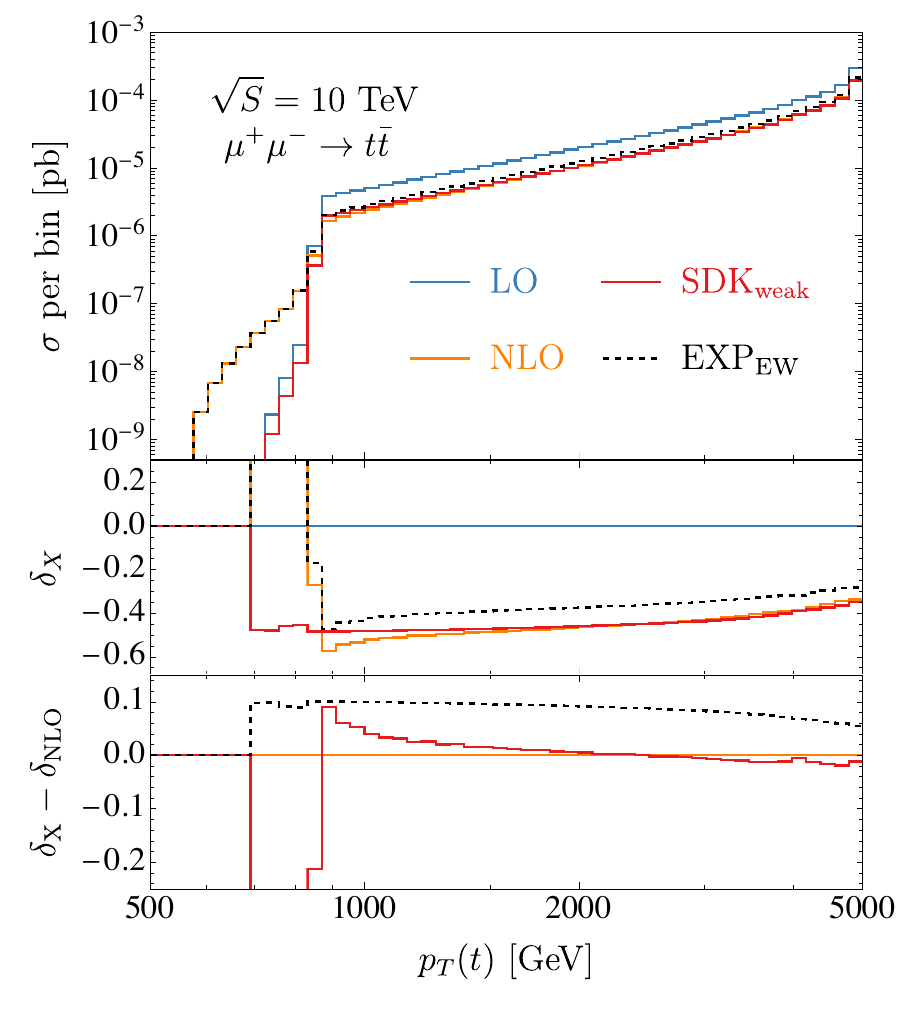}
  \caption{The top quark $p_T$ distribution in $\mu^+\mu^- \to t {\bar t}$.
  The left (right) panel shows results at $\sqrt S=3$ TeV ($\sqrt S=10$ TeV). The histograms show $\SigmaLO$ (blue), $\SigmaNLO$ (orange), the EWSL prediction in the $\SDKw$ approach  and $\SigmaExp$ (black dashed), which corresponds to the approximate resummation of EW corrections.
  } \label{fig:mmtt_NLO_ptt}
\end{figure}

In this section, we investigate if and when the resummation of EWSL of higher order, {\it i.e.}~of $\ord(\alpha^n)$ with $n>1$, is expected to be relevant. We do not perform actual resummation, rather we approximate it via exponentiation of the $\SDKw$ result and match it additively to the $\NLOEW$ prediction.\footnote{This procedure is very similar to what has been done for instance in Refs.~\cite{Bothmann:2020sxm, Bothmann:2021led} in the context of hadronic collisions.}. We have dubbed such approximation as $\Exp$ and it has been properly defined in Eq.~\eqref{eq:resumdef}. We show plots similar to those of Sec.~\ref{sec:NLOvsEWSL}, showing also the $\Exp$ predictions and in the first inset the $\deltaExp$ (black dashed).  In the second inset we show  $\deltaExp-\deltaNLOEW$, in order to emphasise the expected impact of EWSL of higher order and possibly the relevance of resumming them. The term $\deltaExp-\deltaNLOEW$, multiplied by -1, is the finite part of not EWSL origin that would not enter the resummation procedure.

We start showing again the $p_T(t)$ in $t \bar t$ production, as in Fig.~\ref{fig:mmtt_ptt}. For this observable, the EWSL of higher order are expected to be relevant for precision at $\ord(1\%)$ at 3 TeV and at $\ord(10\%)$ at 10 TeV, as can be seen in Fig.~\ref{fig:mmtt_NLO_ptt} by the difference between the $\deltaExp$ and $\deltaNLOEW$ relative corrections. However, resummation appears not to be mandatory for obtaining phenomenologically sensible results at 3 TeV. Moreover, it is of the same order as the non-logarithmic enhanced contributions at $\NLOEW$ accuracy.

\begin{figure}[!t]
  \includegraphics[width=0.48\textwidth]{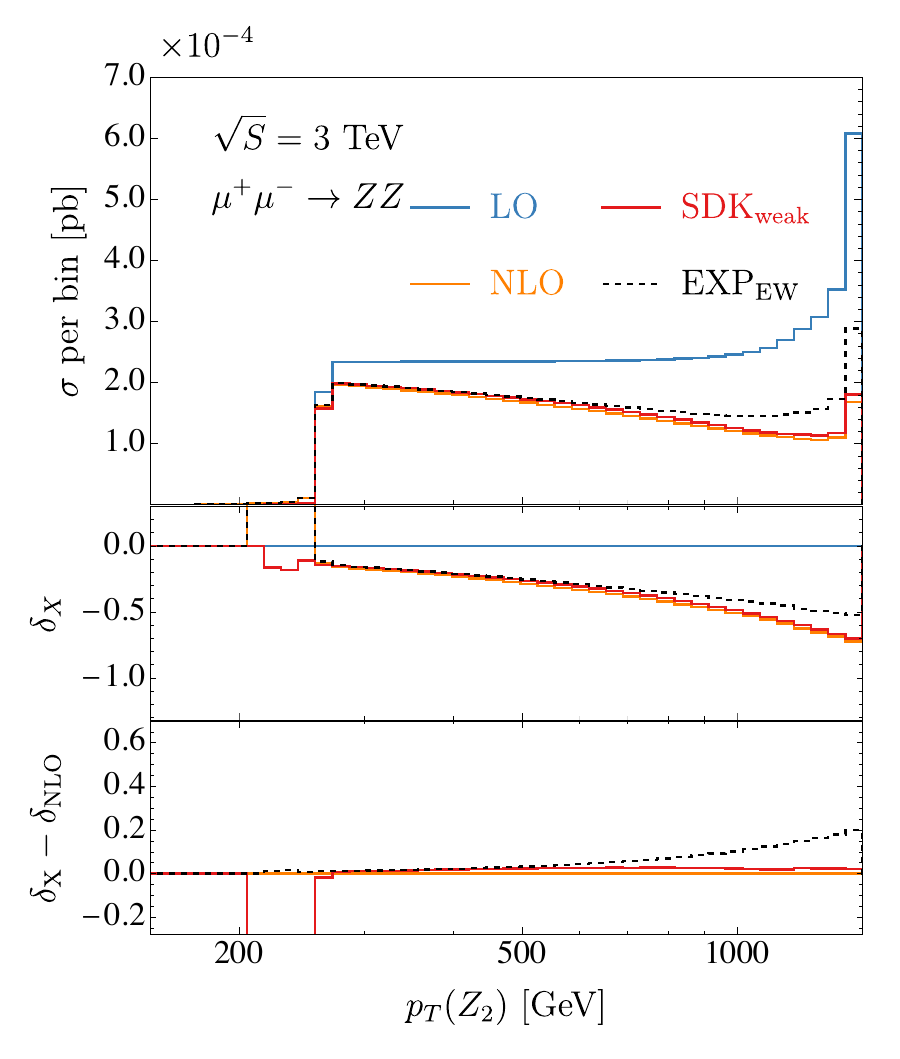}
  \includegraphics[width=0.48\textwidth]{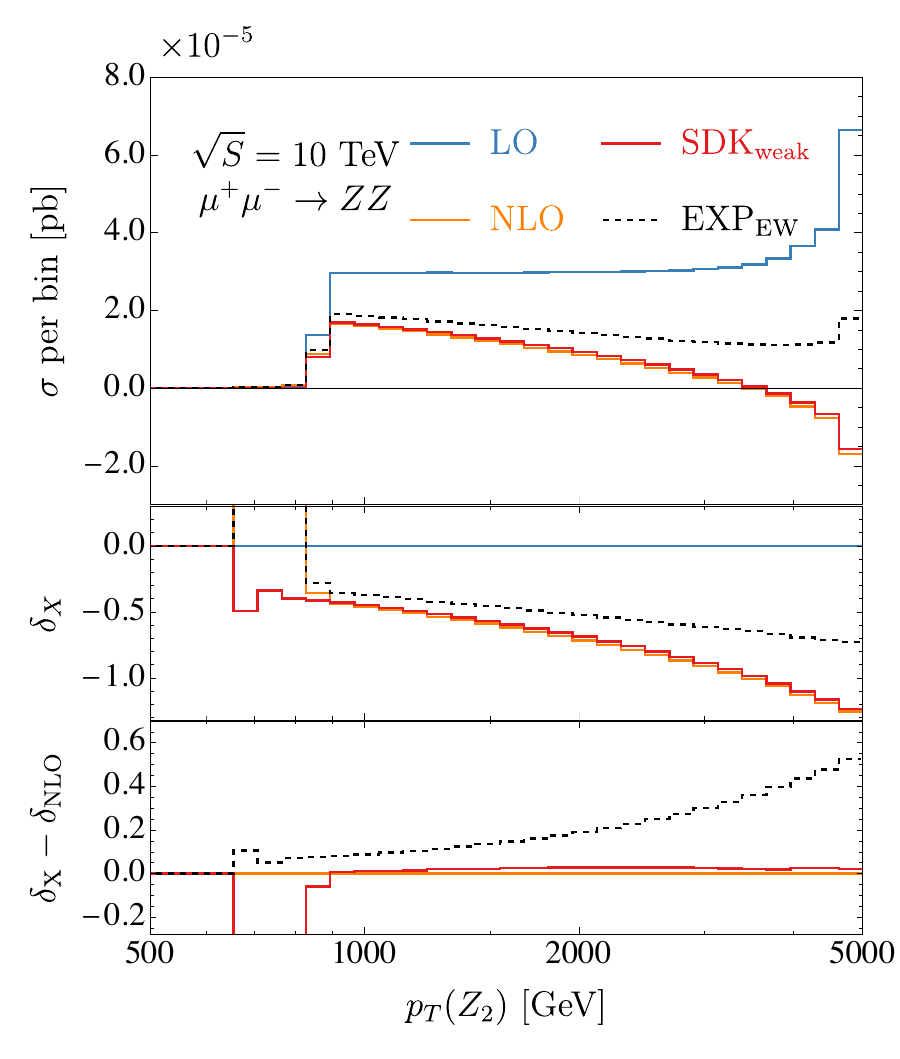} \\
  \caption{Same as Fig.~\ref{fig:mmtt_NLO_ptt}, for the $p_T(Z_2)$ distribution in  $\mu^+\mu^- \to Z Z$} \label{fig:mmzz_NLO_ptz2}
\end{figure}

The case of $p_T(Z_2)$ in $ZZ$ production, already considered in Fig.~\ref{fig:mmzz_ptz2}, is  different. As can be seen in Fig.~\ref{fig:mmzz_NLO_ptz2}, at 3 TeV EWSL of higher order are expected to be relevant for precision at $\ord(10\%)$, they are much larger than what has been observed in Fig.~\ref{fig:mmtt_NLO_ptt} at the same energy for  $p_T(t)$ in $t \bar t$ production, but still not mandatory for phenomenologically sensible results. The situation is completely different at 10 TeV. The EW corrections are so large that the prediction at $\NLOEW$ accuracy is even negative for large values of $p_T(Z_2)$, therefore it is non-physical. The resummation in this case is not concerning only precision studies; it is the only possible way of obtaining phenomenologically sensible results. We observed a similar pattern in $WW$ production, which we do not explicitly show for brevity.

\begin{figure}[!t]
  \includegraphics[width=0.48\textwidth]{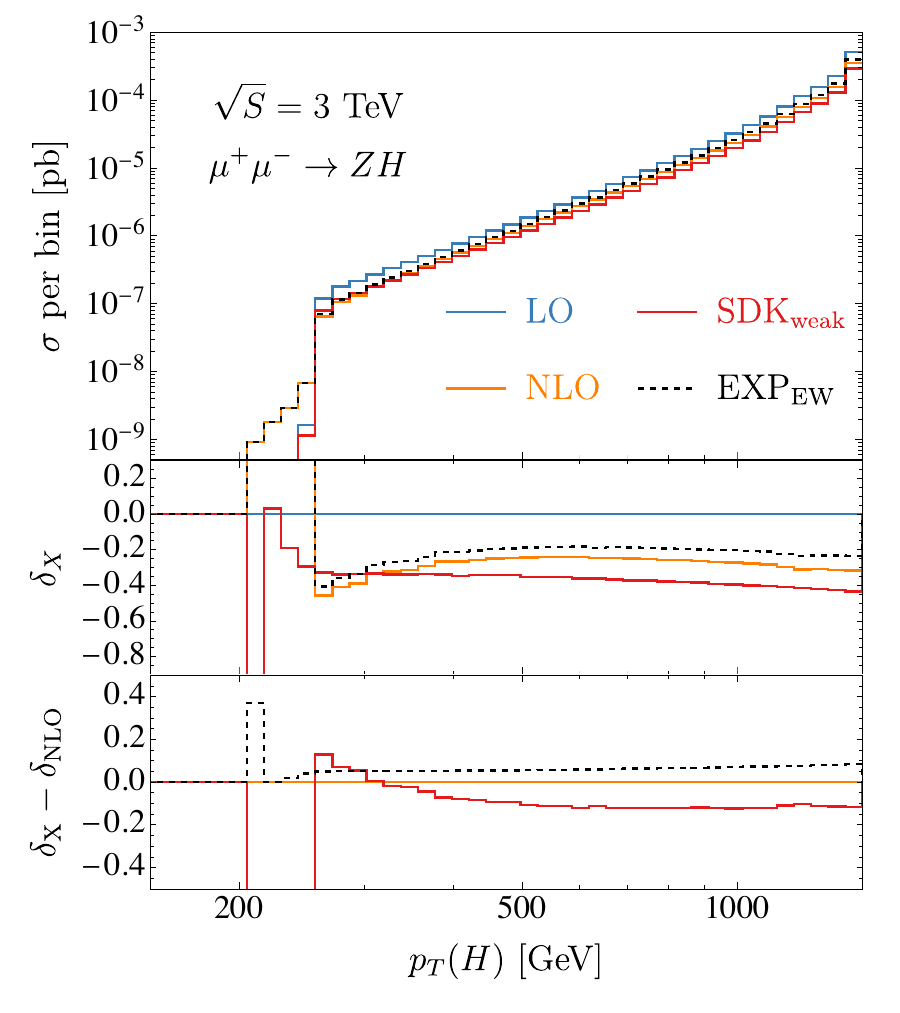}
  \includegraphics[width=0.48\textwidth]{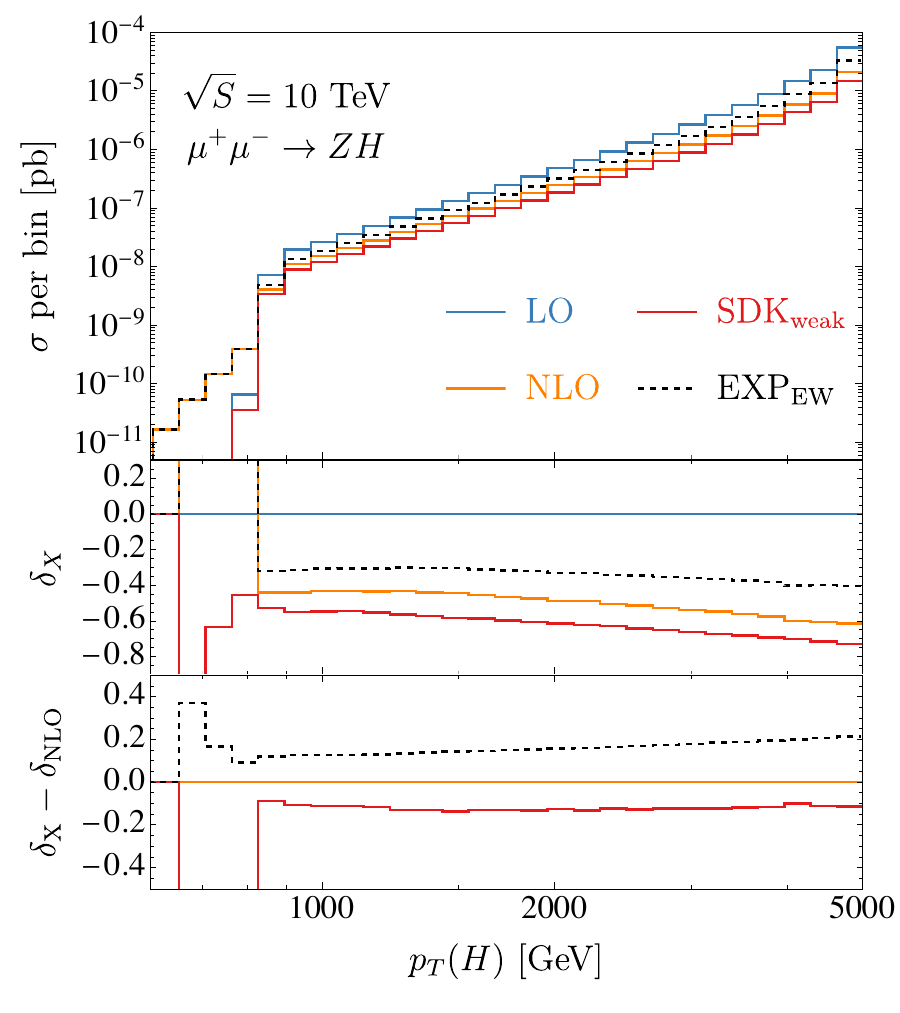} 
  \caption{Same as Fig.~\ref{fig:mmtt_NLO_ptt}, for the $p_T(H)$ distribution in $\mu^+\mu^- \to Z H$} \label{fig:mmzh_NLO_pth}
\end{figure}

The case of $ZH$ production, already shown in Fig.~\ref{fig:mmzh_pth}, is a bit different. As can be seen in Fig.~\ref{fig:mmzh_NLO_pth}, EWSL of higher order are expected to be of $\ord(10\%)$ at 3 TeV and  $\ord(20\%)$ at 10 TeV. Such effects are both of the same order of the $\deltaSDKw-\deltaNLOEW$ result discussed for Fig.~\ref{fig:mmzh_pth} and also visible in the second insets of the plots of Fig.~\ref{fig:mmzh_pth}. If a precision at $\ord(10\%)$ is required, both the resummation of the EWSL and the matching with the exact $\NLOEW$ predictions are expected to be relevant. 

\begin{figure}[!t]
  \includegraphics[width=0.48\textwidth]{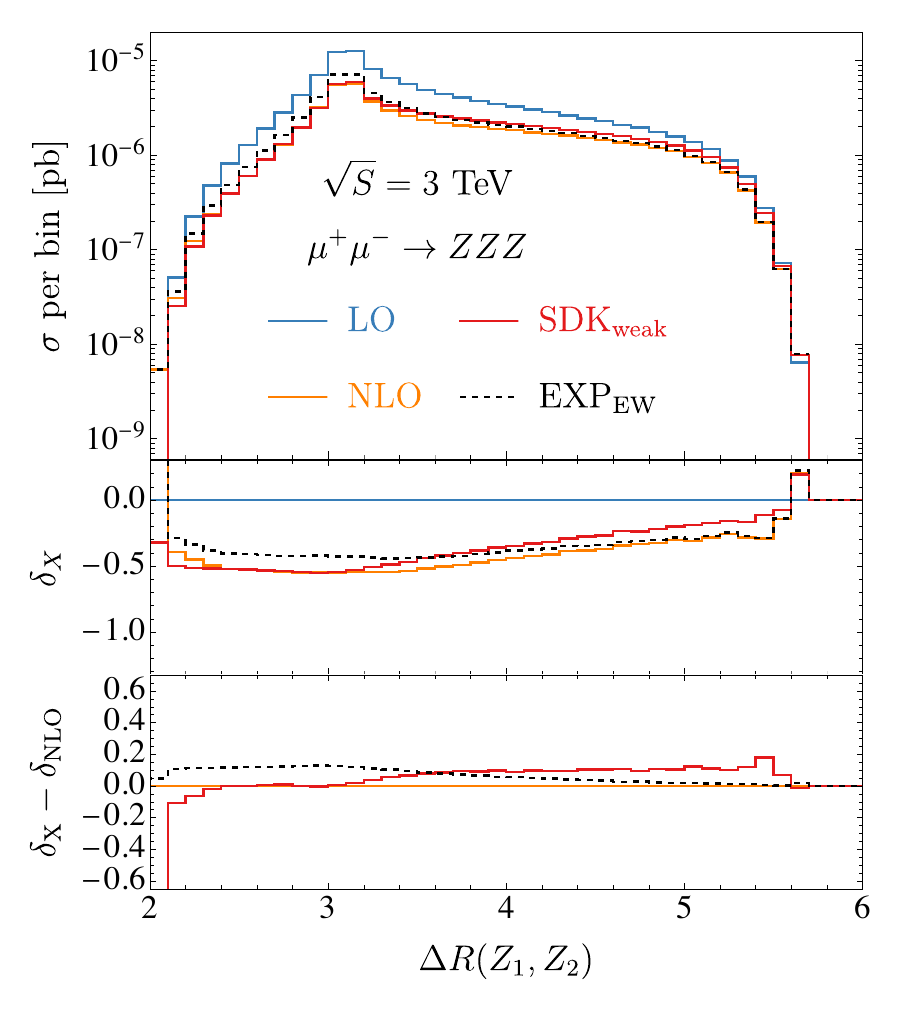}
  \includegraphics[width=0.48\textwidth]{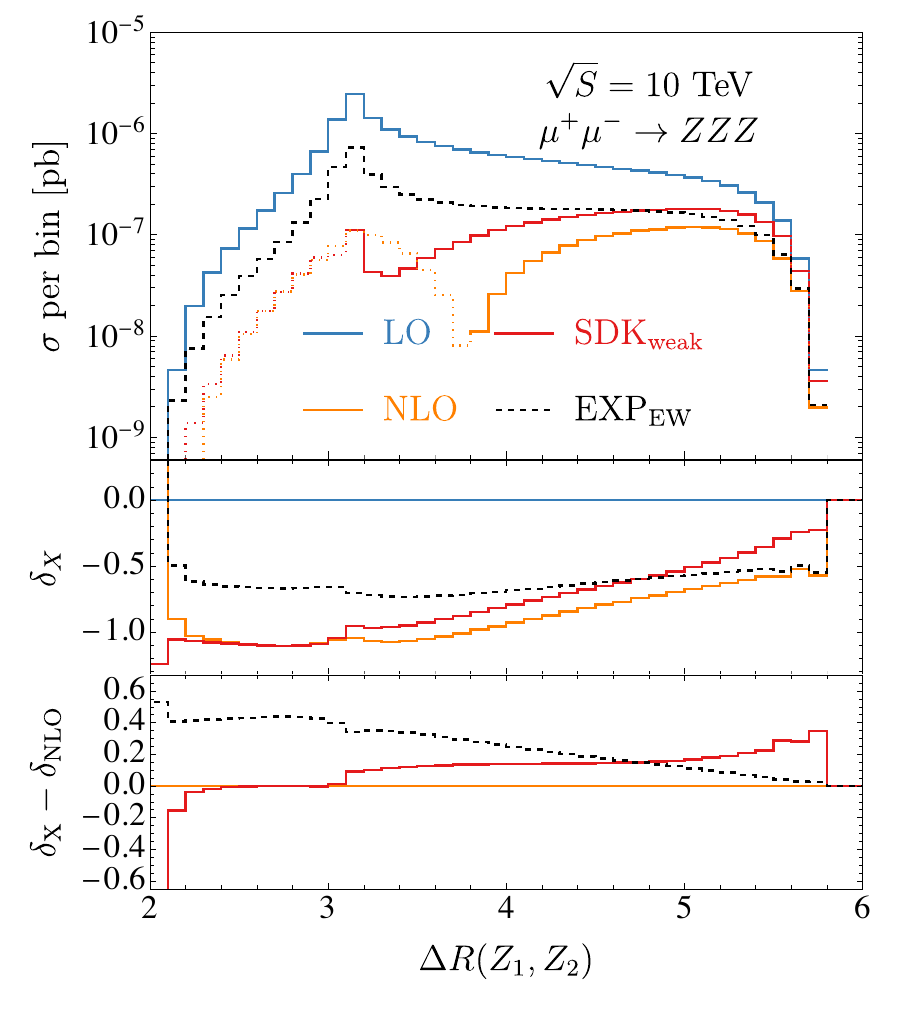}  
  \caption{Same as Fig.~\ref{fig:mmtt_NLO_ptt}, for the $\Delta R(Z_1, Z_2)$ distribution in $\mu^+\mu^- \to Z Z Z$.}\label{fig:mmzzz_NLO_Drz1z2}
\end{figure}

As a last example, before giving our conclusion on the relevance of resummation of EWSL for muon-collider physics, we show the case of $\Delta R(Z_1, Z_2)$ in $ Z Z Z$ production, already shown in Fig.~\ref{fig:mmzzz_Srij_Drz1z2}. In Fig.~\ref{fig:mmzzz_NLO_Drz1z2}  we see that again at 3 TeV resummation is relevant only for $\ord(10\%)$ accuracy studies, while at 10 TeV resummation is mandatory for having positive cross sections and phenomenologically sensible results. It is also interesting to notice that, for both energies, the range where resummation is of particular relevance ($\Delta R(Z_1, Z_2)\lesssim \pi$) is the opposite of the one ($\Delta R(Z_1, Z_2)\gtrsim \pi$) of the range where the difference between $\deltaNLOEW$ and $\deltaSDKw$ are not negligible.

In conclusion, we observe that at 3 TeV resummation is certainly relevant for precise predictions, but it is not mandatory for performing phenomenological studies. The $\NLOEW$ predictions, or equivalently the Sudakov approximation (including $\deltastorkl$ and using the $\SDKw$ scheme as discussed in Sec.~\ref{sec:NLOvsEWSL}), can be sufficient. When precision is the target, other effects not considered in this paper should be also considered: NNLO EW corrections, the accuracy of PDFs, the renormalisation scheme as well as the resummation of multiple-photon emissions. In that direction, several studies and technological advancements are desirable in view of the muon-collider physics program. We want to stress that we considered here processes featuring the $Z$ bosons in the final state, which are related to the largest EW corrections at high energies. Thus, it is difficult that other processes may exhibit even larger effects.

At 10 TeV the picture is different. Some processes, as shown in the case of $t \bar t$ production, are not expected to lead to extremely large corrections and so also at this energy resummation is necessary only for precision. Other processes instead require the resummation of EWSL. We stress again that we did not perform actual resummation, we approximated it via simple exponentiation. However, our results call for the necessity of performing (at least) Next-to-Leading-Logarithmic (NLL) resummation in muon-collider studies for physics at 10 TeV. We stress that, as already discussed in Sec.~\ref{sec:NLOvsEWSL}, for these processes and given the Bjorken-$x$ dependence of the muon PDF, since $s\simeq S$ if $\deltaNLOEW$ is not a constant, that is a clear sign that not only the LL (the double logarithms in $\deltaNLOEW$) but also the NLL (the single logarithms in $\deltaNLOEW$) are relevant; LL resummation is not sufficient. That said, it appears clear that the case of $t \bar t$ production is not unique, {\it e.g.} we observed the same pattern in $e^+ e^-$ production. Intermediate configurations between this case and the processes featuring $Z$ bosons and very large corrections are clearly possible. In other words, case-by-case studies are necessary and the automation of NLL resummation of EWSL would be very helpful. Finally, the same considerations for 3 TeV collisions concerning precision are clearly also valid for 10 TeV collisions.

\subsection{Heavy Boson Radiation (HBR)}
\label{sec:HBR}

One of the widespread assumptions about a high energy muon collider is that soft and/or collinear splittings involving heavy weak bosons ($W$, $Z$ and $H$) will lead to $\ord(1)$ corrections, similarly to the case of QCD at the LHC, see {\it e.g.} \cite{Han:2020uid,Han:2021lnp}. On the one hand, this implies that a muon collider can be treated as a EW boson collider, \cite{Costantini:2020stv,Buttazzo:2020uzc,Han:2020uid,Han:2021lnp,Ruiz:2021tdt,AlAli:2021let,Accettura:2023ked} leading to very large cross sections for VBF processes. On the other hand, the emission of $W, Z$ and possibly $H$ bosons, {\it i.e.}, what has been dubbed in this work as HBR, is in general expected to lead to large effects or even becoming dominant w.r.t.~the case with no radiation. This dynamics has already been studied in detail in particular scenarios, {\it e.g.} showing how it can help to gain sensitivity to new short-distance physical laws \cite{Chen:2022msz}.

In this section, we investigate if $\ord(1)$ corrections from HBR is really a ubiquitous effect. To this purpose, we consider direct production processes as those already studied in the previous sections. Moreover, we compare and combine the relative corrections induced by the HBR, {\it i.e.}~the quantity $\deltaHBR$ defined in Eq.~\eqref{eq:deltaHBR}, and the relative NLO EW corrections ($\deltaNLOEW$), since both of them are of $\ord(\alpha)$ w.r.t.~the inclusive production process considered. In doing so we also inspect the degree of cancellation between  $\deltaNLOEW$, which is typically negative, and $\deltaHBR$, which is positive by definition.

In Sec.~\ref{sec:HBRwwtt} we consider the $\mu^+\mu^- \to F$ processes $F=t \bar t,\, W^+W^-$ and the effects of the additional HBR with $B=H, Z$. In  Sec.~\ref{sec:HBREWj} instead we consider the case of EW jets ($\EWjet$), where  $\EWjet$ is emerging from the clustering of $V=W,Z$. Details on the clustering of the  $\EWjet$'s themselves and of the HBR with the particles in the final state $F$ are described in Sec.~\ref{sec:CalcSetUp} and we will not repeat them through the next two sections.

\subsubsection{Final states: $t \bar t$, $W^+W^-$}
\label{sec:HBRwwtt}

\begin{figure}[!t]
  \includegraphics[width=0.48\textwidth]{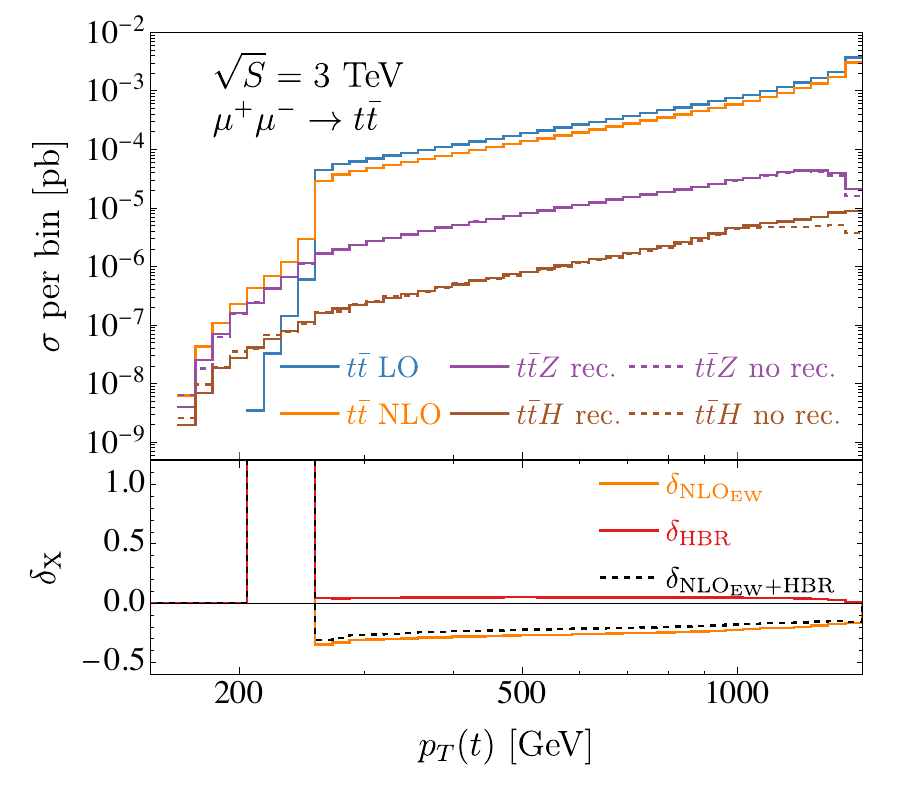}
  \includegraphics[width=0.48\textwidth]{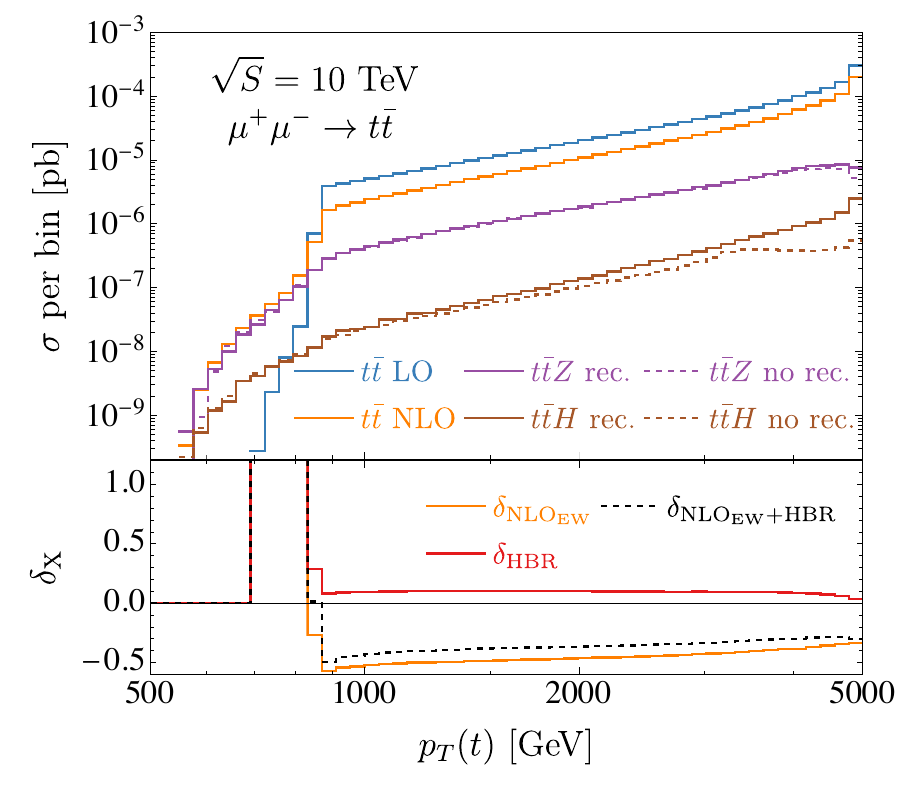} \\
  \caption{The top quark $p_T$ distribution in $\mu^+\mu^- \to t {\bar t}$.
  The left (right) plot shows results at $\sqrt S=3$ TeV ($\sqrt S=10$ TeV). The histograms show $\SigmaLO$ (blue), $\SigmaNLO$ (orange), the HBR contribution due to the $Z$ (violet) and to the $H$ boson (brown), with (solid) or without (dashed) their recombination with top quarks. In the inset, besides the impact of NLO EW corrections (orange), it is shown the total HBR contribution (red) and the sum of NLO EW and HBR (black dashed). The cut
  $m(t \bar{t})> 0.8 \sqrt{S}$ in Eq.~\eqref{eq:minvcut} is imposed.} \label{fig:mmtt_HBR_ptt}
\end{figure}

\begin{figure}[!t]
  \includegraphics[width=0.48\textwidth]{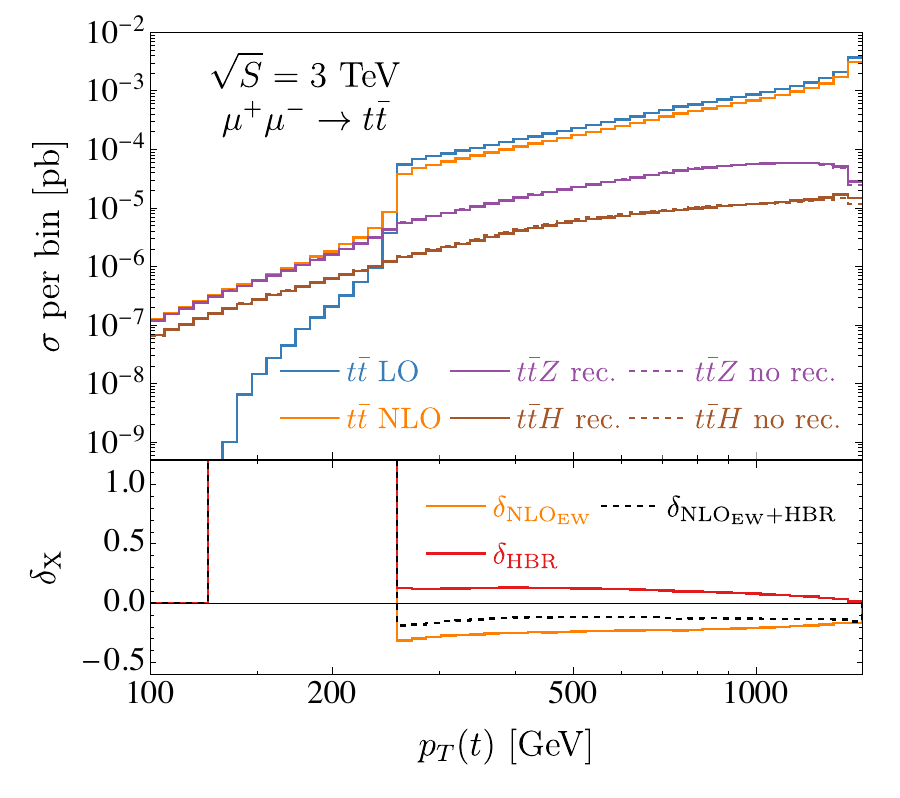}
  \includegraphics[width=0.48\textwidth]{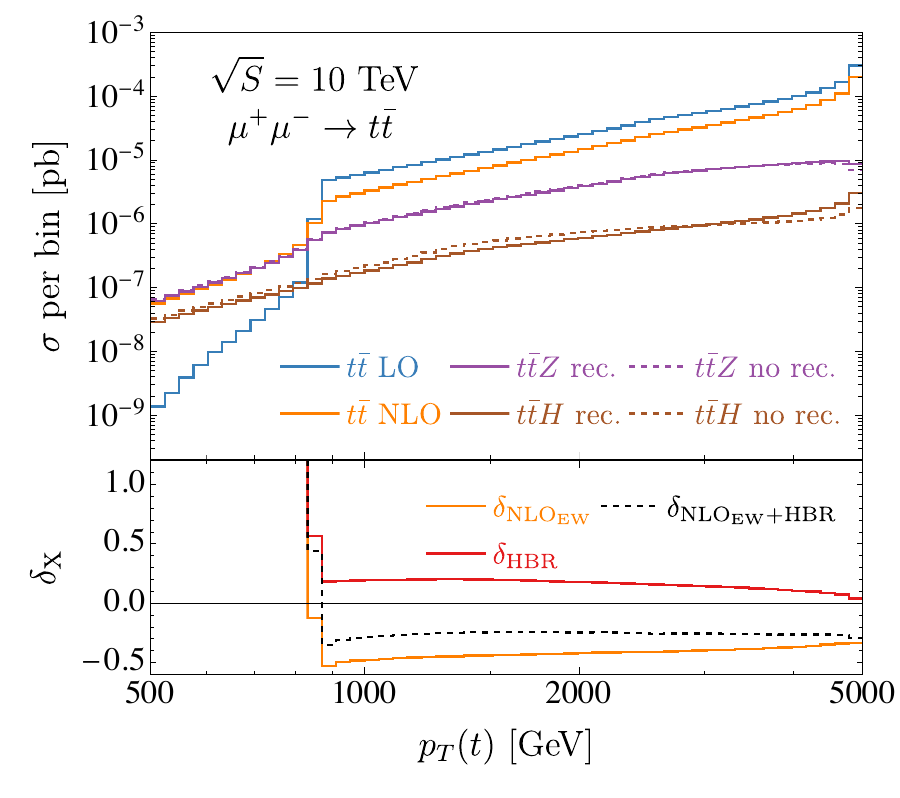} \\
  \caption{Same as Fig.~\ref{fig:mmtt_HBR_ptt}, but imposing the cut $m(t \bar{t})> 0.5 \sqrt{S}$, unlike as done in general, Eq.~\eqref{eq:minvcut}, for the other plots in this work.} \label{fig:mmtt_HBR_ptt_s50}
\end{figure}

\begin{figure}[!t]
  \includegraphics[width=0.48\textwidth]{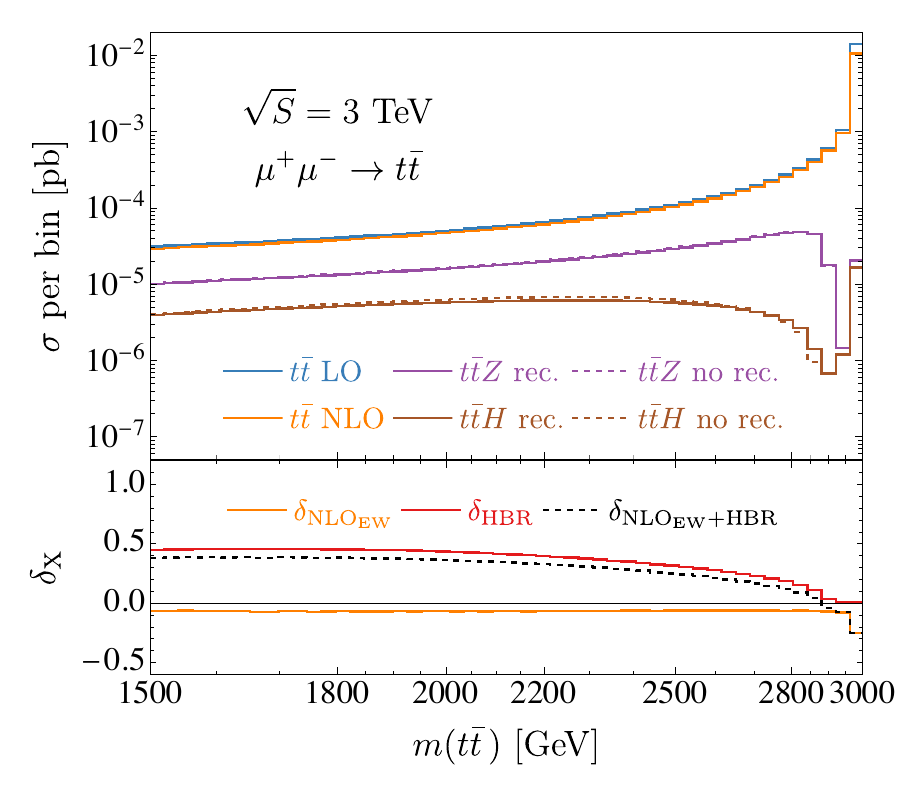}
  \includegraphics[width=0.48\textwidth]{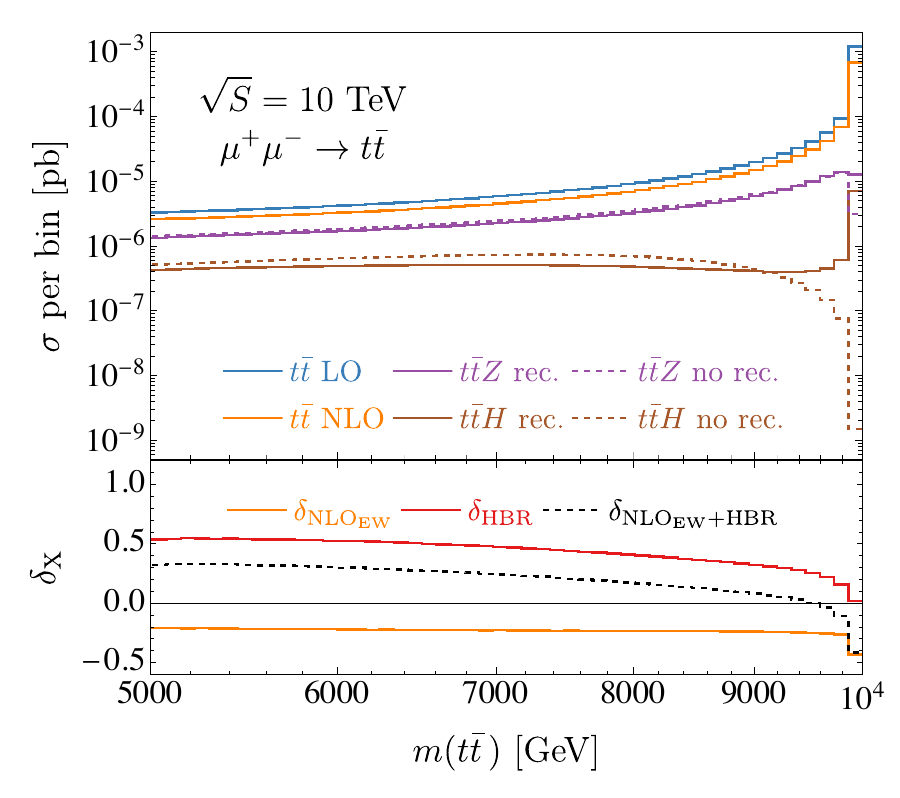} \\
  \caption{
  Same as Fig.~\ref{fig:mmtt_HBR_ptt_s50}, 
  for the $m(t {\bar t})$ distribution in $\mu^+\mu^- \to t {\bar t}$. 
  } \label{fig:mmtt_HBR_mtt_s50}
\end{figure}

We start considering again the $p_T(t)$ distribution in $t \bar t $ production in Fig. \ref{fig:mmtt_HBR_ptt}. In the main panel we show as in the previous sections LO and $\NLOEW$ predictions, but also the contribution to the same observable from the $t \bar t Z$ (violet) and  $t \bar t H$ (brown) production processes. In the solid (dashed) lines correspond to the case ``Recombination'' (``No recombination'') described in Sec.~\ref{sec:CalcSetUp}. In the first inset we show, as usual,  $\deltaNLOEW$ as an orange line but also $\deltaHBR$ (see Eq.~\eqref{eq:deltaHBR}) as a red line and $\deltaNLOHBR$  (see Eq.~\eqref{eq:deltaNLOHBR}) as a black dashed line. It is clear that the HBR contribution is given by the sum of $t \bar t Z$ and $t \bar t H$ and for the sake of simplicity and the purpose of our discussion in the inset we consider only the case where the HBR is recombined with the top quarks.

A couple of features are manifest in Fig.~\ref{fig:mmtt_HBR_ptt}. First, the contribution of  $t \bar t Z$ is much larger than the one from $t \bar t H$, so $\deltaHBR$ is dominated by the former. The origin of this difference is due to the fact that at high energy the soft emission of a Higgs boson is logarithmically enhanced, as in the case of the $Z$ emission, but it is also mass suppressed ($m_t^2/ s$), at variance with the $Z$ emission.\footnote{In the case of loop corrections, this is precisely the same reason which allows to neglect the contribution of Higgs loops in the calculation of EWSL of soft origin.} Second, we see that at very large $p_T(t)$, if top-quarks are recombined with the $B=H,V$ the contribution from both $t \bar t H$ and $t \bar t Z$ is much larger (solid {\it vs.}~dashed). Indeed, only via the recombination of a top with the HBR the $t \bar t$ pair can reach the value $m(t \bar t)\simeq s \simeq S$, which would not be possible otherwise even with soft HBR since $M_B\ne 0$. We will come back with more details on this point later in the discussion. Third, even considering such a recombination procedure, the total HBR contribution is negligible w.r.t.~the effects of the same perturbative order induced by NLO EW corrections, both at 3 and 10 TeV. This can be seen by comparing the absolute values of $\deltaHBR$ and $\deltaNLOEW$ over the spectrum for $p_T(t)\gtrsim250~\gev$ ($p_T(t)\gtrsim800~\gev$) at 3 TeV (10 TeV) collisions. Only for smaller values of  $p_T(t)$ the HBR contribution is very large, similar to the case of the $\NLOEW$ predictions. The same argument presented for explaining the large contribution from the real photon radiation and in turn the  $\NLOEW$ prediction in the discussion of Fig.~\ref{fig:mmtt_ptt} can be repeated here for the HBR.
Finally, we see that, as a consequence of the previous point, $\deltaNLOEW \simeq \deltaNLOHBR$ for $p_T(t)\gtrsim250~\gev$ ($p_T(t)\gtrsim800~\gev$) at 3 TeV (10 TeV) collisions. This implies the EWSL from weak virtual corrections, {\it i.e.} the $\deltaSDKw$ prediction which we have been shown to be the dominant component of $\deltaNLOEW$ in the discussion of  Fig~\ref{fig:mmtt_ptt}, are minimally compensated by those from real radiation.

Let us see how it is possible that the contribution from HBR is much smaller than the one from $\NLOEW$, although $\sqrt S\gg\MW$.
First of all,  in order to select the direct-production mechanism and exclude VBF configuration, the cut in \eqref{eq:minvcut} is present in our simulations. Moreover, also the cuts in $\eqref{eq:cuts}$ are present and have to be satisfied. Thus, the phase space of the $H$ and $Z$ radiation is much more constrained than one could naively expect. 
As a consequence, the double logarithms of the form $L(S,M^2_B)$, as those appearing in the virtual contributions, cannot be present. A larger contribution can be observed if the cut in Eq.~\eqref{eq:minvcut} is relaxed to $m(t \bar t)\ge 0.5\, \sqrt{S}$, as shown in Fig.~\ref{fig:mmtt_HBR_ptt_s50}. However, the effects from HBR are far from being of $\ord(1)$ and are especially much smaller than NLO EW corrections.  

While the EWSL from virtual corrections receive contributions from $W$ and $Z$ bosons, at $\ord(\alpha)$ the HBR receive contributions only from the $Z$ boson (and the $H$ boson) emissions and especially are suppressed due to the phase-space cuts. One cannot simply estimate the HBR rescaling the  LO predictions by the EWSL as done in the case of virtual contributions. Moreover, as pointed out in Ref.~\cite{Bagdatova:2024aem}, the fact that the $W, Z$, and $H$ have a non-zero mass cannot be neglected and has an impact also in the results that are obtained.

 In order to better understand the difference between the case of $\NLOEW$ predictions and the HBR, in Fig.~\ref{fig:mmtt_HBR_mtt_s50} we show the $m(t\bar t)$ distribution in the range $m(t \bar t) \ge 0.5\, \sqrt{S}$.  For such distribution we clearly observe a much larger impact of the HBR w.r.t.~the $p_T(t)$ case. However, one should, first of all, keep in mind that due to the shape of lepton PDFs, as already said, the direct production mechanism is dominated by $s\simeq S$ and therefore all the $p_T(t)$ distribution is mostly correlated to the last bin $m(t \bar t) \simeq \sqrt{S}$, where instead the HBR contribution is minimal or even zero in the case of no recombination of the top quarks with $B=H,Z$. Thus, there is no contradiction between what is observed in the $m(t \bar t)$ and $p_T(t)$ distributions. First, we explain in more detail the case without recombination and then we move to the case where top quarks are recombined with the HBR.
 
When $m(t \bar t)< \sqrt S$, at LO it implies $s<  S$, and therefore the prediction originates from Bjorken-$x<1$ in the muon PDFs, featuring a very large suppression. Conversely, at $\NLOEW$ accuracy, due to the real photon emissions, or thanks to the HBR contributions, if $m(t \bar t)< \sqrt S$ it is still possible to have $\sqrt s \simeq \sqrt S$ due to the $2\to 3$ kinematic and avoid the suppression of the PDFs. Thus the large relative contribution from  HBR ($\deltaHBR$) at $m(t \bar t)\ll \sqrt S$ is due in part to the larger phase space volumes for the radiation, as also observed by the comparison of Figs.~\ref{fig:mmtt_HBR_ptt} and \ref{fig:mmtt_HBR_ptt_s50}, but especially to the lack of PDF suppression that is instead present at LO. For $m(t \bar t)\simeq \sqrt S$ it is the opposite; there is no HBR contribution without performing the recombination. Indeed, due to momentum conservation, the relation $m(t \bar t)< \sqrt S - M_B$ has to be satisfied.

If the recombination of top quarks with $B=H,Z$ is performed, the predictions at $m(t \bar t)< \sqrt S$ are very similar to the case without recombination. The prediction at  $m(t \bar t) \simeq \sqrt{S}$, especially for the case of the $H$ boson emission,\footnote{The Higgs boson can be emitted only from the final state so a larger fraction of them, w.r.t.~the case of the $Z$ bosons, is recombined with the top quarks.} is instead very different.  This is not a surprise since when a top is recombined with a $B$ the requirement $m(t \bar t)\simeq \sqrt S$ translates into $m(t \bar t B)\simeq \sqrt S$ and so the relation $m(t \bar t)< \sqrt S - M_B$ has not to be satisfied. We stress again that $\deltaHBR$ in the inset corresponds to the case with recombination. The case without it therefore would exhibit even smaller predictions for it.

\medskip

\begin{figure}[!t]
  \includegraphics[width=0.48\textwidth]{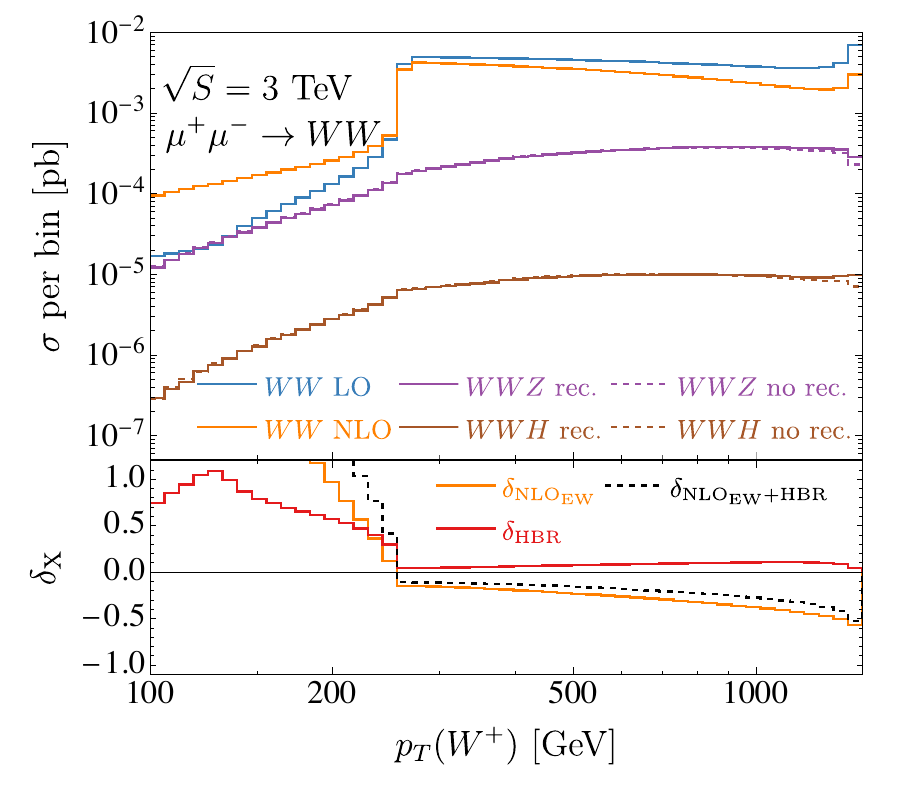}
  \includegraphics[width=0.48\textwidth]{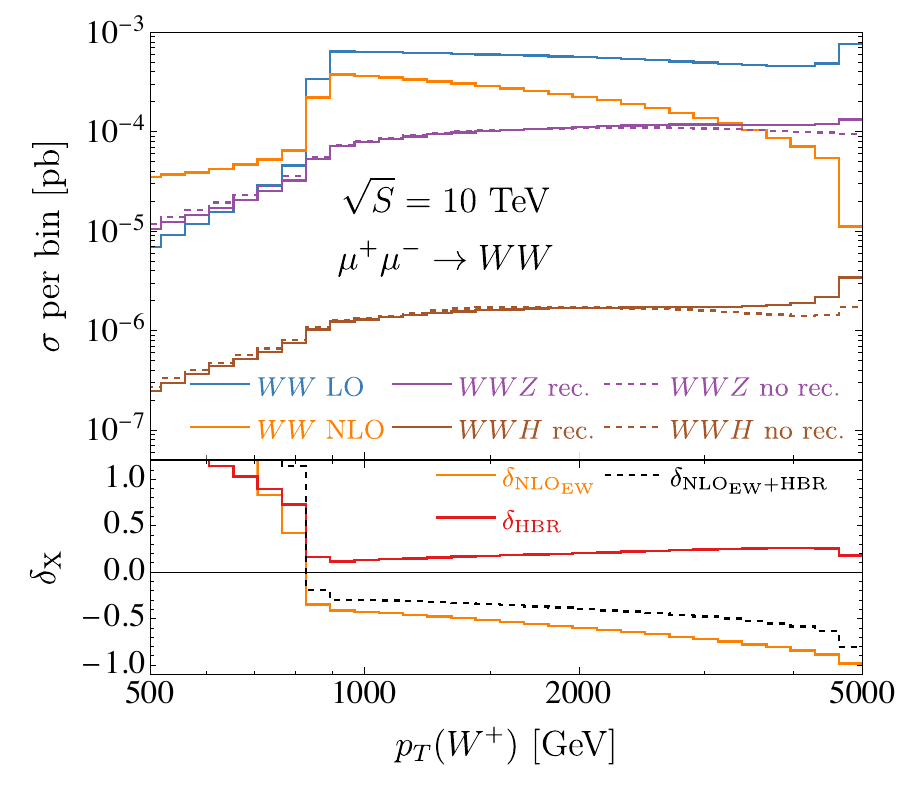} \\
  \caption{
  Same as Fig.~\ref{fig:mmtt_HBR_ptt_s50}, but for 
   the $p_T(W^+)$ distribution in $\mu^+\mu^- \to W^+ W^-$. Also in this case $m(W^+ W^-)> 0.5 \sqrt{S}$. 
  } \label{fig:mmww_HBR_ptw_s50}
\end{figure}

\begin{figure}[!t]
  \includegraphics[width=0.48\textwidth]{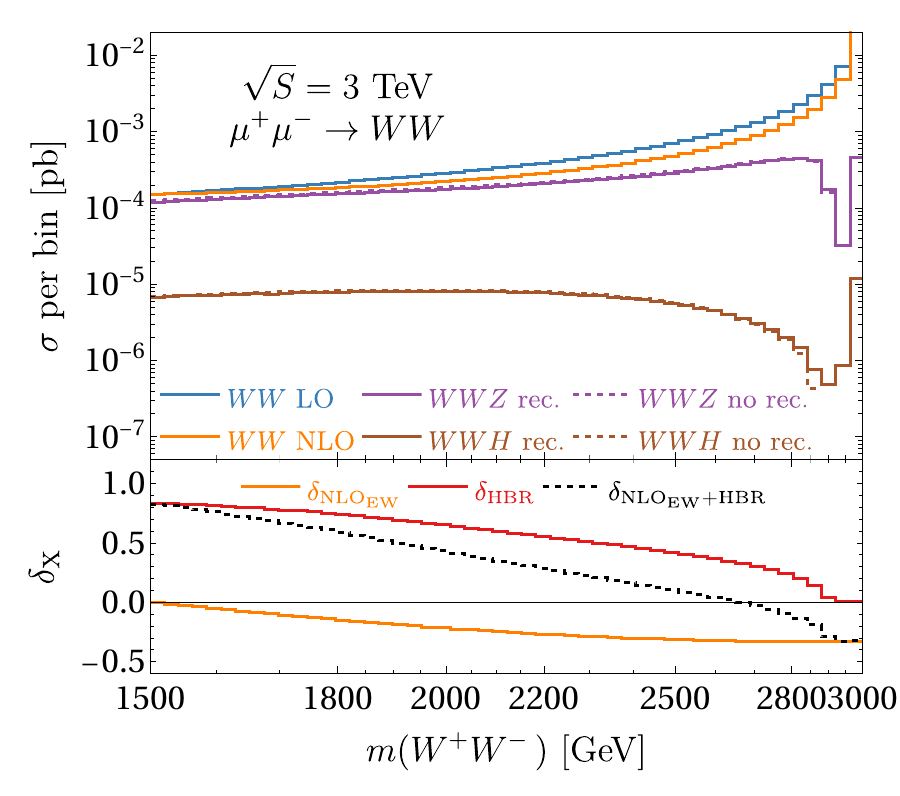}
  \includegraphics[width=0.48\textwidth]{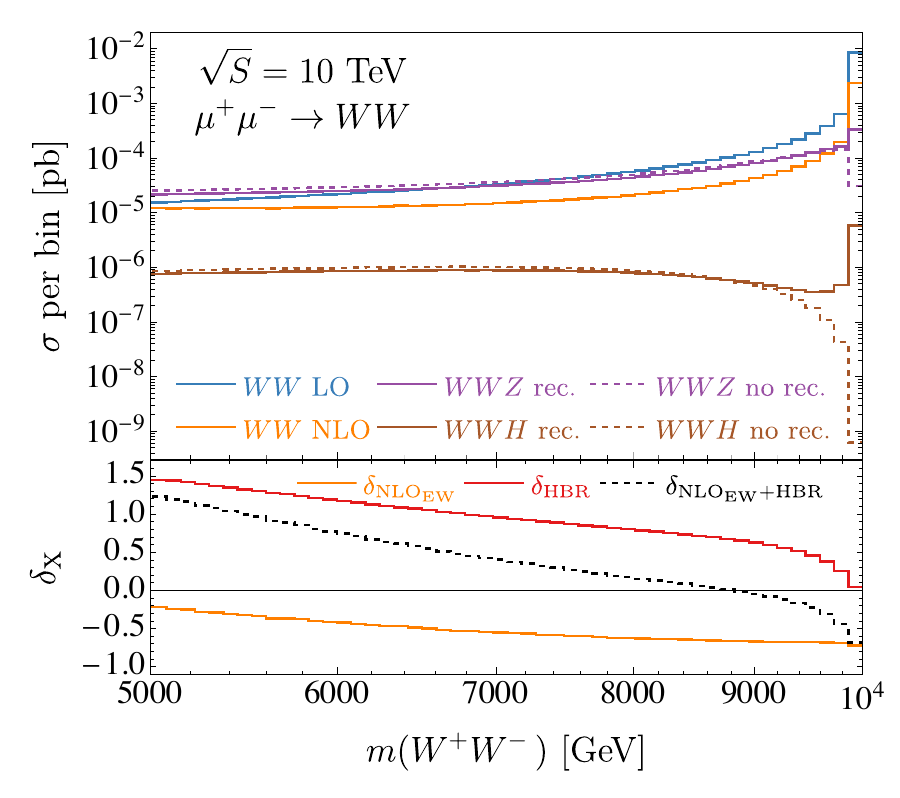} \\
  \caption{Same as Fig.~\ref{fig:mmww_HBR_ptw_s50} 
  for the $m(W^+W^-)$ distribution in $\mu^+\mu^- \to W^+ W^-$.} \label{fig:mmww_HBR_mww_s50}
\end{figure}

The dynamics observed for the HBR in $t \bar t$ production is not peculiar for this process and we show as a further example the case of $WW$ production. In Fig.~\ref{fig:mmww_HBR_ptw_s50} we show the $p_T(W^+)$ distribution and in Fig.~\ref{fig:mmww_HBR_mww_s50} the $m(W^+ W^-)$ distribution, both of them obtained with the cut \eqref{eq:minvcut} replaced by $m(W^+ W^-)> 0.5 \sqrt{S}$. The $WW$ process is very different from the $t \bar t$ one. At LO a $t$-channel diagram is present and the  $p_T(W^+)$ distribution is much flatter than the $p_T(t)$ in $t \bar t$ production. Nevertheless, the same effects observed for $t \bar t$ production are present also for this process.

In conclusion, the HBR is not expected in general to lead to large effects in the bulk of the cross sections of direct production processes. Conversely, it can be relevant for instance in regions of the phase space where the cross section is very suppressed at LO, as in the case shown here for $m(F)<\sqrt S$. However, as already mentioned before in this paper, further effects have to be considered in these cases, such as effects from other PDFs (photon PDF in the muon). The results of this section show also that while the resummation of EWSL from virtual corrections can be essential, as discussed in Sec.~\ref{sec:resum}, in the case of HBR the multiple emission of weak bosons is not expected to be always of primary relevance. That said, we are not claiming that HBR is always leading to a small effect. An example is what is discussed in Ref.~\cite{Buttazzo:2020uzc}, where the $WWH$ production process has been shown to be of the same order as $ZH$ one. In some sense,\footnote{The sense is the same underlying the idea of EW-jets introduced in Sec.~\ref{sec:CalcSetUp} and the corresponding results discussed in Sec.~\ref{sec:HBREWj}. }  the $WWH$ can be considered as one of the HBR corrections to $ZH$ production, if the physical object rather than a $Z$ boson is a generic $V$ bosons $V=W,Z$; thus we compare $VH$ and $VVH$ in this case. However, the origin of the large contribution from $WWH$ is not only due to the presence of the double logarithms. Since $ZH$ is an $s-$channel process and in $WWH$ there are instead $t$-channel configurations, featuring less suppression at high-energy,  a further enhancement not related to EWSL is present for the HBR w.r.t.~the LO prediction. This mechanism is not present in the case of $t \bar t$ and $W^+W^-$ production processes. Rather than a common feature it should be regarded as a special case, similarly to, {\it e.g.}, the giant QCD $K$-factors observed for some hadroproduction processes \cite{Frixione:1992pj,Frixione:1993yp, Rubin:2010xp, Maltoni:2015ena}.

\subsubsection{EW jets in multi-$V$ production}
\label{sec:HBREWj}

We now consider the case of inclusive EW-dijet production, $\mu^+\mu^-\to 2 \EWjet$, where in the EW-jet $\EWjet$  we cluster  $W$ and $Z$ bosons. In other words, the LO originates from the processes  $\mu^+\mu^-\to VV$ with $V=W,Z$, which correspond to the final states $WW$ and $ZZ$. Having found in the previous section that the impact of the Higgs boson radiation is minimal, we consider as a contribution to the HBR only the processes involving an additional $V=W,Z$ radiation, $\mu^+\mu^-\to 3V$, which correspond to the final states $ZZZ$ and $WWZ$. Clearly, the (at least) two $\EWjet$'s that are required may be also associated with $W$ and $Z$, respectively, or also with a $WZ$ pair and a $W$ respectively. Several combinations are possible and some of them, as the ones just mentioned and unlike the cases discussed in the previous section, do not feature the same ``partonic'' final state of the LO, which can feature only $ZZ$ and $WW$. We will consider also the case of a double HBR and therefore the $4V$ final states: $ZZZZ$, $WWWW$ and $WWZZ$.\footnote{As a side comment, such processes have been shown to be very sensitive to possible anomalous interactions of the Higgs boson with the muon \cite{Celada:2023oji, Han:2021lnp} and therefore are of particular relevance in the muon collider physics program.}

\begin{figure}[!t]
  \includegraphics[width=0.48\textwidth]{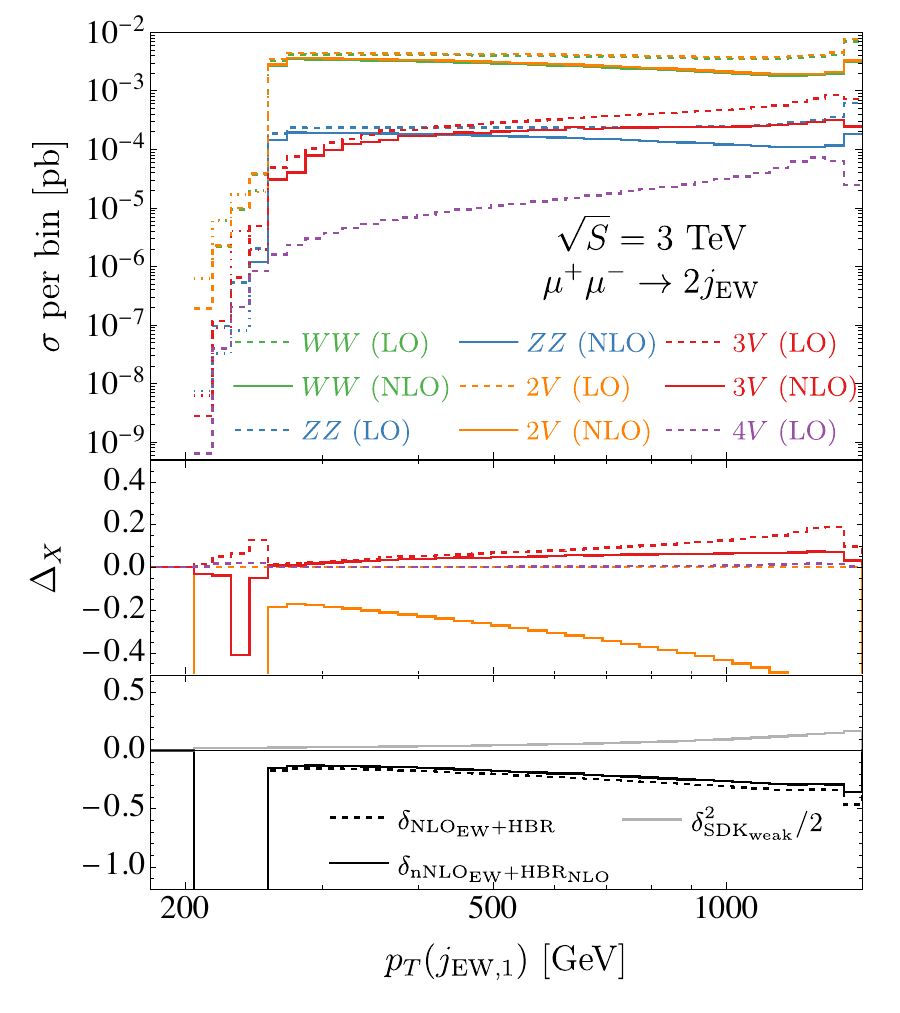}
  \includegraphics[width=0.48\textwidth]{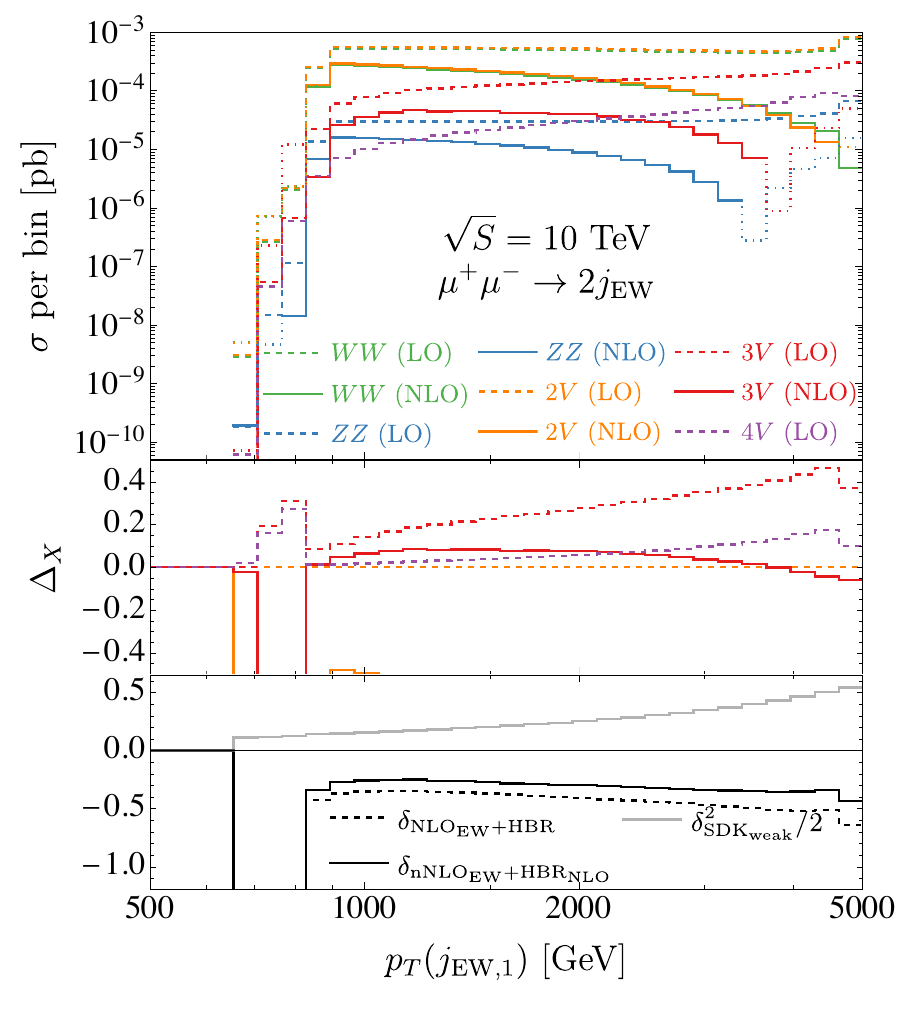} \\
  \caption{The $p_T(j_{{\rm EW},1})$ distribution in $\mu^+\mu^- \to 2j_{\rm EW}$, with $m(2j_{\rm EW})>0.8\sqrt{S}$.
  The left (right) plot shows results at $\sqrt S=3$ TeV ($\sqrt S=10$ TeV). The histograms show LO (dashed) and NLO (solid predictions) for $W^+W^-$ (green), $ZZ$ (blue), $2V =W^+W^-+ZZ$ (orange), $3V$ (red), and $4V$ (violet, only at the LO). In the second inset,  the quantities $\deltaNNLOHBR$ and  $\deltaNLOHBR$ are shown respectively as black-solid and black-dashed lines. These quantities are defined by Eq.~\eqref{eq:deltaX} in terms of respectively  Eq.~\eqref{eq:NNLO} and Eq.~\eqref{eq:NLOHBREWj}.  The solid grey line shows the quantity $\delta_{\SDKw}^2/2$, which enters only in $\deltaNNLOHBR$.} \label{fig:mm2v_ptj1}
\end{figure}

\begin{figure}[!t]
  \includegraphics[width=0.48\textwidth]{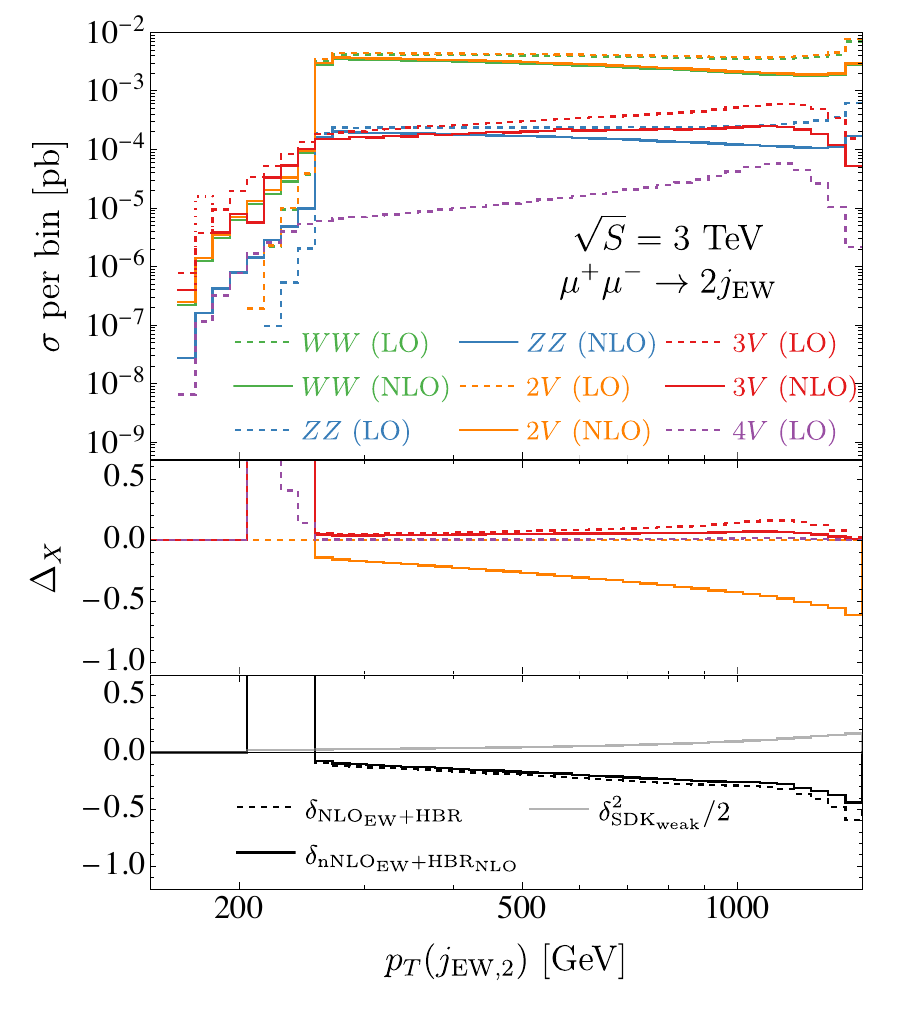}
  \includegraphics[width=0.48\textwidth]{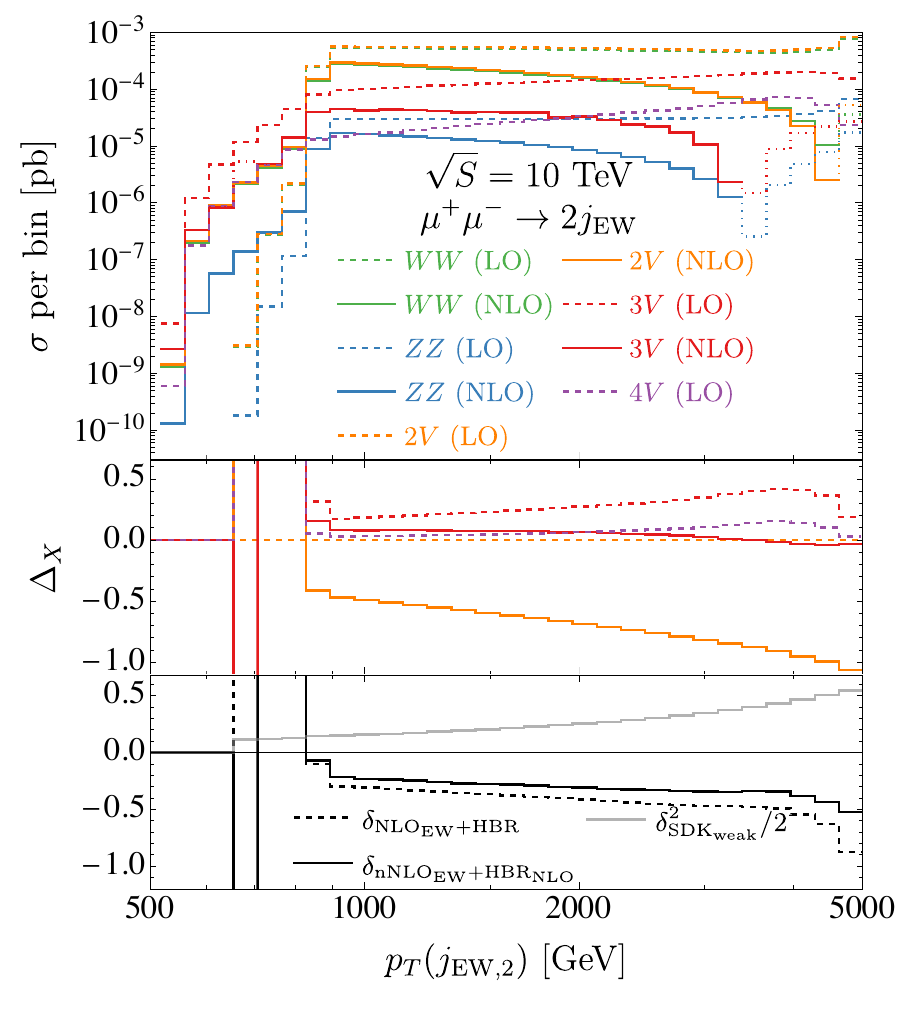} \\
  \caption{Same as Fig.~\ref{fig:mm2v_ptj1} for the $p_T(j_{{\rm EW},2})$ distribution for $\mu^+\mu^- \to 2j_{\rm EW}$, with $m(2j_{\rm EW})> 0.8 \sqrt{S}$.}\label{fig:mm2v_ptj2}
\end{figure}

In the plot in Fig.~\ref{fig:mm2v_ptj1} we show the transverse momentum distribution of the hardest $\EWjet$, $p_T(\EWjeto)$, while in  Fig.~\ref{fig:mm2v_ptj2}   the same distribution for the second-hardest $\EWjet$, $p_T(\EWjett)$. The plots have a different colour code with respect to those shown in the previous sections and we describe them in the following. In the main panel we show the contribution from the $WW$ final state (green) and ZZ (blue) which once summed leads to the $2V$ prediction (orange). The total $3V$ contribution is in red and the $4V$ one in violet. All LO contributions are shown as dashed lines while those at $\NLOEW$ accuracy as solid lines. 

In the first inset we plot the quantities
\bea
\Delta_X(2V)  &\equiv& \frac{\sigma_X(2V)-\sigma_{\rm LO}(2V)}{\sigma_{\rm LO}(2V)}\,. \label{eq:DeltaX}\\
\Delta_X(3V)  &\equiv& \frac{\sigma_X(3V)}{\sigma_{\rm LO}(2V)}\\
\Delta_X(4V)  &\equiv& \frac{\sigma_X(4V)}{\sigma_{\rm LO}(2V)}
\eea
where $\sigma_{\rm LO}(2V)$ corresponds to the LO predictions for $2\EWjet$ production. Similarly,  $\sigma_{\NLOEW}(2V)$ corresponds to the $\NLOEW$ prediction for $2\EWjet$ production, and $\sigma_{\rm LO}(3V)$ to the HBR contribution. Thus, $\Delta_{\NLOEW}(2V)=\deltaNLOEW(2\EWjet)$ and $\Delta_{\LO}(3V)=\deltaHBR(2\EWjet)$. 
In the second inset we instead plot the quantities defined in Eqs.~\eqref{eq:NLOHBREWj} and \eqref{eq:NNLO}, respectively, as $\deltaNLOHBR$ and $\deltaNNLOHBR$. The former (the black-dashed line) accounts for all the possible corrections of $\ord(\alpha)$, which are calculated exactly. The latter (the black-solid line) accounts for the former effects plus all the $\ord(\alpha^2)$ corrections:  the double HBR and NLO EW corrections to single HBR, which are evaluated exactly, and the NNLO EW corrections, which are approximated via the EWSL. In particular the contribution from the approximated NNLO EW corrections, {\it i.e.}, the quantity $\deltaSDKw^2/{2}$ in Eq.~\eqref{eq:NNLO}, is displayed as a grey solid line in the second inset.

In the main panel we notice that the LO is dominated by the $WW$ ``partonic'' process, which has a much larger cross section than the $ZZ$ ``partonic'' process and therefore coincides with $2V$. Especially at 10 TeV, both $2V$ and $3V$ processes receive very large and negative contributions from NLO EW corrections. The coloured dotted lines in the main inset of the right plot of Fig.~\ref{fig:mm2v_ptj1} indicates negative values for the lines depicted as solid, which we plot in absolute value in order to be visible on a logarithmic scale. It means that for both $2V$ and $3V$ processes, the relative NLO EW corrections are negative and larger than $100\%$ in absolute value. At 3 TeV they are smaller, but also in this case they are much larger than the contribution from double HBR production at LO, {\it i.e.} $4V$ at LO. Similarly to the case of the HBR, {\it i.e.} $3V$ at LO, which is smaller than the NLO EW corrections, the double HBR,  {\it i.e.} $4V$ at LO, is smaller than the NLO EW corrections to  $3V$ and so to the single HBR. Estimating the two-loop corrections, {\it i.e.}~the NNLO to $2V$, as $\deltaNLOEW^2/2\simeq\deltaSDKw^2/2$, we see that this contribution (the grey solid line) is even larger than the NLO EW corrections to the HBR contribution. In other words, looking at the case $\ord(\alpha)$ and $\ord(\alpha^2)$, it appears that for a  given $n$ the leading component of the $\ord(\alpha^n)$ corrections to LO in $2 \EWjet$ production is the $n$-loop corrections,  the subleading component is the $(n-1)$-loop corrections to HBR and so on; loops win over HBR.  

In the second inset, we see the relative corrections induced by all $\ord(\alpha)$ contributions, $\deltaNLOHBR$, are negative and larger in absolute value than the $\deltaNNLOHBR$ quantity, which is also negative and includes also the $\ord(\alpha^2)$ contributions. The NLO EW corrections to HBR are negative but they are completely compensated by the positive $\deltaSDKw^2/2$ contribution (see Eq.~\eqref{eq:NNLO}) and in part by the double HBR. Without taking into account such contribution, we would see a different picture: $\deltaNNLOHBR$ would be much larger than $\deltaNLOHBR$ in absolute value. It is also interesting to note how close $\deltaNLOHBR$ and $\deltaNNLOHBR$ are over the full spectrum.

Unlike what has been done in Sec.~\ref{sec:resum}, here we did not approximate via exponentiation the resummation of higher-order EWSL; we have only retained those of $\ord(\alpha^2)$ arising from such exponentiation in order to approximate the NNLO EW.  However, a proper resummation of such effects is also in this case clearly necessary for reliable results at 10 TeV collisions.\footnote{If $\deltaNLOEW\sim100\%$, as in the last bins of the distributions shown here, one could expect $\simeq-15\%$ effects just from the EWSL of $\ord(\alpha^3)$. }

We have also inspected the same results allowing $m(\EWjeto\EWjett)>0.5 \sqrt S$. Contributions from HBR are slightly more relevant, similarly to what was observed in Sec.~\ref{sec:HBRwwtt}, but otherwise the qualitative picture is not altered. Given the results shown in Ref.~\cite{Buttazzo:2020uzc} a different picture may arise if instead of the $2\EWjet$ processes the $H\EWjet$ is considered. We reckon this would be very interesting, but it would not be representative of the typical impact of HBR at muon colliders.

\section{Conclusions and outlook}
\label{sec:conclusions}

We have performed a comprehensive study of NLO EW corrections at a future multi-TeV muon collider, using the {\aNLO} framework. The focus of our work has been on direct-production processes. For these colliders, NLO EW corrections are sizeable, and their inclusion is mandatory not just for precision studies, but even for proper estimates of scattering rates. Within {\aNLOs}, EW corrections can be computed either exactly at NLO, or in the Sudakov (high-energy) approximation, exploiting the {\denpoz} algorithm. The latter case has clear advantages in terms of speed and numerical stability, however its accuracy must always be validated against the exact NLO EW computation. Indeed, in this respect, we have shown the benefit of employing a purely-weak version of the {\denpoz} approach, and the importance of including extra angular-dependent terms. Both aspects were first highlighted in Ref.~\cite{Pagani:2021vyk}. Still, we have discussed exceptional cases when the high-energy approximation as obtained from the {\denpoz} method fails, sometimes in a rather spectacular way, as in the case of $ZHH$ production.

Besides, we have discussed when, due to their large size, the resummation of EW Sudakov logarithms is necessary in order to have sensible predictions or even simply positive cross sections. Although based on an approximate formula, our findings show that resummation is mandatory for multi-boson processes at 10 TeV or in general whenever EW corrections approach -100\% with respect to the LO rate. However, there are also  processes, as, {\it e.g.}, the top-quark production, where resummation is definitely necessary for precision studies, but its impact on top of NLO EW predictions is below the 5\%(10\%) level at 3(10) TeV. Thus, dedicated studies on the resummation of Sudakov logarithms for specific processes and at different energies would be desirable.  

Finally, we have discussed the impact of the real-emission counterpart of the Sudakov logarithms, {\it i.e.}, the radiation of a heavy boson. Generally, the impact of these processes is subdominant w.r.t.~their virtual counterpart. This is somehow in disagreement with existing studies on the subject which, however, have focused on specific processes. Indeed, enhancements due to the HBR can be found only for particular processes or regions of the phase space, either when the HBR processes are kinematically favourable w.r.t. the LO ({\it e.g.}~appearance of $t$ channels), or because the emission of an extra particle modifies phase-space boundaries.

While the study has been carried out focusing on SM processes, many of the conclusions can be considered generic and can be extended to BSM scenarios. Studies aimed at specific extensions of the SM are envisaged.

A natural follow-up of this work
is the study of the low-invariant mass region, where direct production is not
the dominant mechanism. In this region, photon-initiated processes and VBF topologies are relevant, and a number of aspects need to be under control in order to have reliable predictions. For example,  small-$x$ effects in the parton distributions, and the interplay between power corrections and logarithmic enhancements related to the vector-boson mass.

\begin{acknowledgments}
Computational resources have been provided by the supercomputing facilities of the Universit\'e catholique de Louvain (CISM/UCL) and the Consortium des \'Equipements de Calcul Intensif en 
F\'ed\'eration Wallonie Bruxelles (C\'ECI) funded by the Fond de la Recherche Scientifique de Belgique (F.R.S.-FNRS) under convention 2.5020.11 and by the Walloon Region. YM acknowledges the support as a Postdoctoral Fellow of the Fond de la Recherche Scientifique de Belgique (F.R.S.-FNRS), Belgium. DP acknowledges support from the DESY computing infrastructures provided through
the DESY theory group for the initial stages of this work. The work of MZ has been partly supported by the ``Programma per Giovani Ricercatori Rita Levi Montalcini'' granted by the Italian Ministero dell’Universit\`a e della Ricerca (MUR). We also acknowledge support from the COMETA COST Action CA22130.
\end{acknowledgments}

\appendix

\section{ISR parton distribution}
\label{sec:isr}

In this Appendix we discuss the modelling of ISR effects and therefore of the muon PDF in our simulations. As it should be clear from the discussion in Sec.~\ref{sec:CalcSetUp}
our interest is in the direct-production processes, characterised by an invariant mass of the produced final state close to the nominal collider energy, $\sqrt s \simeq \sqrt S$. In this region, the dominant parton luminosity is the muonic one, while all other ones (photon, the other fermions, {\it etc.}) are suppressed by large factors.
This can be clearly seen, {\it e.g.}, in several plots that have been shown in Refs.~\cite{Han:2020uid,Han:2021kes}.

Having established that the only relevant parton is the muon, the next question to address is how to model its luminosity density. At the moment, at variance with
the electron case, only LL-accurate muon densities are available. Furthermore,
given the results presented in Ref.~\cite{Bertone:2022ktl} for the case of $e^+ e^-$ collisions, one can appreciate that
the size of effects on physical observables, due to the NLL evolution or the various factorisation schemes at this order, is
at the percent level or below.  EW corrections instead, as documented also in Sec.~\ref{sec:numres},
 lie in the ball-park of several tens of percent for direct production processes at a high-energy muon collider. Moreover, the modelling of ISR, although being of primary relevance for precision studies, is not one of the several aspects (see Sec.~\ref{sec:list}) that we are investigating in this work. 

Given all the previous considerations, a LL description for the muon PDF is sufficient for our simulations. 
We reckon that, quite recently, a fully-fledged description of the muonic content in terms of massless partons (leptons, the photon, but also quarks and the gluon) has appeared in Ref.~\cite{Frixione:2023gmf}, employing LL-accurate evolution in QCD and QED. For the purpose of this work, however, we opt for a simpler approach. Specifically, we minimally modify the electron ISR PDF at Leading-Logarithmic (LL) accuracy~\cite{Gribov:1972ri, Lipatov:1974qm, Altarelli:1977zs, Dokshitzer:1977sg} in the so-called $\beta$ scheme, including up to $\mathcal O (\alpha^3)$ terms~\cite{Skrzypek:1990qs,Skrzypek:1992vk,Cacciari:1992pz}:
\begin{equation}
\Gamma^{\rm LL}_{\mu^\pm/\mu^\pm}(x, Q^2)=\frac{\exp (-\gamma_E \beta + \frac{3}{4}\beta)}{\Gamma(1+\beta)}
\beta (1-x)^{\beta-1} 
-\sum_{n=1}^3\frac{1}{2^n n!} \beta^n h_n(x)\,,
\end{equation}
where 
\begin{equation}
    \beta = \frac{\alpha}{\pi}\left(\log(Q^2/m^2)-1\right)\,.
\end{equation} 

We then set $m$ to the muon mass, and we neglect the running of $\alpha$. Rather, we fix its value to 
\beq
\alpha=\alpha_{G_\mu}\equiv \frac{G_\mu}{\pi}  \sqrt 2 \left(1-\frac{\MW^2}{\MZ^2}\right)\MW^2\, ,
\eeq
 consistently with 
the renormalisation scheme that we will employ in the computations, with $G_\mu$ measured from the muon decay. 
Finally, as it is well known, the usage of LL-accurate ISR in NLO EW computations requires the inclusion, at the level of short-distance cross section, of 
additional terms required to attain formal NLO accuracy (treated on the same 
footage as change of factorisation-scheme contributions)~\footnote{
For more details, see {\it e.g.} Sec.~2 of Ref.~\cite{Denner:2003iy} or  Appendix A 
of~Ref.~\cite{Bertone:2022ktl}. 
}. All our results at NLO will be computed including these terms, which are already available 
inside \aNLO.

\bibliographystyle{JHEP}
\bibliography{ref}

\providecommand{\href}[2]{#2}\begingroup\raggedright\begin{thebibliography}{100}

\bibitem{MuonCollider:2022xlm}
{\scshape Muon Collider} collaboration, \emph{{The physics case of a 3 TeV muon
  collider stage}},  \href{https://arxiv.org/abs/2203.07261}{{\ttfamily
  2203.07261}}.

\bibitem{Aime:2022flm}
C.~Aime et~al., \emph{{Muon Collider Physics Summary}},
  \href{https://arxiv.org/abs/2203.07256}{{\ttfamily 2203.07256}}.

\bibitem{Black:2022cth}
K.M.~Black et~al., \emph{{Muon Collider Forum report}},
  \href{https://doi.org/10.1088/1748-0221/19/02/T02015}{\emph{JINST} {\bfseries
  19} (2024) T02015} [\href{https://arxiv.org/abs/2209.01318}{{\ttfamily
  2209.01318}}].

\bibitem{Maltoni:2022bqs}
F.~Maltoni et~al., \emph{{TF07 Snowmass Report: Theory of Collider Phenomena}},
   \href{https://arxiv.org/abs/2210.02591}{{\ttfamily 2210.02591}}.

\bibitem{Belloni:2022due}
A.~Belloni et~al., \emph{{Report of the Topical Group on Electroweak Precision
  Physics and Constraining New Physics for Snowmass 2021}},
  \href{https://arxiv.org/abs/2209.08078}{{\ttfamily 2209.08078}}.

\bibitem{Accettura:2023ked}
C.~Accettura et~al., \emph{{Towards a muon collider}},
  \href{https://doi.org/10.1140/epjc/s10052-023-11889-x}{\emph{Eur. Phys. J. C}
  {\bfseries 83} (2023) 864}
  [\href{https://arxiv.org/abs/2303.08533}{{\ttfamily 2303.08533}}].

\bibitem{Delahaye:2019omf}
J.P.~Delahaye, M.~Diemoz, K.~Long, B.~Mansouli\'e, N.~Pastrone, L.~Rivkin
  et~al., \emph{{Muon Colliders}},
  \href{https://arxiv.org/abs/1901.06150}{{\ttfamily 1901.06150}}.

\bibitem{Bartosik:2020xwr}
N.~Bartosik et~al., \emph{{Detector and Physics Performance at a Muon
  Collider}},
  \href{https://doi.org/10.1088/1748-0221/15/05/P05001}{\emph{JINST} {\bfseries
  15} (2020) P05001} [\href{https://arxiv.org/abs/2001.04431}{{\ttfamily
  2001.04431}}].

\bibitem{Schulte:2021hgo}
D.~Schulte, J.-P.~Delahaye, M.~Diemoz, K.~Long, B.~Mansouli\'e, N.~Pastrone
  et~al., \emph{{Prospects on Muon Colliders}},
  \href{https://doi.org/10.22323/1.390.0703}{\emph{PoS} {\bfseries ICHEP2020}
  (2021) 703}.

\bibitem{Long:2020wfp}
K.~Long, D.~Lucchesi, M.~Palmer, N.~Pastrone, D.~Schulte and V.~Shiltsev,
  \emph{{Muon colliders to expand frontiers of particle physics}},
  \href{https://doi.org/10.1038/s41567-020-01130-x}{\emph{Nature Phys.}
  {\bfseries 17} (2021) 289}
  [\href{https://arxiv.org/abs/2007.15684}{{\ttfamily 2007.15684}}].

\bibitem{MuonCollider:2022nsa}
{\scshape Muon Collider} collaboration, \emph{{A Muon Collider Facility for
  Physics Discovery}},  \href{https://arxiv.org/abs/2203.08033}{{\ttfamily
  2203.08033}}.

\bibitem{MuonCollider:2022ded}
{\scshape Muon Collider} collaboration, \emph{{Simulated Detector Performance
  at the Muon Collider}},  \href{https://arxiv.org/abs/2203.07964}{{\ttfamily
  2203.07964}}.

\bibitem{MuonCollider:2022glg}
{\scshape Muon Collider} collaboration, \emph{{Promising Technologies and R\&D
  Directions for the Future Muon Collider Detectors}},
  \href{https://arxiv.org/abs/2203.07224}{{\ttfamily 2203.07224}}.

\bibitem{Han:2020pif}
T.~Han, D.~Liu, I.~Low and X.~Wang, \emph{{Electroweak couplings of the Higgs
  boson at a multi-TeV muon collider}},
  \href{https://doi.org/10.1103/PhysRevD.103.013002}{\emph{Phys. Rev. D}
  {\bfseries 103} (2021) 013002}
  [\href{https://arxiv.org/abs/2008.12204}{{\ttfamily 2008.12204}}].

\bibitem{Buttazzo:2020uzc}
D.~Buttazzo, R.~Franceschini and A.~Wulzer, \emph{{Two Paths Towards Precision
  at a Very High Energy Lepton Collider}},
  \href{https://doi.org/10.1007/JHEP05(2021)219}{\emph{JHEP} {\bfseries 05}
  (2021) 219} [\href{https://arxiv.org/abs/2012.11555}{{\ttfamily
  2012.11555}}].

\bibitem{Han:2021lnp}
T.~Han, W.~Kilian, N.~Kreher, Y.~Ma, J.~Reuter, T.~Striegl et~al.,
  \emph{{Precision test of the muon-Higgs coupling at a high-energy muon
  collider}}, \href{https://doi.org/10.1007/JHEP12(2021)162}{\emph{JHEP}
  {\bfseries 12} (2021) 162}
  [\href{https://arxiv.org/abs/2108.05362}{{\ttfamily 2108.05362}}].

\bibitem{Reuter:2022zuv}
J.~Reuter, T.~Han, W.~Kilian, N.~Kreher, Y.~Ma, T.~Striegl et~al.,
  \emph{{Precision test of the muon-Higgs coupling at a high-energy muon
  collider}}, \href{https://doi.org/10.22323/1.414.1239}{\emph{PoS} {\bfseries
  ICHEP2022} (2022) 1239} [\href{https://arxiv.org/abs/2212.01323}{{\ttfamily
  2212.01323}}].

\bibitem{Celada:2023oji}
E.~Celada, T.~Han, W.~Kilian, N.~Kreher, Y.~Ma, F.~Maltoni et~al.,
  \emph{{Probing Higgs-muon interactions at a multi-TeV muon collider}},
  \href{https://doi.org/10.1007/JHEP08(2024)021}{\emph{JHEP} {\bfseries 08}
  (2024) 021} [\href{https://arxiv.org/abs/2312.13082}{{\ttfamily
  2312.13082}}].

\bibitem{Forslund:2022xjq}
M.~Forslund and P.~Meade, \emph{{High precision higgs from high energy muon
  colliders}}, \href{https://doi.org/10.1007/JHEP08(2022)185}{\emph{JHEP}
  {\bfseries 08} (2022) 185}
  [\href{https://arxiv.org/abs/2203.09425}{{\ttfamily 2203.09425}}].

\bibitem{Chen:2022yiu}
M.~Chen and D.~Liu, \emph{{Top Yukawa coupling measurement at the muon
  collider}}, \href{https://doi.org/10.1103/PhysRevD.109.075020}{\emph{Phys.
  Rev. D} {\bfseries 109} (2024) 075020}
  [\href{https://arxiv.org/abs/2212.11067}{{\ttfamily 2212.11067}}].

\bibitem{Ruhdorfer:2023uea}
M.~Ruhdorfer, E.~Salvioni and A.~Wulzer, \emph{{Invisible Higgs boson decay
  from forward muons at a muon collider}},
  \href{https://doi.org/10.1103/PhysRevD.107.095038}{\emph{Phys. Rev. D}
  {\bfseries 107} (2023) 095038}
  [\href{https://arxiv.org/abs/2303.14202}{{\ttfamily 2303.14202}}].

\bibitem{Han:2023njx}
T.~Han, D.~Liu, I.~Low and X.~Wang, \emph{{Electroweak scattering at the muon
  shot}}, \href{https://doi.org/10.1103/PhysRevD.110.013005}{\emph{Phys. Rev.
  D} {\bfseries 110} (2024) 013005}
  [\href{https://arxiv.org/abs/2312.07670}{{\ttfamily 2312.07670}}].

\bibitem{Dermisek:2023rvv}
R.~Dermisek, K.~Hermanek, T.~Lee, N.~McGinnis and S.~Yoon, \emph{{Multi-Higgs
  boson signals of a modified muon Yukawa coupling at a muon collider}},
  \href{https://doi.org/10.1103/PhysRevD.109.095003}{\emph{Phys. Rev. D}
  {\bfseries 109} (2024) 095003}
  [\href{https://arxiv.org/abs/2311.05078}{{\ttfamily 2311.05078}}].

\bibitem{Liu:2023yrb}
Z.~Liu, K.-F.~Lyu, I.~Mahbub and L.-T.~Wang, \emph{{Top Yukawa coupling
  determination at high energy muon collider}},
  \href{https://doi.org/10.1103/PhysRevD.109.035021}{\emph{Phys. Rev. D}
  {\bfseries 109} (2024) 035021}
  [\href{https://arxiv.org/abs/2308.06323}{{\ttfamily 2308.06323}}].

\bibitem{Li:2024joa}
P.~Li, Z.~Liu and K.-F.~Lyu, \emph{{Higgs boson width and couplings at high
  energy muon colliders with forward muon detection}},
  \href{https://doi.org/10.1103/PhysRevD.109.073009}{\emph{Phys. Rev. D}
  {\bfseries 109} (2024) 073009}
  [\href{https://arxiv.org/abs/2401.08756}{{\ttfamily 2401.08756}}].

\bibitem{Cassidy:2023lwd}
M.E.~Cassidy, Z.~Dong, K.~Kong, I.M.~Lewis, Y.~Zhang and Y.-J.~Zheng,
  \emph{{Probing the CP structure of the top quark Yukawa at the future muon
  collider}}, \href{https://doi.org/10.1007/JHEP05(2024)176}{\emph{JHEP}
  {\bfseries 05} (2024) 176}
  [\href{https://arxiv.org/abs/2311.07645}{{\ttfamily 2311.07645}}].

\bibitem{Han:2022edd}
T.~Han, S.~Li, S.~Su, W.~Su and Y.~Wu, \emph{{BSM Higgs Production at a Muon
  Collider}},  in \emph{{Snowmass 2021}}, 5, 2022
  [\href{https://arxiv.org/abs/2205.11730}{{\ttfamily 2205.11730}}].

\bibitem{Bandyopadhyay:2020otm}
P.~Bandyopadhyay and A.~Costantini, \emph{{Obscure Higgs boson at Colliders}},
  \href{https://doi.org/10.1103/PhysRevD.103.015025}{\emph{Phys. Rev. D}
  {\bfseries 103} (2021) 015025}
  [\href{https://arxiv.org/abs/2010.02597}{{\ttfamily 2010.02597}}].

\bibitem{Han:2021udl}
T.~Han, S.~Li, S.~Su, W.~Su and Y.~Wu, \emph{{Heavy Higgs bosons in 2HDM at a
  muon collider}},
  \href{https://doi.org/10.1103/PhysRevD.104.055029}{\emph{Phys. Rev. D}
  {\bfseries 104} (2021) 055029}
  [\href{https://arxiv.org/abs/2102.08386}{{\ttfamily 2102.08386}}].

\bibitem{Bandyopadhyay:2024plc}
P.~Bandyopadhyay, S.~Parashar, C.~Sen and J.~Song, \emph{{Probing Inert Triplet
  Model at a multi-TeV muon collider via vector boson fusion with forward muon
  tagging}}, \href{https://doi.org/10.1007/JHEP07(2024)253}{\emph{JHEP}
  {\bfseries 07} (2024) 253}
  [\href{https://arxiv.org/abs/2401.02697}{{\ttfamily 2401.02697}}].

\bibitem{Ouazghour:2023plc}
B.A.~Ouazghour, A.~Arhrib, K.~Cheung, E.-s.~Ghourmin and L.~Rahili,
  \emph{{Comparison between \ensuremath{\mu}-\ensuremath{\mu}+ and e-e+
  colliders for charged Higgs production in the 2HDM}},
  \href{https://doi.org/10.1103/PhysRevD.109.115009}{\emph{Phys. Rev. D}
  {\bfseries 109} (2024) 115009}
  [\href{https://arxiv.org/abs/2308.15664}{{\ttfamily 2308.15664}}].

\bibitem{Jueid:2023qcf}
A.~Jueid, T.A.~Chowdhury, S.~Nasri and S.~Saad, \emph{{Probing Zee-Babu states
  at muon colliders}},
  \href{https://doi.org/10.1103/PhysRevD.109.075011}{\emph{Phys. Rev. D}
  {\bfseries 109} (2024) 075011}
  [\href{https://arxiv.org/abs/2306.01255}{{\ttfamily 2306.01255}}].

\bibitem{Han:2020uak}
T.~Han, Z.~Liu, L.-T.~Wang and X.~Wang, \emph{{WIMPs at High Energy Muon
  Colliders}}, \href{https://doi.org/10.1103/PhysRevD.103.075004}{\emph{Phys.
  Rev. D} {\bfseries 103} (2021) 075004}
  [\href{https://arxiv.org/abs/2009.11287}{{\ttfamily 2009.11287}}].

\bibitem{Capdevilla:2021fmj}
R.~Capdevilla, F.~Meloni, R.~Simoniello and J.~Zurita, \emph{{Hunting wino and
  higgsino dark matter at the muon collider with disappearing tracks}},
  \href{https://doi.org/10.1007/JHEP06(2021)133}{\emph{JHEP} {\bfseries 06}
  (2021) 133} [\href{https://arxiv.org/abs/2102.11292}{{\ttfamily
  2102.11292}}].

\bibitem{Han:2022ubw}
T.~Han, Z.~Liu, L.-T.~Wang and X.~Wang, \emph{{WIMP Dark Matter at High Energy
  Muon Colliders $-$A White Paper for Snowmass 2021}},  in \emph{{Snowmass
  2021}}, 3, 2022 [\href{https://arxiv.org/abs/2203.07351}{{\ttfamily
  2203.07351}}].

\bibitem{Cesarotti:2024rbh}
C.~Cesarotti and G.~Krnjaic, \emph{{Hitting the Thermal Target for Leptophilic
  Dark Matter}},  \href{https://arxiv.org/abs/2404.02906}{{\ttfamily
  2404.02906}}.

\bibitem{Asadi:2023csb}
P.~Asadi, A.~Radick and T.-T.~Yu, \emph{{Interplay of freeze-in and freeze-out:
  Lepton-flavored dark matter and muon colliders}},
  \href{https://doi.org/10.1103/PhysRevD.110.035022}{\emph{Phys. Rev. D}
  {\bfseries 110} (2024) 035022}
  [\href{https://arxiv.org/abs/2312.03826}{{\ttfamily 2312.03826}}].

\bibitem{Belfkir:2023vpo}
M.~Belfkir, A.~Jueid and S.~Nasri, \emph{{Boosting dark matter searches at muon
  colliders with machine learning: The mono-Higgs channel as a case study}},
  \href{https://doi.org/10.1093/ptep/ptad144}{\emph{PTEP} {\bfseries 2023}
  (2023) 123B03} [\href{https://arxiv.org/abs/2309.11241}{{\ttfamily
  2309.11241}}].

\bibitem{Jueid:2023zxx}
A.~Jueid and S.~Nasri, \emph{{Lepton portal dark matter at muon colliders:
  Total rates and generic features for phenomenologically viable scenarios}},
  \href{https://doi.org/10.1103/PhysRevD.107.115027}{\emph{Phys. Rev. D}
  {\bfseries 107} (2023) 115027}
  [\href{https://arxiv.org/abs/2301.12524}{{\ttfamily 2301.12524}}].

\bibitem{Huang:2021nkl}
G.-y.~Huang, F.S.~Queiroz and W.~Rodejohann, \emph{{Gauged
  $L^{}_{\mu}{-}L^{}_{\tau}$ at a muon collider}},
  \href{https://doi.org/10.1103/PhysRevD.103.095005}{\emph{Phys. Rev. D}
  {\bfseries 103} (2021) 095005}
  [\href{https://arxiv.org/abs/2101.04956}{{\ttfamily 2101.04956}}].

\bibitem{Huang:2021biu}
G.-y.~Huang, S.~Jana, F.S.~Queiroz and W.~Rodejohann, \emph{{Probing the RK(*)
  anomaly at a muon collider}},
  \href{https://doi.org/10.1103/PhysRevD.105.015013}{\emph{Phys. Rev. D}
  {\bfseries 105} (2022) 015013}
  [\href{https://arxiv.org/abs/2103.01617}{{\ttfamily 2103.01617}}].

\bibitem{He:2024dwh}
R.-Y.~He, J.-Q.~Huang, J.-Y.~Xu, F.-X.~Yang, Z.-L.~Han and F.-L.~Shao,
  \emph{{Heavy neutral leptons in gauged U(1)$_{L \mu -L \tau}$ at muon
  collider}}, \href{https://doi.org/10.1088/1674-1137/ad4d61}{\emph{Chin. Phys.
  C} {\bfseries 48} (2024) 093102}
  [\href{https://arxiv.org/abs/2401.14687}{{\ttfamily 2401.14687}}].

\bibitem{Buttazzo:2020ibd}
D.~Buttazzo and P.~Paradisi, \emph{{Probing the muon $g-2$ anomaly with the
  Higgs boson at a muon collider}},
  \href{https://doi.org/10.1103/PhysRevD.104.075021}{\emph{Phys. Rev. D}
  {\bfseries 104} (2021) 075021}
  [\href{https://arxiv.org/abs/2012.02769}{{\ttfamily 2012.02769}}].

\bibitem{Capdevilla:2020qel}
R.~Capdevilla, D.~Curtin, Y.~Kahn and G.~Krnjaic, \emph{{Discovering the
  physics of $(g-2)_\mu$ at future muon colliders}},
  \href{https://doi.org/10.1103/PhysRevD.103.075028}{\emph{Phys. Rev. D}
  {\bfseries 103} (2021) 075028}
  [\href{https://arxiv.org/abs/2006.16277}{{\ttfamily 2006.16277}}].

\bibitem{Yin:2020afe}
W.~Yin and M.~Yamaguchi, \emph{{Muon g-2 at a multi-TeV muon collider}},
  \href{https://doi.org/10.1103/PhysRevD.106.033007}{\emph{Phys. Rev. D}
  {\bfseries 106} (2022) 033007}
  [\href{https://arxiv.org/abs/2012.03928}{{\ttfamily 2012.03928}}].

\bibitem{Capdevilla:2021rwo}
R.~Capdevilla, D.~Curtin, Y.~Kahn and G.~Krnjaic, \emph{{No-lose theorem for
  discovering the new physics of~$(g-2)_\mu$ at muon colliders}},
  \href{https://doi.org/10.1103/PhysRevD.105.015028}{\emph{Phys. Rev. D}
  {\bfseries 105} (2022) 015028}
  [\href{https://arxiv.org/abs/2101.10334}{{\ttfamily 2101.10334}}].

\bibitem{Gu:2020ldn}
J.~Gu, L.-T.~Wang and C.~Zhang, \emph{{Unambiguously Testing Positivity at
  Lepton Colliders}},
  \href{https://doi.org/10.1103/PhysRevLett.129.011805}{\emph{Phys. Rev. Lett.}
  {\bfseries 129} (2022) 011805}
  [\href{https://arxiv.org/abs/2011.03055}{{\ttfamily 2011.03055}}].

\bibitem{Costantini:2020stv}
A.~Costantini, F.~De~Lillo, F.~Maltoni, L.~Mantani, O.~Mattelaer, R.~Ruiz
  et~al., \emph{{Vector boson fusion at multi-TeV muon colliders}},
  \href{https://doi.org/10.1007/JHEP09(2020)080}{\emph{JHEP} {\bfseries 09}
  (2020) 080} [\href{https://arxiv.org/abs/2005.10289}{{\ttfamily
  2005.10289}}].

\bibitem{Bandyopadhyay:2021pld}
P.~Bandyopadhyay, A.~Karan, R.~Mandal and S.~Parashar, \emph{{Distinguishing
  signatures of scalar leptoquarks at hadron and muon colliders}},
  \href{https://doi.org/10.1140/epjc/s10052-022-10809-9}{\emph{Eur. Phys. J. C}
  {\bfseries 82} (2022) 916}
  [\href{https://arxiv.org/abs/2108.06506}{{\ttfamily 2108.06506}}].

\bibitem{Qian:2021ihf}
S.~Qian, C.~Li, Q.~Li, F.~Meng, J.~Xiao, T.~Yang et~al., \emph{{Searching for
  heavy leptoquarks at a muon collider}},
  \href{https://doi.org/10.1007/JHEP12(2021)047}{\emph{JHEP} {\bfseries 12}
  (2021) 047} [\href{https://arxiv.org/abs/2109.01265}{{\ttfamily
  2109.01265}}].

\bibitem{Bao:2022onq}
Y.~Bao, J.~Fan and L.~Li, \emph{{Electroweak ALP searches at a muon collider}},
  \href{https://doi.org/10.1007/JHEP08(2022)276}{\emph{JHEP} {\bfseries 08}
  (2022) 276} [\href{https://arxiv.org/abs/2203.04328}{{\ttfamily
  2203.04328}}].

\bibitem{Ghosh:2023xbj}
N.~Ghosh, S.K.~Rai and T.~Samui, \emph{{Search for a leptoquark and vector-like
  lepton in a muon collider}},
  \href{https://doi.org/10.1016/j.nuclphysb.2024.116564}{\emph{Nucl. Phys. B}
  {\bfseries 1004} (2024) 116564}
  [\href{https://arxiv.org/abs/2309.07583}{{\ttfamily 2309.07583}}].

\bibitem{Dermisek:2023tgq}
R.~Dermisek, K.~Hermanek, N.~McGinnis and S.~Yoon, \emph{{Predictions for muon
  electric and magnetic dipole moments from
  h\textrightarrow{}\ensuremath{\mu}+\ensuremath{\mu}- in two-Higgs-doublet
  models with new leptons}},
  \href{https://doi.org/10.1103/PhysRevD.108.055019}{\emph{Phys. Rev. D}
  {\bfseries 108} (2023) 055019}
  [\href{https://arxiv.org/abs/2306.13212}{{\ttfamily 2306.13212}}].

\bibitem{Altmannshofer:2023uci}
W.~Altmannshofer, S.A.~Gadam and S.~Profumo, \emph{{Probing new physics with
  \ensuremath{\mu}+\ensuremath{\mu}-\textrightarrow{}bs at a muon collider}},
  \href{https://doi.org/10.1103/PhysRevD.108.115033}{\emph{Phys. Rev. D}
  {\bfseries 108} (2023) 115033}
  [\href{https://arxiv.org/abs/2306.15017}{{\ttfamily 2306.15017}}].

\bibitem{Liu:2023jta}
D.~Liu, L.-T.~Wang and K.-P.~Xie, \emph{{Composite resonances at a 10 TeV muon
  collider}}, \href{https://doi.org/10.1007/JHEP04(2024)084}{\emph{JHEP}
  {\bfseries 04} (2024) 084}
  [\href{https://arxiv.org/abs/2312.09117}{{\ttfamily 2312.09117}}].

\bibitem{Sun:2023cuf}
S.~Sun, Q.-S.~Yan, X.~Zhao and Z.~Zhao, \emph{{Constraining rare B decays by
  \ensuremath{\mu}+\ensuremath{\mu}-\textrightarrow{}tc at future lepton
  colliders}}, \href{https://doi.org/10.1103/PhysRevD.108.075016}{\emph{Phys.
  Rev. D} {\bfseries 108} (2023) 075016}
  [\href{https://arxiv.org/abs/2302.01143}{{\ttfamily 2302.01143}}].

\bibitem{Kwok:2023dck}
T.H.~Kwok, L.~Li, T.~Liu and A.~Rock, \emph{{Searching for heavy neutral
  leptons at a future muon collider}},
  \href{https://doi.org/10.1103/PhysRevD.110.075009}{\emph{Phys. Rev. D}
  {\bfseries 110} (2024) 075009}
  [\href{https://arxiv.org/abs/2301.05177}{{\ttfamily 2301.05177}}].

\bibitem{Cesarotti:2023sje}
C.~Cesarotti and R.~Gambhir, \emph{{The new physics case for beam-dump
  experiments with accelerated muon beams}},
  \href{https://doi.org/10.1007/JHEP05(2024)283}{\emph{JHEP} {\bfseries 05}
  (2024) 283} [\href{https://arxiv.org/abs/2310.16110}{{\ttfamily
  2310.16110}}].

\bibitem{Ema:2023buz}
Y.~Ema, Z.~Liu, K.-F.~Lyu and M.~Pospelov, \emph{{Heavy Neutral Leptons from
  Stopped Muons and Pions}},
  \href{https://doi.org/10.1007/JHEP08(2023)169}{\emph{JHEP} {\bfseries 08}
  (2023) 169} [\href{https://arxiv.org/abs/2306.07315}{{\ttfamily
  2306.07315}}].

\bibitem{Li:2023tbx}
P.~Li, Z.~Liu and K.-F.~Lyu, \emph{{Heavy neutral leptons at muon colliders}},
  \href{https://doi.org/10.1007/JHEP03(2023)231}{\emph{JHEP} {\bfseries 03}
  (2023) 231} [\href{https://arxiv.org/abs/2301.07117}{{\ttfamily
  2301.07117}}].

\bibitem{Bhattacharya:2023beo}
S.~Bhattacharya, S.~Jahedi, S.~Nandi and A.~Sarkar, \emph{{Probing flavor
  constrained SMEFT operators through tc production at the muon collider}},
  \href{https://doi.org/10.1007/JHEP07(2024)061}{\emph{JHEP} {\bfseries 07}
  (2024) 061} [\href{https://arxiv.org/abs/2312.14872}{{\ttfamily
  2312.14872}}].

\bibitem{Das:2024ekt}
N.~Das and N.~Ghosh, \emph{{Unveiling the CP-odd Higgs boson in a generalized
  2HDM at a muon collider}},
  \href{https://doi.org/10.1103/PhysRevD.111.015035}{\emph{Phys. Rev. D}
  {\bfseries 111} (2025) 015035}
  [\href{https://arxiv.org/abs/2406.18698}{{\ttfamily 2406.18698}}].

\bibitem{FCC:2018evy}
{\scshape FCC} collaboration, \emph{{FCC-ee: The Lepton Collider}: {Future
  Circular Collider Conceptual Design Report Volume 2}},
  \href{https://doi.org/10.1140/epjst/e2019-900045-4}{\emph{Eur. Phys. J. ST}
  {\bfseries 228} (2019) 261}.

\bibitem{Bernardi:2022hny}
G.~Bernardi et~al., \emph{{The Future Circular Collider: a Summary for the US
  2021 Snowmass Process}},  \href{https://arxiv.org/abs/2203.06520}{{\ttfamily
  2203.06520}}.

\bibitem{CEPCStudyGroup:2018rmc}
{\scshape CEPC Study Group} collaboration, \emph{{CEPC Conceptual Design
  Report: Volume 1 - Accelerator}},
  \href{https://arxiv.org/abs/1809.00285}{{\ttfamily 1809.00285}}.

\bibitem{CEPCStudyGroup:2018ghi}
{\scshape CEPC Study Group} collaboration, \emph{{CEPC Conceptual Design
  Report: Volume 2 - Physics \& Detector}},
  \href{https://arxiv.org/abs/1811.10545}{{\ttfamily 1811.10545}}.

\bibitem{An:2018dwb}
F.~An et~al., \emph{{Precision Higgs physics at the CEPC}},
  \href{https://doi.org/10.1088/1674-1137/43/4/043002}{\emph{Chin. Phys. C}
  {\bfseries 43} (2019) 043002}
  [\href{https://arxiv.org/abs/1810.09037}{{\ttfamily 1810.09037}}].

\bibitem{CEPCAcceleratorStudyGroup:2019myu}
{\scshape CEPC Accelerator Study Group} collaboration, \emph{{CEPC Input to the
  ESPP 2018 -Accelerator}},  \href{https://arxiv.org/abs/1901.03169}{{\ttfamily
  1901.03169}}.

\bibitem{CEPCPhysicsStudyGroup:2022uwl}
{\scshape CEPC Physics Study Group} collaboration, \emph{{The Physics potential
  of the CEPC. Prepared for the US Snowmass Community Planning Exercise
  (Snowmass 2021)}},  in \emph{{Snowmass 2021}}, 5, 2022
  [\href{https://arxiv.org/abs/2205.08553}{{\ttfamily 2205.08553}}].

\bibitem{CEPCStudyGroup:2023quu}
{\scshape CEPC Study Group} collaboration, \emph{{CEPC Technical Design Report:
  Accelerator}},
  \href{https://doi.org/10.1007/s41605-024-00463-y}{\emph{Radiat. Detect.
  Technol. Methods} {\bfseries 8} (2024) 1}
  [\href{https://arxiv.org/abs/2312.14363}{{\ttfamily 2312.14363}}].

\bibitem{Kuhn:2007ca}
J.H.~K\"uhn, F.~Metzler and A.A.~Penin, \emph{{Next-to-next-to-leading
  electroweak logarithms in W-pair production at ILC}},
  \href{https://doi.org/10.1016/j.nuclphysb.2007.11.019}{\emph{Nucl. Phys. B}
  {\bfseries 795} (2008) 277}
  [\href{https://arxiv.org/abs/0709.4055}{{\ttfamily 0709.4055}}].

\bibitem{ILC:2013jhg}
{\scshape ILC} collaboration, H.~Baer et~al., eds., \emph{{The International
  Linear Collider Technical Design Report - Volume 2: Physics}},
  \href{https://arxiv.org/abs/1306.6352}{{\ttfamily 1306.6352}}.

\bibitem{Behnke:2013lya}
H.~Abramowicz et~al., \emph{{The International Linear Collider Technical Design
  Report - Volume 4: Detectors}},
  \href{https://arxiv.org/abs/1306.6329}{{\ttfamily 1306.6329}}.

\bibitem{ILCInternationalDevelopmentTeam:2022izu}
{\scshape ILC International Development Team} collaboration, \emph{{The
  International Linear Collider: Report to Snowmass 2021}},
  \href{https://arxiv.org/abs/2203.07622}{{\ttfamily 2203.07622}}.

\bibitem{Linssen:2012hp}
L.~Linssen, A.~Miyamoto, M.~Stanitzki and H.~Weerts, eds., \emph{{Physics and
  Detectors at CLIC: CLIC Conceptual Design Report}},
  \href{https://arxiv.org/abs/1202.5940}{{\ttfamily 1202.5940}}.

\bibitem{Lebrun:2012hj}
P.~Lebrun, L.~Linssen, A.~Lucaci-Timoce, D.~Schulte, F.~Simon, S.~Stapnes
  et~al., \emph{{The CLIC Programme: Towards a Staged e+e- Linear Collider
  Exploring the Terascale : CLIC Conceptual Design Report}},
  \href{https://arxiv.org/abs/1209.2543}{{\ttfamily 1209.2543}}.

\bibitem{CLIC:2016zwp}
{\scshape CLIC, CLICdp} collaboration, \emph{{Updated baseline for a staged
  Compact Linear Collider}},
  \href{https://arxiv.org/abs/1608.07537}{{\ttfamily 1608.07537}}.

\bibitem{Brunner:2022usy}
O.~Brunner et~al., \emph{{The CLIC project}},
  \href{https://arxiv.org/abs/2203.09186}{{\ttfamily 2203.09186}}.

\bibitem{Bai:2021rdg}
M.~Bai et~al., \emph{{C$^3$: A ''Cool'' Route to the Higgs Boson and Beyond}},
  in \emph{{Snowmass 2021}}, 10, 2021
  [\href{https://arxiv.org/abs/2110.15800}{{\ttfamily 2110.15800}}].

\bibitem{Vernieri:2022fae}
C.~Vernieri et~al., \emph{{Strategy for Understanding the Higgs Physics: The
  Cool Copper Collider}},
  \href{https://doi.org/10.1088/1748-0221/18/07/P07053}{\emph{JINST} {\bfseries
  18} (2023) P07053} [\href{https://arxiv.org/abs/2203.07646}{{\ttfamily
  2203.07646}}].

\bibitem{Chen:2022msz}
S.~Chen, A.~Glioti, R.~Rattazzi, L.~Ricci and A.~Wulzer, \emph{{Learning from
  radiation at a very high energy lepton collider}},
  \href{https://doi.org/10.1007/JHEP05(2022)180}{\emph{JHEP} {\bfseries 05}
  (2022) 180} [\href{https://arxiv.org/abs/2202.10509}{{\ttfamily
  2202.10509}}].

\bibitem{AlAli:2021let}
H.~Al~Ali et~al., \emph{{The muon Smasher\textquoteright{}s guide}},
  \href{https://doi.org/10.1088/1361-6633/ac6678}{\emph{Rept. Prog. Phys.}
  {\bfseries 85} (2022) 084201}
  [\href{https://arxiv.org/abs/2103.14043}{{\ttfamily 2103.14043}}].

\bibitem{Ruiz:2021tdt}
R.~Ruiz, A.~Costantini, F.~Maltoni and O.~Mattelaer, \emph{{The Effective
  Vector Boson Approximation in high-energy muon collisions}},
  \href{https://doi.org/10.1007/JHEP06(2022)114}{\emph{JHEP} {\bfseries 06}
  (2022) 114} [\href{https://arxiv.org/abs/2111.02442}{{\ttfamily
  2111.02442}}].

\bibitem{Han:2020uid}
T.~Han, Y.~Ma and K.~Xie, \emph{{High energy leptonic collisions and
  electroweak parton distribution functions}},
  \href{https://doi.org/10.1103/PhysRevD.103.L031301}{\emph{Phys. Rev. D}
  {\bfseries 103} (2021) L031301}
  [\href{https://arxiv.org/abs/2007.14300}{{\ttfamily 2007.14300}}].

\bibitem{Han:2021kes}
T.~Han, Y.~Ma and K.~Xie, \emph{{Quark and gluon contents of a lepton at high
  energies}}, \href{https://doi.org/10.1007/JHEP02(2022)154}{\emph{JHEP}
  {\bfseries 02} (2022) 154}
  [\href{https://arxiv.org/abs/2103.09844}{{\ttfamily 2103.09844}}].

\bibitem{Garosi:2023bvq}
F.~Garosi, D.~Marzocca and S.~Trifinopoulos, \emph{{LePDF: Standard Model PDFs
  for high-energy lepton colliders}},
  \href{https://doi.org/10.1007/JHEP09(2023)107}{\emph{JHEP} {\bfseries 09}
  (2023) 107} [\href{https://arxiv.org/abs/2303.16964}{{\ttfamily
  2303.16964}}].

\bibitem{Sudakov:1954sw}
V.V.~Sudakov, \emph{{Vertex parts at very high-energies in quantum
  electrodynamics}}, {\emph{Sov. Phys. JETP} {\bfseries 3} (1956) 65}.

\bibitem{Denner:2000jv}
A.~Denner and S.~Pozzorini, \emph{{One loop leading logarithms in electroweak
  radiative corrections. 1. Results}},
  \href{https://doi.org/10.1007/s100520100551}{\emph{Eur. Phys. J. C}
  {\bfseries 18} (2001) 461}
  [\href{https://arxiv.org/abs/hep-ph/0010201}{{\ttfamily hep-ph/0010201}}].

\bibitem{Denner:2001gw}
A.~Denner and S.~Pozzorini, \emph{{One loop leading logarithms in electroweak
  radiative corrections. 2. Factorization of collinear singularities}},
  \href{https://doi.org/10.1007/s100520100721}{\emph{Eur. Phys. J. C}
  {\bfseries 21} (2001) 63}
  [\href{https://arxiv.org/abs/hep-ph/0104127}{{\ttfamily hep-ph/0104127}}].

\bibitem{Denner:2003wi}
A.~Denner, M.~Melles and S.~Pozzorini, \emph{{Two loop electroweak angular
  dependent logarithms at high-energies}},
  \href{https://doi.org/10.1016/S0550-3213(03)00307-9}{\emph{Nucl. Phys. B}
  {\bfseries 662} (2003) 299}
  [\href{https://arxiv.org/abs/hep-ph/0301241}{{\ttfamily hep-ph/0301241}}].

\bibitem{Denner:2004iz}
A.~Denner and S.~Pozzorini, \emph{{An Algorithm for the high-energy expansion
  of multi-loop diagrams to next-to-leading logarithmic accuracy}},
  \href{https://doi.org/10.1016/j.nuclphysb.2005.03.036}{\emph{Nucl. Phys. B}
  {\bfseries 717} (2005) 48}
  [\href{https://arxiv.org/abs/hep-ph/0408068}{{\ttfamily hep-ph/0408068}}].

\bibitem{Denner:2006jr}
A.~Denner, B.~Jantzen and S.~Pozzorini, \emph{{Two-loop electroweak
  next-to-leading logarithmic corrections to massless fermionic processes}},
  \href{https://doi.org/10.1016/j.nuclphysb.2006.10.014}{\emph{Nucl. Phys. B}
  {\bfseries 761} (2007) 1}
  [\href{https://arxiv.org/abs/hep-ph/0608326}{{\ttfamily hep-ph/0608326}}].

\bibitem{Denner:2008yn}
A.~Denner, B.~Jantzen and S.~Pozzorini, \emph{{Two-loop electroweak
  next-to-leading logarithms for processes involving heavy quarks}},
  \href{https://doi.org/10.1088/1126-6708/2008/11/062}{\emph{JHEP} {\bfseries
  11} (2008) 062} [\href{https://arxiv.org/abs/0809.0800}{{\ttfamily
  0809.0800}}].

\bibitem{Bothmann:2020sxm}
E.~Bothmann and D.~Napoletano, \emph{{Automated evaluation of electroweak
  Sudakov logarithms in Sherpa}},
  \href{https://doi.org/10.1140/epjc/s10052-020-08596-2}{\emph{Eur. Phys. J. C}
  {\bfseries 80} (2020) 1024}
  [\href{https://arxiv.org/abs/2006.14635}{{\ttfamily 2006.14635}}].

\bibitem{Sherpa:2019gpd}
{\scshape Sherpa} collaboration, \emph{{Event Generation with Sherpa 2.2}},
  \href{https://doi.org/10.21468/SciPostPhys.7.3.034}{\emph{SciPost Phys.}
  {\bfseries 7} (2019) 034} [\href{https://arxiv.org/abs/1905.09127}{{\ttfamily
  1905.09127}}].

\bibitem{Pagani:2021vyk}
D.~Pagani and M.~Zaro, \emph{{One-loop electroweak Sudakov logarithms: a
  revisitation and automation}},
  \href{https://doi.org/10.1007/JHEP02(2022)161}{\emph{JHEP} {\bfseries 02}
  (2022) 161} [\href{https://arxiv.org/abs/2110.03714}{{\ttfamily
  2110.03714}}].

\bibitem{Alwall:2014hca}
J.~Alwall, R.~Frederix, S.~Frixione, V.~Hirschi, F.~Maltoni, O.~Mattelaer
  et~al., \emph{{The automated computation of tree-level and next-to-leading
  order differential cross sections, and their matching to parton shower
  simulations}}, \href{https://doi.org/10.1007/JHEP07(2014)079}{\emph{JHEP}
  {\bfseries 07} (2014) 079} [\href{https://arxiv.org/abs/1405.0301}{{\ttfamily
  1405.0301}}].

\bibitem{Frederix:2018nkq}
R.~Frederix, S.~Frixione, V.~Hirschi, D.~Pagani, H.S.~Shao and M.~Zaro,
  \emph{{The automation of next-to-leading order electroweak calculations}},
  \href{https://doi.org/10.1007/JHEP11(2021)085}{\emph{JHEP} {\bfseries 07}
  (2018) 185} [\href{https://arxiv.org/abs/1804.10017}{{\ttfamily
  1804.10017}}].

\bibitem{Pagani:2023wgc}
D.~Pagani, T.~Vitos and M.~Zaro, \emph{{Improving NLO QCD event generators with
  high-energy EW corrections}},
  \href{https://doi.org/10.1140/epjc/s10052-024-12836-0}{\emph{Eur. Phys. J. C}
  {\bfseries 84} (2024) 514}
  [\href{https://arxiv.org/abs/2309.00452}{{\ttfamily 2309.00452}}].

\bibitem{Lindert:2023fcu}
J.M.~Lindert and L.~Mai, \emph{{Logarithmic EW corrections at one-loop}},
  \href{https://doi.org/10.1140/epjc/s10052-024-13430-0}{\emph{Eur. Phys. J. C}
  {\bfseries 84} (2024) 1084}
  [\href{https://arxiv.org/abs/2312.07927}{{\ttfamily 2312.07927}}].

\bibitem{Cascioli:2011va}
F.~Cascioli, P.~Maierhofer and S.~Pozzorini, \emph{{Scattering Amplitudes with
  Open Loops}},
  \href{https://doi.org/10.1103/PhysRevLett.108.111601}{\emph{Phys. Rev. Lett.}
  {\bfseries 108} (2012) 111601}
  [\href{https://arxiv.org/abs/1111.5206}{{\ttfamily 1111.5206}}].

\bibitem{Buccioni:2019sur}
F.~Buccioni, J.-N.~Lang, J.M.~Lindert, P.~Maierh\"ofer, S.~Pozzorini, H.~Zhang
  et~al., \emph{{OpenLoops 2}},
  \href{https://doi.org/10.1140/epjc/s10052-019-7306-2}{\emph{Eur. Phys. J. C}
  {\bfseries 79} (2019) 866}
  [\href{https://arxiv.org/abs/1907.13071}{{\ttfamily 1907.13071}}].

\bibitem{Chiu:2007yn}
J.-y.~Chiu, F.~Golf, R.~Kelley and A.V.~Manohar, \emph{{Electroweak Sudakov
  corrections using effective field theory}},
  \href{https://doi.org/10.1103/PhysRevLett.100.021802}{\emph{Phys. Rev. Lett.}
  {\bfseries 100} (2008) 021802}
  [\href{https://arxiv.org/abs/0709.2377}{{\ttfamily 0709.2377}}].

\bibitem{Chiu:2008vv}
J.-y.~Chiu, R.~Kelley and A.V.~Manohar, \emph{{Electroweak Corrections using
  Effective Field Theory: Applications to the LHC}},
  \href{https://doi.org/10.1103/PhysRevD.78.073006}{\emph{Phys. Rev. D}
  {\bfseries 78} (2008) 073006}
  [\href{https://arxiv.org/abs/0806.1240}{{\ttfamily 0806.1240}}].

\bibitem{Manohar:2018kfx}
A.V.~Manohar and W.J.~Waalewijn, \emph{{Electroweak Logarithms in Inclusive
  Cross Sections}}, \href{https://doi.org/10.1007/JHEP08(2018)137}{\emph{JHEP}
  {\bfseries 08} (2018) 137}
  [\href{https://arxiv.org/abs/1802.08687}{{\ttfamily 1802.08687}}].

\bibitem{Bauer:2000ew}
C.W.~Bauer, S.~Fleming and M.E.~Luke, \emph{{Summing Sudakov logarithms in $B
  \to  X_s \gamma $in effective field theory.}},
  \href{https://doi.org/10.1103/PhysRevD.63.014006}{\emph{Phys. Rev. D}
  {\bfseries 63} (2000) 014006}
  [\href{https://arxiv.org/abs/hep-ph/0005275}{{\ttfamily hep-ph/0005275}}].

\bibitem{Bauer:2000yr}
C.W.~Bauer, S.~Fleming, D.~Pirjol and I.W.~Stewart, \emph{{An Effective field
  theory for collinear and soft gluons: Heavy to light decays}},
  \href{https://doi.org/10.1103/PhysRevD.63.114020}{\emph{Phys. Rev. D}
  {\bfseries 63} (2001) 114020}
  [\href{https://arxiv.org/abs/hep-ph/0011336}{{\ttfamily hep-ph/0011336}}].

\bibitem{Bauer:2001ct}
C.W.~Bauer and I.W.~Stewart, \emph{{Invariant operators in collinear effective
  theory}}, \href{https://doi.org/10.1016/S0370-2693(01)00902-9}{\emph{Phys.
  Lett. B} {\bfseries 516} (2001) 134}
  [\href{https://arxiv.org/abs/hep-ph/0107001}{{\ttfamily hep-ph/0107001}}].

\bibitem{Bauer:2001yt}
C.W.~Bauer, D.~Pirjol and I.W.~Stewart, \emph{{Soft collinear factorization in
  effective field theory}},
  \href{https://doi.org/10.1103/PhysRevD.65.054022}{\emph{Phys. Rev. D}
  {\bfseries 65} (2002) 054022}
  [\href{https://arxiv.org/abs/hep-ph/0109045}{{\ttfamily hep-ph/0109045}}].

\bibitem{Denner:2024yut}
A.~Denner and S.~Rode, \emph{{Automated resummation of electroweak Sudakov
  logarithms in diboson production at future colliders}},
  \href{https://doi.org/10.1140/epjc/s10052-024-12879-3}{\emph{Eur. Phys. J. C}
  {\bfseries 84} (2024) 542}
  [\href{https://arxiv.org/abs/2402.10503}{{\ttfamily 2402.10503}}].

\bibitem{Aicheler:2012bya}
M.~Aicheler, P.~Burrows, M.~Draper, T.~Garvey, P.~Lebrun, K.~Peach et~al.,
  eds., \emph{{A Multi-TeV Linear Collider Based on CLIC Technology}: {CLIC
  Conceptual Design Report}}, .

\bibitem{CLICdp:2018cto}
{\scshape CLICdp, CLIC} collaboration, \emph{{The Compact Linear Collider
  (CLIC) - 2018 Summary Report}},
  \href{https://arxiv.org/abs/1812.06018}{{\ttfamily 1812.06018}}.

\bibitem{FCC:2018byv}
{\scshape FCC} collaboration, \emph{{FCC Physics Opportunities}: {Future
  Circular Collider Conceptual Design Report Volume 1}},
  \href{https://doi.org/10.1140/epjc/s10052-019-6904-3}{\emph{Eur. Phys. J. C}
  {\bfseries 79} (2019) 474}.

\bibitem{FCC:2018vvp}
{\scshape FCC} collaboration, \emph{{FCC-hh: The Hadron Collider}: {Future
  Circular Collider Conceptual Design Report Volume 3}},
  \href{https://doi.org/10.1140/epjst/e2019-900087-0}{\emph{Eur. Phys. J. ST}
  {\bfseries 228} (2019) 755}.

\bibitem{Benedikt:2022kan}
M.~Benedikt et~al., \emph{{Future Circular Hadron Collider FCC-hh: Overview and
  Status}},  \href{https://arxiv.org/abs/2203.07804}{{\ttfamily 2203.07804}}.

\bibitem{Ciafaloni:2010ti}
P.~Ciafaloni, D.~Comelli, A.~Riotto, F.~Sala, A.~Strumia and A.~Urbano,
  \emph{{Weak Corrections are Relevant for Dark Matter Indirect Detection}},
  \href{https://doi.org/10.1088/1475-7516/2011/03/019}{\emph{JCAP} {\bfseries
  03} (2011) 019} [\href{https://arxiv.org/abs/1009.0224}{{\ttfamily
  1009.0224}}].

\bibitem{Kallweit:2014xda}
S.~Kallweit, J.M.~Lindert, P.~Maierh\"ofer, S.~Pozzorini and M.~Sch\"onherr,
  \emph{{NLO electroweak automation and precise predictions for W+multijet
  production at the LHC}},
  \href{https://doi.org/10.1007/JHEP04(2015)012}{\emph{JHEP} {\bfseries 04}
  (2015) 012} [\href{https://arxiv.org/abs/1412.5157}{{\ttfamily 1412.5157}}].

\bibitem{Frixione:2015zaa}
S.~Frixione, V.~Hirschi, D.~Pagani, H.S.~Shao and M.~Zaro, \emph{{Electroweak
  and QCD corrections to top-pair hadroproduction in association with heavy
  bosons}}, \href{https://doi.org/10.1007/JHEP06(2015)184}{\emph{JHEP}
  {\bfseries 06} (2015) 184}
  [\href{https://arxiv.org/abs/1504.03446}{{\ttfamily 1504.03446}}].

\bibitem{Chiesa:2015mya}
M.~Chiesa, N.~Greiner and F.~Tramontano, \emph{{Automation of electroweak
  corrections for LHC processes}},
  \href{https://doi.org/10.1088/0954-3899/43/1/013002}{\emph{J. Phys. G}
  {\bfseries 43} (2016) 013002}
  [\href{https://arxiv.org/abs/1507.08579}{{\ttfamily 1507.08579}}].

\bibitem{Biedermann:2017yoi}
B.~Biedermann, S.~Br\"auer, A.~Denner, M.~Pellen, S.~Schumann and
  J.M.~Thompson, \emph{{Automation of NLO QCD and EW corrections with Sherpa
  and Recola}},
  \href{https://doi.org/10.1140/epjc/s10052-017-5054-8}{\emph{Eur. Phys. J. C}
  {\bfseries 77} (2017) 492}
  [\href{https://arxiv.org/abs/1704.05783}{{\ttfamily 1704.05783}}].

\bibitem{Chiesa:2017gqx}
M.~Chiesa, N.~Greiner, M.~Sch\"onherr and F.~Tramontano, \emph{{Electroweak
  corrections to diphoton plus jets}},
  \href{https://doi.org/10.1007/JHEP10(2017)181}{\emph{JHEP} {\bfseries 10}
  (2017) 181} [\href{https://arxiv.org/abs/1706.09022}{{\ttfamily
  1706.09022}}].

\bibitem{Pagani:2021iwa}
D.~Pagani, H.-S.~Shao, I.~Tsinikos and M.~Zaro, \emph{{Automated EW corrections
  with isolated photons: t$ \overline{t} $\ensuremath{\gamma}, t$ \overline{t}
  $\ensuremath{\gamma}\ensuremath{\gamma} and t\ensuremath{\gamma}j as case
  studies}}, \href{https://doi.org/10.1007/JHEP09(2021)155}{\emph{JHEP}
  {\bfseries 09} (2021) 155}
  [\href{https://arxiv.org/abs/2106.02059}{{\ttfamily 2106.02059}}].

\bibitem{Bertone:2022ktl}
V.~Bertone, M.~Cacciari, S.~Frixione, G.~Stagnitto, M.~Zaro and X.~Zhao,
  \emph{{Improving methods and predictions at high-energy e$^{+}$e$^{-}$
  colliders within collinear factorisation}},
  \href{https://doi.org/10.1007/JHEP10(2022)089}{\emph{JHEP} {\bfseries 10}
  (2022) 089} [\href{https://arxiv.org/abs/2207.03265}{{\ttfamily
  2207.03265}}].

\bibitem{Bredt:2022dmm}
P.M.~Bredt, W.~Kilian, J.~Reuter and P.~Stienemeier, \emph{{NLO electroweak
  corrections to multi-boson processes at a muon collider}},
  \href{https://doi.org/10.1007/JHEP12(2022)138}{\emph{JHEP} {\bfseries 12}
  (2022) 138} [\href{https://arxiv.org/abs/2208.09438}{{\ttfamily
  2208.09438}}].

\bibitem{Frixione:2014qaa}
S.~Frixione, V.~Hirschi, D.~Pagani, H.S.~Shao and M.~Zaro, \emph{{Weak
  corrections to Higgs hadroproduction in association with a top-quark pair}},
  \href{https://doi.org/10.1007/JHEP09(2014)065}{\emph{JHEP} {\bfseries 09}
  (2014) 065} [\href{https://arxiv.org/abs/1407.0823}{{\ttfamily 1407.0823}}].

\bibitem{Pagani:2016caq}
D.~Pagani, I.~Tsinikos and M.~Zaro, \emph{{The impact of the photon PDF and
  electroweak corrections on $t \bar{t}$ distributions}},
  \href{https://doi.org/10.1140/epjc/s10052-016-4318-z}{\emph{Eur. Phys. J. C}
  {\bfseries 76} (2016) 479}
  [\href{https://arxiv.org/abs/1606.01915}{{\ttfamily 1606.01915}}].

\bibitem{Frederix:2016ost}
R.~Frederix, S.~Frixione, V.~Hirschi, D.~Pagani, H.-S.~Shao and M.~Zaro,
  \emph{{The complete NLO corrections to dijet hadroproduction}},
  \href{https://doi.org/10.1007/JHEP04(2017)076}{\emph{JHEP} {\bfseries 04}
  (2017) 076} [\href{https://arxiv.org/abs/1612.06548}{{\ttfamily
  1612.06548}}].

\bibitem{Czakon:2017wor}
M.~Czakon, D.~Heymes, A.~Mitov, D.~Pagani, I.~Tsinikos and M.~Zaro,
  \emph{{Top-pair production at the LHC through NNLO QCD and NLO EW}},
  \href{https://doi.org/10.1007/JHEP10(2017)186}{\emph{JHEP} {\bfseries 10}
  (2017) 186} [\href{https://arxiv.org/abs/1705.04105}{{\ttfamily
  1705.04105}}].

\bibitem{Frederix:2017wme}
R.~Frederix, D.~Pagani and M.~Zaro, \emph{{Large NLO corrections in
  $t\bar{t}W^{\pm}$ and $t\bar{t}t\bar{t}$ hadroproduction from supposedly
  subleading EW contributions}},
  \href{https://doi.org/10.1007/JHEP02(2018)031}{\emph{JHEP} {\bfseries 02}
  (2018) 031} [\href{https://arxiv.org/abs/1711.02116}{{\ttfamily
  1711.02116}}].

\bibitem{Broggio:2019ewu}
A.~Broggio, A.~Ferroglia, R.~Frederix, D.~Pagani, B.D.~Pecjak and I.~Tsinikos,
  \emph{{Top-quark pair hadroproduction in association with a heavy boson at
  NLO+NNLL including EW corrections}},
  \href{https://doi.org/10.1007/JHEP08(2019)039}{\emph{JHEP} {\bfseries 08}
  (2019) 039} [\href{https://arxiv.org/abs/1907.04343}{{\ttfamily
  1907.04343}}].

\bibitem{Frederix:2019ubd}
R.~Frederix, D.~Pagani and I.~Tsinikos, \emph{{Precise predictions for
  single-top production: the impact of EW corrections and QCD shower on the
  $t$-channel signature}},
  \href{https://doi.org/10.1007/JHEP09(2019)122}{\emph{JHEP} {\bfseries 09}
  (2019) 122} [\href{https://arxiv.org/abs/1907.12586}{{\ttfamily
  1907.12586}}].

\bibitem{Pagani:2020rsg}
D.~Pagani, H.-S.~Shao and M.~Zaro, \emph{{RIP $ Hb\overline{b} $: how other
  Higgs production modes conspire to kill a rare signal at the LHC}},
  \href{https://doi.org/10.1007/JHEP11(2020)036}{\emph{JHEP} {\bfseries 11}
  (2020) 036} [\href{https://arxiv.org/abs/2005.10277}{{\ttfamily
  2005.10277}}].

\bibitem{Pagani:2020mov}
D.~Pagani, I.~Tsinikos and E.~Vryonidou, \emph{{NLO QCD+EW predictions for
  $tHj$ and $tZj$ production at the LHC}},
  \href{https://doi.org/10.1007/JHEP08(2020)082}{\emph{JHEP} {\bfseries 08}
  (2020) 082} [\href{https://arxiv.org/abs/2006.10086}{{\ttfamily
  2006.10086}}].

\bibitem{Maltoni:2024wyh}
F.~Maltoni, D.~Pagani and S.~Tentori, \emph{{Top-quark pair production as a
  probe of light top-philic scalars and anomalous Higgs interactions}},
  \href{https://doi.org/10.1007/JHEP09(2024)098}{\emph{JHEP} {\bfseries 09}
  (2024) 098} [\href{https://arxiv.org/abs/2406.06694}{{\ttfamily
  2406.06694}}].

\bibitem{ElFaham:2024egs}
H.~El~Faham, K.~Mimasu, D.~Pagani, C.~Severi, E.~Vryonidou and M.~Zaro,
  \emph{{Electroweak corrections in the SMEFT: four-fermion operators at high
  energies}},  \href{https://arxiv.org/abs/2412.16076}{{\ttfamily 2412.16076}}.

\bibitem{Frixione:1995ms}
S.~Frixione, Z.~Kunszt and A.~Signer, \emph{{Three jet cross-sections to
  next-to-leading order}},
  \href{https://doi.org/10.1016/0550-3213(96)00110-1}{\emph{Nucl. Phys. B}
  {\bfseries 467} (1996) 399}
  [\href{https://arxiv.org/abs/hep-ph/9512328}{{\ttfamily hep-ph/9512328}}].

\bibitem{Frixione:1997np}
S.~Frixione, \emph{{A General approach to jet cross-sections in QCD}},
  \href{https://doi.org/10.1016/S0550-3213(97)00574-9}{\emph{Nucl. Phys. B}
  {\bfseries 507} (1997) 295}
  [\href{https://arxiv.org/abs/hep-ph/9706545}{{\ttfamily hep-ph/9706545}}].

\bibitem{Frederix:2009yq}
R.~Frederix, S.~Frixione, F.~Maltoni and T.~Stelzer, \emph{{Automation of
  next-to-leading order computations in QCD: The FKS subtraction}},
  \href{https://doi.org/10.1088/1126-6708/2009/10/003}{\emph{JHEP} {\bfseries
  10} (2009) 003} [\href{https://arxiv.org/abs/0908.4272}{{\ttfamily
  0908.4272}}].

\bibitem{Frederix:2016rdc}
R.~Frederix, S.~Frixione, A.S.~Papanastasiou, S.~Prestel and P.~Torrielli,
  \emph{{Off-shell single-top production at NLO matched to parton showers}},
  \href{https://doi.org/10.1007/JHEP06(2016)027}{\emph{JHEP} {\bfseries 06}
  (2016) 027} [\href{https://arxiv.org/abs/1603.01178}{{\ttfamily
  1603.01178}}].

\bibitem{Ossola:2006us}
G.~Ossola, C.G.~Papadopoulos and R.~Pittau, \emph{{Reducing full one-loop
  amplitudes to scalar integrals at the integrand level}},
  \href{https://doi.org/10.1016/j.nuclphysb.2006.11.012}{\emph{Nucl. Phys. B}
  {\bfseries 763} (2007) 147}
  [\href{https://arxiv.org/abs/hep-ph/0609007}{{\ttfamily hep-ph/0609007}}].

\bibitem{Passarino:1978jh}
G.~Passarino and M.J.G.~Veltman, \emph{{One Loop Corrections for e+ e-
  Annihilation Into mu+ mu- in the Weinberg Model}},
  \href{https://doi.org/10.1016/0550-3213(79)90234-7}{\emph{Nucl. Phys. B}
  {\bfseries 160} (1979) 151}.

\bibitem{Davydychev:1991va}
A.I.~Davydychev, \emph{{A Simple formula for reducing Feynman diagrams to
  scalar integrals}},
  \href{https://doi.org/10.1016/0370-2693(91)91715-8}{\emph{Phys. Lett. B}
  {\bfseries 263} (1991) 107}.

\bibitem{Denner:2005nn}
A.~Denner and S.~Dittmaier, \emph{{Reduction schemes for one-loop tensor
  integrals}},
  \href{https://doi.org/10.1016/j.nuclphysb.2005.11.007}{\emph{Nucl. Phys. B}
  {\bfseries 734} (2006) 62}
  [\href{https://arxiv.org/abs/hep-ph/0509141}{{\ttfamily hep-ph/0509141}}].

\bibitem{Mastrolia:2012bu}
P.~Mastrolia, E.~Mirabella and T.~Peraro, \emph{{Integrand reduction of
  one-loop scattering amplitudes through Laurent series expansion}},
  \href{https://doi.org/10.1007/JHEP11(2012)128}{\emph{JHEP} {\bfseries 06}
  (2012) 095} [\href{https://arxiv.org/abs/1203.0291}{{\ttfamily 1203.0291}}].

\bibitem{Ossola:2007ax}
G.~Ossola, C.G.~Papadopoulos and R.~Pittau, \emph{{CutTools: A Program
  implementing the OPP reduction method to compute one-loop amplitudes}},
  \href{https://doi.org/10.1088/1126-6708/2008/03/042}{\emph{JHEP} {\bfseries
  03} (2008) 042} [\href{https://arxiv.org/abs/0711.3596}{{\ttfamily
  0711.3596}}].

\bibitem{ShaoIREGI}
H.-S.~Shao, \emph{{IREGI user's manual}}, {\emph{Unpublished} }.

\bibitem{Peraro:2014cba}
T.~Peraro, \emph{{Ninja: Automated Integrand Reduction via Laurent Expansion
  for One-Loop Amplitudes}},
  \href{https://doi.org/10.1016/j.cpc.2014.06.017}{\emph{Comput. Phys. Commun.}
  {\bfseries 185} (2014) 2771}
  [\href{https://arxiv.org/abs/1403.1229}{{\ttfamily 1403.1229}}].

\bibitem{Hirschi:2016mdz}
V.~Hirschi and T.~Peraro, \emph{{Tensor integrand reduction via Laurent
  expansion}}, \href{https://doi.org/10.1007/JHEP06(2016)060}{\emph{JHEP}
  {\bfseries 06} (2016) 060}
  [\href{https://arxiv.org/abs/1604.01363}{{\ttfamily 1604.01363}}].

\bibitem{Denner:2014gla}
A.~Denner, S.~Dittmaier and L.~Hofer, \emph{{COLLIER - A fortran-library for
  one-loop integrals}}, \href{https://doi.org/10.22323/1.211.0071}{\emph{PoS}
  {\bfseries LL2014} (2014) 071}
  [\href{https://arxiv.org/abs/1407.0087}{{\ttfamily 1407.0087}}].

\bibitem{Denner:2016kdg}
A.~Denner, S.~Dittmaier and L.~Hofer, \emph{{Collier: a fortran-based Complex
  One-Loop LIbrary in Extended Regularizations}},
  \href{https://doi.org/10.1016/j.cpc.2016.10.013}{\emph{Comput. Phys. Commun.}
  {\bfseries 212} (2017) 220}
  [\href{https://arxiv.org/abs/1604.06792}{{\ttfamily 1604.06792}}].

\bibitem{Hirschi:2011pa}
V.~Hirschi, R.~Frederix, S.~Frixione, M.V.~Garzelli, F.~Maltoni and R.~Pittau,
  \emph{{Automation of one-loop QCD corrections}},
  \href{https://doi.org/10.1007/JHEP05(2011)044}{\emph{JHEP} {\bfseries 05}
  (2011) 044} [\href{https://arxiv.org/abs/1103.0621}{{\ttfamily 1103.0621}}].

\bibitem{Frixione:2002ik}
S.~Frixione and B.R.~Webber, \emph{{Matching NLO QCD computations and parton
  shower simulations}},
  \href{https://doi.org/10.1088/1126-6708/2002/06/029}{\emph{JHEP} {\bfseries
  06} (2002) 029} [\href{https://arxiv.org/abs/hep-ph/0204244}{{\ttfamily
  hep-ph/0204244}}].

\bibitem{Czakon:2017lgo}
M.~Czakon, D.~Heymes, A.~Mitov, D.~Pagani, I.~Tsinikos and M.~Zaro,
  \emph{{Top-quark charge asymmetry at the LHC and Tevatron through NNLO QCD
  and NLO EW}}, \href{https://doi.org/10.1103/PhysRevD.98.014003}{\emph{Phys.
  Rev. D} {\bfseries 98} (2018) 014003}
  [\href{https://arxiv.org/abs/1711.03945}{{\ttfamily 1711.03945}}].

\bibitem{Frederix:2021agh}
R.~Frederix and I.~Tsinikos, \emph{{On improving NLO merging for $
  \mathrm{t}\overline{\mathrm{t}}\mathrm{W} $ production}},
  \href{https://doi.org/10.1007/JHEP11(2021)029}{\emph{JHEP} {\bfseries 11}
  (2021) 029} [\href{https://arxiv.org/abs/2108.07826}{{\ttfamily
  2108.07826}}].

\bibitem{Frederix:2021zsh}
R.~Frederix, I.~Tsinikos and T.~Vitos, \emph{{Probing the spin correlations of
  $t{\bar{t}} $ production at NLO QCD+EW}},
  \href{https://doi.org/10.1140/epjc/s10052-021-09612-9}{\emph{Eur. Phys. J. C}
  {\bfseries 81} (2021) 817}
  [\href{https://arxiv.org/abs/2105.11478}{{\ttfamily 2105.11478}}].

\bibitem{Frixione:2019lga}
S.~Frixione, \emph{{Initial conditions for electron and photon structure and
  fragmentation functions}},
  \href{https://doi.org/10.1007/JHEP11(2019)158}{\emph{JHEP} {\bfseries 11}
  (2019) 158} [\href{https://arxiv.org/abs/1909.03886}{{\ttfamily
  1909.03886}}].

\bibitem{Bertone:2019hks}
V.~Bertone, M.~Cacciari, S.~Frixione and G.~Stagnitto, \emph{{The partonic
  structure of the electron at the next-to-leading logarithmic accuracy in
  QED}}, \href{https://doi.org/10.1007/JHEP03(2020)135}{\emph{JHEP} {\bfseries
  03} (2020) 135} [\href{https://arxiv.org/abs/1911.12040}{{\ttfamily
  1911.12040}}].

\bibitem{Frixione:2012wtz}
S.~Frixione, \emph{{On factorisation schemes for the electron parton
  distribution functions in QED}},
  \href{https://doi.org/10.1007/JHEP07(2021)180}{\emph{JHEP} {\bfseries 07}
  (2021) 180} [\href{https://arxiv.org/abs/2105.06688}{{\ttfamily
  2105.06688}}].

\bibitem{Frixione:2021zdp}
S.~Frixione, O.~Mattelaer, M.~Zaro and X.~Zhao, \emph{{Lepton collisions in
  MadGraph5\_aMC@NLO}},  \href{https://arxiv.org/abs/2108.10261}{{\ttfamily
  2108.10261}}.

\bibitem{BuarqueFranzosi:2021wrv}
D.~Buarque~Franzosi et~al., \emph{{Vector boson scattering processes: Status
  and prospects}},
  \href{https://doi.org/10.1016/j.revip.2022.100071}{\emph{Rev. Phys.}
  {\bfseries 8} (2022) 100071}
  [\href{https://arxiv.org/abs/2106.01393}{{\ttfamily 2106.01393}}].

\bibitem{Bothmann:2021led}
E.~Bothmann, D.~Napoletano, M.~Sch\"onherr, S.~Schumann and S.L.~Villani,
  \emph{{Higher-order EW corrections in ZZ and ZZj production at the LHC}},
  \href{https://doi.org/10.1007/JHEP06(2022)064}{\emph{JHEP} {\bfseries 06}
  (2022) 064} [\href{https://arxiv.org/abs/2111.13453}{{\ttfamily
  2111.13453}}].

\bibitem{Dawson:1984gx}
S.~Dawson, \emph{{The Effective W Approximation}},
  \href{https://doi.org/10.1016/0550-3213(85)90038-0}{\emph{Nucl. Phys. B}
  {\bfseries 249} (1985) 42}.

\bibitem{Kane:1984bb}
G.L.~Kane, W.W.~Repko and W.B.~Rolnick, \emph{{The Effective W+-, Z0
  Approximation for High-Energy Collisions}},
  \href{https://doi.org/10.1016/0370-2693(84)90105-9}{\emph{Phys. Lett. B}
  {\bfseries 148} (1984) 367}.

\bibitem{vonWeizsacker:1934nji}
C.F.~von Weizsacker, \emph{{Radiation emitted in collisions of very fast
  electrons}}, \href{https://doi.org/10.1007/BF01333110}{\emph{Z. Phys.}
  {\bfseries 88} (1934) 612}.

\bibitem{Williams:1934ad}
E.J.~Williams, \emph{{Nature of the high-energy particles of penetrating
  radiation and status of ionization and radiation formulae}},
  \href{https://doi.org/10.1103/PhysRev.45.729}{\emph{Phys. Rev.} {\bfseries
  45} (1934) 729}.

\bibitem{Kilian:2007gr}
W.~Kilian, T.~Ohl and J.~Reuter, \emph{{WHIZARD: Simulating Multi-Particle
  Processes at LHC and ILC}},
  \href{https://doi.org/10.1140/epjc/s10052-011-1742-y}{\emph{Eur. Phys. J. C}
  {\bfseries 71} (2011) 1742}
  [\href{https://arxiv.org/abs/0708.4233}{{\ttfamily 0708.4233}}].

\bibitem{Cacciari:2011ma}
M.~Cacciari, G.P.~Salam and G.~Soyez, \emph{{FastJet User Manual}},
  \href{https://doi.org/10.1140/epjc/s10052-012-1896-2}{\emph{Eur. Phys. J. C}
  {\bfseries 72} (2012) 1896}
  [\href{https://arxiv.org/abs/1111.6097}{{\ttfamily 1111.6097}}].

\bibitem{Dokshitzer:1997in}
Y.L.~Dokshitzer, G.D.~Leder, S.~Moretti and B.R.~Webber, \emph{{Better jet
  clustering algorithms}},
  \href{https://doi.org/10.1088/1126-6708/1997/08/001}{\emph{JHEP} {\bfseries
  08} (1997) 001} [\href{https://arxiv.org/abs/hep-ph/9707323}{{\ttfamily
  hep-ph/9707323}}].

\bibitem{Kuhn:1999nn}
J.H.~Kuhn, A.A.~Penin and V.A.~Smirnov, \emph{{Summing up subleading Sudakov
  logarithms}}, \href{https://doi.org/10.1007/s100520000462}{\emph{Eur. Phys.
  J. C} {\bfseries 17} (2000) 97}
  [\href{https://arxiv.org/abs/hep-ph/9912503}{{\ttfamily hep-ph/9912503}}].

\bibitem{Fadin:1999bq}
V.S.~Fadin, L.N.~Lipatov, A.D.~Martin and M.~Melles, \emph{{Resummation of
  double logarithms in electroweak high-energy processes}},
  \href{https://doi.org/10.1103/PhysRevD.61.094002}{\emph{Phys. Rev. D}
  {\bfseries 61} (2000) 094002}
  [\href{https://arxiv.org/abs/hep-ph/9910338}{{\ttfamily hep-ph/9910338}}].

\bibitem{Ciafaloni:1999ub}
P.~Ciafaloni and D.~Comelli, \emph{{Electroweak Sudakov form-factors and
  nonfactorizable soft QED effects at NLC energies}},
  \href{https://doi.org/10.1016/S0370-2693(00)00121-0}{\emph{Phys. Lett. B}
  {\bfseries 476} (2000) 49}
  [\href{https://arxiv.org/abs/hep-ph/9910278}{{\ttfamily hep-ph/9910278}}].

\bibitem{Beccaria:2000jz}
M.~Beccaria, F.M.~Renard and C.~Verzegnassi, \emph{{Top quark production at
  future lepton colliders in the asymptotic regime}},
  \href{https://doi.org/10.1103/PhysRevD.63.053013}{\emph{Phys. Rev. D}
  {\bfseries 63} (2001) 053013}
  [\href{https://arxiv.org/abs/hep-ph/0010205}{{\ttfamily hep-ph/0010205}}].

\bibitem{Hori:2000tm}
M.~Hori, H.~Kawamura and J.~Kodaira, \emph{{Electroweak Sudakov at two loop
  level}}, \href{https://doi.org/10.1016/S0370-2693(00)01027-3}{\emph{Phys.
  Lett. B} {\bfseries 491} (2000) 275}
  [\href{https://arxiv.org/abs/hep-ph/0007329}{{\ttfamily hep-ph/0007329}}].

\bibitem{Ciafaloni:2000df}
M.~Ciafaloni, P.~Ciafaloni and D.~Comelli, \emph{{Bloch-Nordsieck violating
  electroweak corrections to inclusive TeV scale hard processes}},
  \href{https://doi.org/10.1103/PhysRevLett.84.4810}{\emph{Phys. Rev. Lett.}
  {\bfseries 84} (2000) 4810}
  [\href{https://arxiv.org/abs/hep-ph/0001142}{{\ttfamily hep-ph/0001142}}].

\bibitem{Melles:2001ye}
M.~Melles, \emph{{Electroweak radiative corrections in high-energy processes}},
  \href{https://doi.org/10.1016/S0370-1573(02)00550-1}{\emph{Phys. Rept.}
  {\bfseries 375} (2003) 219}
  [\href{https://arxiv.org/abs/hep-ph/0104232}{{\ttfamily hep-ph/0104232}}].

\bibitem{Beenakker:2001kf}
W.~Beenakker and A.~Werthenbach, \emph{{Electroweak two loop Sudakov logarithms
  for on-shell fermions and bosons}},
  \href{https://doi.org/10.1016/S0550-3213(02)00171-2}{\emph{Nucl. Phys. B}
  {\bfseries 630} (2002) 3}
  [\href{https://arxiv.org/abs/hep-ph/0112030}{{\ttfamily hep-ph/0112030}}].

\bibitem{Pozzorini:2004rm}
S.~Pozzorini, \emph{{Next to leading mass singularities in two loop electroweak
  singlet form-factors}},
  \href{https://doi.org/10.1016/j.nuclphysb.2004.05.025}{\emph{Nucl. Phys. B}
  {\bfseries 692} (2004) 135}
  [\href{https://arxiv.org/abs/hep-ph/0401087}{{\ttfamily hep-ph/0401087}}].

\bibitem{Feucht:2004rp}
B.~Feucht, J.H.~Kuhn, A.A.~Penin and V.A.~Smirnov, \emph{{Two loop Sudakov
  form-factor in a theory with mass gap}},
  \href{https://doi.org/10.1103/PhysRevLett.93.101802}{\emph{Phys. Rev. Lett.}
  {\bfseries 93} (2004) 101802}
  [\href{https://arxiv.org/abs/hep-ph/0404082}{{\ttfamily hep-ph/0404082}}].

\bibitem{Jantzen:2005xi}
B.~Jantzen, J.H.~Kuhn, A.A.~Penin and V.A.~Smirnov, \emph{{Two-loop electroweak
  logarithms}}, \href{https://doi.org/10.1103/PhysRevD.74.019901}{\emph{Phys.
  Rev. D} {\bfseries 72} (2005) 051301}
  [\href{https://arxiv.org/abs/hep-ph/0504111}{{\ttfamily hep-ph/0504111}}].

\bibitem{Jantzen:2005az}
B.~Jantzen, J.H.~Kuhn, A.A.~Penin and V.A.~Smirnov, \emph{{Two-loop electroweak
  logarithms in four-fermion processes at high energy}},
  \href{https://doi.org/10.1016/j.nuclphysb.2005.10.010}{\emph{Nucl. Phys. B}
  {\bfseries 731} (2005) 188}
  [\href{https://arxiv.org/abs/hep-ph/0509157}{{\ttfamily hep-ph/0509157}}].

\bibitem{Jantzen:2006jv}
B.~Jantzen and V.A.~Smirnov, \emph{{The Two-loop vector form-factor in the
  Sudakov limit}}, \href{https://doi.org/10.1140/epjc/s2006-02583-9}{\emph{Eur.
  Phys. J. C} {\bfseries 47} (2006) 671}
  [\href{https://arxiv.org/abs/hep-ph/0603133}{{\ttfamily hep-ph/0603133}}].

\bibitem{Manohar:2012rs}
A.V.~Manohar and M.~Trott, \emph{{Electroweak Sudakov Corrections and the Top
  Quark Forward-Backward Asymmetry}},
  \href{https://doi.org/10.1016/j.physletb.2012.04.013}{\emph{Phys. Lett. B}
  {\bfseries 711} (2012) 313}
  [\href{https://arxiv.org/abs/1201.3926}{{\ttfamily 1201.3926}}].

\bibitem{Bauer:2017bnh}
C.W.~Bauer, N.~Ferland and B.R.~Webber, \emph{{Combining initial-state
  resummation with fixed-order calculations of electroweak corrections}},
  \href{https://doi.org/10.1007/JHEP04(2018)125}{\emph{JHEP} {\bfseries 04}
  (2018) 125} [\href{https://arxiv.org/abs/1712.07147}{{\ttfamily
  1712.07147}}].

\bibitem{Mangano:2016jyj}
M.L.~Mangano et~al., \emph{{Physics at a 100 TeV pp Collider: Standard Model
  Processes}},  \href{https://arxiv.org/abs/1607.01831}{{\ttfamily
  1607.01831}}.

\bibitem{Azzi:2019yne}
P.~Azzi et~al., \emph{{Report from Working Group 1}: {Standard Model Physics at
  the HL-LHC and HE-LHC}},
  \href{https://doi.org/10.23731/CYRM-2019-007.1}{\emph{CERN Yellow Rep.
  Monogr.} {\bfseries 7} (2019) 1}
  [\href{https://arxiv.org/abs/1902.04070}{{\ttfamily 1902.04070}}].

\bibitem{Bagdatova:2024aem}
A.G.~Bagdatova and S.P.~Baranov, \emph{{Polarization and kinematic properties
  of the splitting functions $q\to W^\pm +q'$ and $q\to Z^0 +q$}},
  \href{https://arxiv.org/abs/2404.10832}{{\ttfamily 2404.10832}}.

\bibitem{Frixione:1992pj}
S.~Frixione, P.~Nason and G.~Ridolfi, \emph{{Strong corrections to W Z
  production at hadron colliders}},
  \href{https://doi.org/10.1016/0550-3213(92)90668-2}{\emph{Nucl. Phys. B}
  {\bfseries 383} (1992) 3}.

\bibitem{Frixione:1993yp}
S.~Frixione, \emph{{A Next-to-leading order calculation of the cross-section
  for the production of W+ W- pairs in hadronic collisions}},
  \href{https://doi.org/10.1016/0550-3213(93)90435-R}{\emph{Nucl. Phys. B}
  {\bfseries 410} (1993) 280}.

\bibitem{Rubin:2010xp}
M.~Rubin, G.P.~Salam and S.~Sapeta, \emph{{Giant QCD K-factors beyond NLO}},
  \href{https://doi.org/10.1007/JHEP09(2010)084}{\emph{JHEP} {\bfseries 09}
  (2010) 084} [\href{https://arxiv.org/abs/1006.2144}{{\ttfamily 1006.2144}}].

\bibitem{Maltoni:2015ena}
F.~Maltoni, D.~Pagani and I.~Tsinikos, \emph{{Associated production of a
  top-quark pair with vector bosons at NLO in QCD: impact on $
  \mathrm{t}\overline{\mathrm{t}}\mathrm{H} $ searches at the LHC}},
  \href{https://doi.org/10.1007/JHEP02(2016)113}{\emph{JHEP} {\bfseries 02}
  (2016) 113} [\href{https://arxiv.org/abs/1507.05640}{{\ttfamily
  1507.05640}}].

\bibitem{Frixione:2023gmf}
S.~Frixione and G.~Stagnitto, \emph{{The muon parton distribution functions}},
  \href{https://doi.org/10.1007/JHEP12(2023)170}{\emph{JHEP} {\bfseries 12}
  (2023) 170} [\href{https://arxiv.org/abs/2309.07516}{{\ttfamily
  2309.07516}}].

\bibitem{Gribov:1972ri}
V.N.~Gribov and L.N.~Lipatov, \emph{{Deep inelastic e p scattering in
  perturbation theory}}, {\emph{Sov. J. Nucl. Phys.} {\bfseries 15} (1972)
  438}.

\bibitem{Lipatov:1974qm}
L.N.~Lipatov, \emph{{The parton model and perturbation theory}}, {\emph{Yad.
  Fiz.} {\bfseries 20} (1974) 181}.

\bibitem{Altarelli:1977zs}
G.~Altarelli and G.~Parisi, \emph{{Asymptotic Freedom in Parton Language}},
  \href{https://doi.org/10.1016/0550-3213(77)90384-4}{\emph{Nucl. Phys. B}
  {\bfseries 126} (1977) 298}.

\bibitem{Dokshitzer:1977sg}
Y.L.~Dokshitzer, \emph{{Calculation of the Structure Functions for Deep
  Inelastic Scattering and $e^+ e^-$ Annihilation by Perturbation Theory in
  Quantum Chromodynamics.}}, {\emph{Sov. Phys. JETP} {\bfseries 46} (1977)
  641}.

\bibitem{Skrzypek:1990qs}
M.~Skrzypek and S.~Jadach, \emph{{Exact and approximate solutions for the
  electron nonsinglet structure function in QED}},
  \href{https://doi.org/10.1007/BF01483573}{\emph{Z. Phys. C} {\bfseries 49}
  (1991) 577}.

\bibitem{Skrzypek:1992vk}
M.~Skrzypek, \emph{{Leading logarithmic calculations of QED corrections at
  LEP}}, {\emph{Acta Phys. Polon. B} {\bfseries 23} (1992) 135}.

\bibitem{Cacciari:1992pz}
M.~Cacciari, A.~Deandrea, G.~Montagna and O.~Nicrosini, \emph{{QED structure
  functions: A Systematic approach}},
  \href{https://doi.org/10.1209/0295-5075/17/2/007}{\emph{EPL} {\bfseries 17}
  (1992) 123}.

\bibitem{Denner:2003iy}
A.~Denner, S.~Dittmaier, M.~Roth and M.M.~Weber, \emph{{Electroweak radiative
  corrections to $e^+ e^- \to \nu \bar{\nu} H$}},
  \href{https://doi.org/10.1016/S0550-3213(03)00269-4}{\emph{Nucl. Phys. B}
  {\bfseries 660} (2003) 289}
  [\href{https://arxiv.org/abs/hep-ph/0302198}{{\ttfamily hep-ph/0302198}}].

\end{thebibliography}\endgroup

\end{document}